\documentclass[11pt,a4paper]{article}

\usepackage{jheppub}

\usepackage[english]{babel}

\usepackage{widebar}
\usepackage{subfigure}

\allowdisplaybreaks[2]

\newcommand{\rev}[1]{#1}

\newcommand{\half}{{\tfrac{1}{2}}}

\newcommand{\tvec}[1]{\boldsymbol{#1}}

\newcommand{\ms}{\mskip 1.5mu}
\newcommand{\bs}{\mskip -1.5mu}

\newcommand{\lsim}{\raisebox{-4pt}{%
    $\,\stackrel{\textstyle <}{\sim}\,$}}
\newcommand{\gsim}{\raisebox{-4pt}{%
    $\,\stackrel{\textstyle >}{\sim}\,$}}

\newcommand{\gev}{\operatorname{GeV}}
\newcommand{\tev}{\operatorname{TeV}}

\subheader{\hfill DESY 17-013, NIKHEF 2017-007}
\title{Double hard scattering without double counting}

\author[a]{Markus Diehl,}
\author[b]{Jonathan R. Gaunt}
\author[c]{and Kay Sch\"onwald}

\affiliation[a]{Deutsches Elektronen-Synchroton DESY, Notkestra{\ss}e~85,
  22607 Hamburg, Germany}
\affiliation[b]{Nikhef Theory Group and VU University Amsterdam, De
  Boelelaan 1081, 1081 HV Amsterdam, The Netherlands}
\affiliation[c]{Deutsches Elektronen-Synchrotron DESY, Platanenallee 6,
  15738 Zeuthen, Germany}

\emailAdd{markus.diehl@desy.de}
\emailAdd{jgaunt@nikhef.nl}
\emailAdd{kay.schoenwald@desy.de}

\abstract{Double parton scattering in proton-proton collisions includes
  kinematic regions in which two partons inside a proton originate from
  the perturbative splitting of a single parton.  This leads to a double
  counting problem between single and double hard scattering.  We present
  a solution to this problem, which allows for the definition of double
  parton distributions as operator matrix elements in a proton, and which
  can be used at higher orders in perturbation theory.  We show how the
  evaluation of double hard scattering in this framework can provide a
  rough estimate for the size of the higher-order contributions to single
  hard scattering that are affected by double counting.  In a numeric
  study, we identify situations in which these higher-order contributions
  must be explicitly calculated and included if one wants to attain an
  accuracy at which double hard scattering becomes relevant, and other
  situations where such contributions may be neglected.
}

\begin{document}

\maketitle

\section{Introduction}
\label{sec:intro}

The precise description of high-energy proton-proton collisions in QCD is
imperative for maximising the physics potential of the LHC and of possible
future hadron colliders.  An important issue in this context is to
understand the mechanism of double parton scattering (DPS), in which two
pairs of partons undergo a hard scattering in one and the same
proton-proton collision.  In many situations DPS is suppressed compared
with single parton scattering (SPS), but this suppression generically
becomes weaker with increasing collision energy.  For specific kinematics
or specific final states, DPS can become comparable to or even larger than
SPS.  An overview of recent experimental and theoretical activities in
this area can for instance be found in
\cite{Astalos:2015ivw,Proceedings:2016tff}.

Consider a DPS process $pp \to Y_1 + Y_2 + X$, where $Y_1$ and $Y_2$ are
observed particles or groups of particles produced in two separate
hard scattering processes, whilst $X$ denotes all unobserved particles in
the final state.  A cross section formula has been put forward long ago in
the framework of collinear factorisation, where the transverse momenta
$\tvec{q}_1$ and $\tvec{q}_2$ of $Y_1$ and $Y_2$ are integrated over
\cite{Paver:1982yp}.  A corresponding expression has been given in
\cite{Diehl:2011tt,Diehl:2011yj} for transverse momentum dependent (TMD)
factorisation, where $\tvec{q}_1$ and $\tvec{q}_2$ are small compared with
the hard scales in $Y_1$ and $Y_2$.

Inside a proton, the two partons that take part in the two hard scatters
can originate from the perturbative splitting of one parton.  The
relevance of this splitting mechanism for the evolution equations of
double parton distributions (DPDs) has been realised long ago
\cite{Kirschner:1979im,Shelest:1982dg} and studied more recently in
\cite{Snigirev:2003cq,Gaunt:2009re,Ceccopieri:2010kg}.  However, it was
only noted in \cite{Diehl:2011tt,Diehl:2011yj} that the same mechanism
dominates DPDs in the limit of small transverse distance between the two
partons, and that the splitting contribution leads to infinities when
inserted into the DPS cross section formula.  These infinities are closely
connected with double counting between DPS and SPS in particular Feynman
graphs, a problem that had been pointed out earlier in the context of
multi-jet production \cite{Cacciari:2009dp}.

Different ways of dealing with this issue have been proposed in
\cite{Blok:2011bu,Blok:2013bpa}, \cite{Ryskin:2011kk,Ryskin:2012qx} and
\cite{Manohar:2012pe}.  As discussed later in this paper, we find that
these proposals have shortcomings either of theoretical or of practical
nature.  In the present work we present an alternative scheme for
computing the cross section in a consistent way, including both DPS and
SPS (as well as other contributions).  Our scheme allows for a
nonperturbative definition of DPDs in terms of operator matrix elements,
and it is suitable for pushing the limit of theoretical accuracy to higher
orders in $\alpha_s$.

Our paper is structured as follows.  In section~\ref{sec:scene} we recall
the theoretical framework for describing DPS and specify the theoretical
problems mentioned above.  The short-distance behaviour of DPDs is
discussed in section~\ref{sec:short-dist}, since it is essential for the
scheme we propose in section~\ref{sec:scheme}.  We first show how this
scheme works at leading order (LO) in $\alpha_s$, before giving examples
of its application at next-to-leading order (NLO) in
section~\ref{sec:higher-order}.  Collinear DPDs evolve according to DGLAP
equations, and in section~\ref{sec:dglap} we discuss several consequences
of this scale evolution.  Our scheme is naturally formulated with DPDs
that depend on the transverse distance $\tvec{y}$ between the two partons,
but we show in section~\ref{sec:mom-space-dpds} how one may instead use
DPDs depending on the transverse momentum conjugate to $\tvec{y}$.  This
allows us to compare our results with those of
\cite{Blok:2011bu,Blok:2013bpa} and \cite{Ryskin:2011kk,Ryskin:2012qx},
which we do in section~\ref{sec:compare}.  Whilst the focus of the present
work is theoretical rather than phenomenological, we give in
section~\ref{sec:numerics} some quantitative illustrations of our scheme,
obtained with a relatively simple ansatz for the DPDs.  We summarise our
findings in section~\ref{sec:summary}.  Some Fourier integrals required in
the main text are given in an appendix.


\section{Setting the scene}
\label{sec:scene}

In this section we recall theoretical issues originating from the
perturbative splitting mechanism in double parton distributions, namely
the appearance of ultraviolet divergences in the naive cross section for
double parton scattering, the problem of double counting between DPS and
single parton scattering, and the treatment of the so-called 2v1 (two
versus one) contributions to DPS.\footnote{%
We follow here the nomenclature of \cite{Gaunt:2012dd}.  The 2v1
contribution is referred to as $4 \times 2$ in \cite{Diehl:2011yj}, as
$3\to 4$ in \cite{Blok:2011bu,Blok:2013bpa} (where four-jet production is
considered), and as $1 \times 2$ in \cite{Ryskin:2011kk,Ryskin:2012qx}.}
We also give some basic definitions and results.  Throughout this work we
use light-cone coordinates $v^{\pm} = (v^0 + v^3)/\sqrt{2}$ for any
four-vector $v$.  We write $\tvec{v} = (v^1, v^2)$ for the transverse
components and $v = \sqrt{\tvec{v}^2}$ for the length of the transverse
vector.

Since the perturbative splitting mechanism in DPDs leads to issues in the
ultraviolet (UV) region, renormalisation plays a crucial role in our
context.  We work in dimensional regularisation and extend the definitions
of \cite{Diehl:2011yj} to $D= 4-2\epsilon$ dimensions, changing phase
space factors $(2\pi)^4$ into $(2\pi)^D$.  Bare two-parton distributions
are then given by
\begin{align}
  \label{dpd-y-def}
F_{\text{bare}}(x_i,\tvec{z}_i,\tvec{y})
 &= 2p^+\!\! \int dy^-
    \int \frac{dz_1^-}{2\pi}\, \frac{dz_2^-}{2\pi}\,
  e^{i (x_1^{} z_1^- + x_2^{} z_2^-) p^+}
  \bigl\langle p \ms|\ms
     \mathcal{O}_2(0,z_2)\, \mathcal{O}_1(y,z_1) \ms|\ms
  p \bigr\rangle
\end{align}
with twist-two operators
\begin{align}
\mathcal{O}_i(y,z_i) &= \bar{q}\bigl( y - \half z_i \bigr)\ms 
    W^\dagger\bigl( y - \half z_i \bigr) \Gamma_i\ms
    W\bigl( y+ \half z_i \bigr)\ms q\bigl( y + \half z_i \bigr)
 \big|_{z_i^+ = y_{\phantom{z}}^+ = 0}
\end{align}
for quark distributions, where $q$ denotes the bare field.  $W(z)$ is a
Wilson line from $z$ to infinity along a prescribed path, which we do not
recall here.  The Dirac matrices $\Gamma_i$ select different quark
polarisations.  Analogous definitions hold for distributions involving
gluons, with quark fields replaced by gluon field strength operators.  For
ease of writing, we omit colour labels on the operators and distributions
throughout this work, bearing in mind that different colour couplings are
possible for the four parton fields in
\eqref{dpd-y-def}.

In the process of deriving factorisation, one finds that the proton matrix
element in \eqref{dpd-y-def} needs to be multiplied by a combination of
so-called soft factors, which are vacuum expectation values of products of
Wilson lines.  More information for the case of single parton
distributions can be found in chapter~13.7 of \cite{Collins:2011zzd} and
in the recent overview~\cite{Rogers:2015sqa}.  A brief account for DPDs is
given in section~2.1 of \cite{Diehl:2015bca}, and a more detailed
discussion will be provided in \cite{Buffing:2016wip}.  The product
$F_{\text{bare}} \times \{\text{soft factors}\}$ depends on a parameter
$\zeta$ which regulates rapidity divergences.  \rev{A scheme in which a
  soft factor appears explicitly in the cross section formula was
  presented in \cite{Manohar:2012jr}.  Since the treatment of soft gluons
  does not affect parton splitting in any special way, we will not discuss
  it further in the present paper.  Correspondingly, we will suppress the
  argument $\zeta$ in all DPDs.}

As a final step one performs UV renormalisation, which we assume to be
done in the $\overline{\text{MS}}$ scheme.  The DPDs obtained from
\eqref{dpd-y-def} are appropriate for TMD factorisation, with the
transverse positions $\tvec{z}_i$ being Fourier conjugate to the
transverse parton momenta that determine the final state kinematics.
These distributions renormalise multiplicatively, with one
renormalisation factor for each product of a quark field (or gluon
field strength) with the Wilson line at the same transverse position
and one factor for each pair of Wilson lines at the same transverse
position in the soft factors.  Denoting the product of these factors
with $Z$, one obtains the final DPD as
$F = \lim_{\epsilon\to 0} \bigl( Z \times F_{\text{bare}} \times
\{\text{soft factors}\} \bigr)$.
It obeys a renormalisation group equation which is a straightforward
generalisation of the one for TMDs (given e.g.\ in
~\cite{Rogers:2015sqa}).

The DPDs needed for collinear factorisation are obtained by
setting $\tvec{z}_i = \tvec{0}$ in $F_{\text{bare}}$ and in the
associated soft factors, before renormalisation.  Setting
$\tvec{z}_i = \tvec{0}$ introduces ultraviolet divergences in the
operators $\mathcal{O}_1$ and $\mathcal{O}_2$, and in the associated
soft factors.  The renormalised DPDs are then obtained as
$F = \lim_{\epsilon\to 0} \bigl( Z_1 \otimes Z_2 \otimes
F_{\text{bare}} \times \{\text{soft factors}\} \bigr)$,
where $Z_i$ renormalises the operators associated with parton $i$ and
where the convolution products are in the momentum fractions $x_i$.
In the colour singlet channel, where both operators $\mathcal{O}_i$ in
\eqref{dpd-y-def} are colour singlets, the soft factors reduce to
unity and one obtains the renormalised twist-two operators that appear
in single parton densities.

Since the operators associated with partons $1$ and $2$ renormalise
independently (both for the TMD and the collinear case) one may choose
different renormalisation scales $\mu_1$ and $\mu_2$ in each of them.
This is useful when the two hard subprocesses in double parton
scattering have widely different hard scales.  In particular, one can
then approach the kinematics of the so-called underlying event, with a
very hard scattering at scale $\mu_1$ and additional jet production at
a much lower scale $\mu_2$ (of course $\mu_2$ needs to remain in the
perturbative region for our factorisation approach to be justified).

With different scales $\mu_1, \mu_2$ in the collinear DPDs, we have a
homogeneous evolution equation
\begin{align}
  \label{DGLAP-mu1}
& \frac{d}{d\log\mu_1^2}\,
   F_{a_1 a_2}(x_1,x_2,\tvec{y}; \mu_1,\mu_2)
\nonumber \\
 &\qquad = \sum_{b_1} \int_{x_1}^{1-x_2} \frac{dx_1'}{x_1'}\;
      P_{a_1 b_1}\Bigl( \frac{x_1}{x_1'}, \alpha_s(\mu_1^{}) \Bigr)\,
      F_{b_1 a_2}(x_1',x_2^{},\tvec{y}; \mu_1,\mu_2)
\end{align}
in $\mu_1$ and its analogue for $\mu_2$.  For colour singlet DPDs, the
kernels $P_{a_1 b_1}$ on the r.h.s.\ are the usual DGLAP splitting
functions for single parton densities.  In colour non-singlet channels,
both the DPDs and the splitting kernels have an additional dependence on
the rapidity parameter $\zeta$ \cite{Buffing:2016wip}.  \rev{Following the
  notation of \cite{Diehl:2011yj}, the labels $a_i, b_i$ in
  \eqref{DGLAP-mu1} denote both the species and the polarisation of the
  partons.  The relevant labels are $q, \Delta q, \delta q$ for
  unpolarised, longitudinally polarised and transversely polarised quarks,
  likewise $\bar{q}, \Delta\bar{q}, \delta\bar{q}$ for antiquarks, and $g,
  \Delta g, \delta g$ for unpolarised, longitudinally polarised and
  linearly polarised gluons, respectively.  Note that the polarisations of
  the two partons can be correlated with each other, even in an
  unpolarised proton.}

To simplify our presentation, we will consider the production $pp\to V_1 +
V_2 + X$ of two electroweak gauge bosons $V_i = \gamma^*, Z, W$.  Our
results readily generalise to other processes for which DPS factorisation
can be established; in the case of TMD factorisation this requires that
the produced particles are colour singlets.  We denote the four-momenta of
the two bosons by $q_i$, their squared invariant masses by $Q_i^2 =
q_i^\mu\ms q_{i \mu}^{}$ and their rapidities by $Y_i = \half \log(q_i^+
/q_i^-)$.  We work in the proton-proton centre-of-mass frame, taking the
proton with momentum $p$ ($\bar{p}$) to move in the positive (negative)
$3$ direction.  Furthermore we define
\begin{align}
  \label{little-x-def}
x_i &= \sqrt{\frac{Q_i^2}{s}}\, e^{Y_i} \,,
&
\bar{x}_i &= \sqrt{\frac{Q_i^2}{s}}\, e^{-Y_i} \,,
\end{align}
with $s = (p+\bar{p})^2$.  For the phase space of each gauge boson we
have
\begin{align}
d^4 q_i &= \frac{s}{2}\, dx_i^{}\, d\bar{x}_i^{}\, d^2\tvec{q}_i
         = \frac{1}{2}\, dY_i^{}\, dQ_i^2\, d^2\tvec{q}_i \,.
\end{align}
The ``naive'' cross section formulae (not taking into account the UV
problems discussed below) read
\begin{align}
  \label{TMD-Xsect-naive}
& \frac{d\sigma_{\text{DPS}}}{dx_1\, dx_2\,
          d\bar{x}_1\, d\bar{x}_2\, d^2\tvec{q}_1\, d^2\tvec{q}_2}
  = \frac{1}{C}\, \sum_{a_1 a_2 b_1 b_2} \!\!\!
     \hat{\sigma}_{a_1 b_1}(Q_1^2, \mu_1^2)\,
     \hat{\sigma}_{a_2 b_2}(Q_2^2, \mu_2^2)\,
\nonumber \\
 & \qquad \times
     \int \frac{d^2\tvec{z}_1}{(2\pi)^2}\,
          \frac{d^2\tvec{z}_2}{(2\pi)^2}\; d^2\tvec{y}\;
       e^{-i (\tvec{q}_1^{} \tvec{z}_1^{} +\tvec{q}_2^{} \tvec{z}_2^{})}\,
     F_{b_1 b_2}(\bar{x}_i,\tvec{z}_i,\tvec{y};\mu_i)\,
     F_{a_1 a_2}(x_i,\tvec{z}_i,\tvec{y};\mu_i)
\end{align}
for TMD factorisation and
\begin{align}
  \label{coll-Xsect-naive}
& \frac{d\sigma_{\text{DPS}}}{dx_1\, dx_2\,
          d\bar{x}_1\, d\bar{x}_2}
  = \frac{1}{C}\, \sum_{a_1 a_2 b_1 b_2}
    \int_{x_1}^{1-x_2} \frac{dx_1'}{x_1'} 
    \int_{x_2}^{1-x_1'} \frac{dx_2'}{x_2'}
    \int_{\bar{x}_1}^{1-\bar{x}_2} \frac{d\bar{x}_1'}{\bar{x}_1'} 
    \int_{\bar{x}_2}^{1-\bar{x}_1'} \frac{d\bar{x}_2'}{\bar{x}_2'}
\nonumber \\
 & \qquad \times
     \hat{\sigma}_{a_1 b_1}(x_1' \bar{x}_1' s, \mu_1^2)\,
     \hat{\sigma}_{a_2 b_2}(x_2' \bar{x}_2'\ms s, \mu_2^2)\,
     \int d^2\tvec{y}\;
     F_{b_1 b_2}(\bar{x}_i',\tvec{y};\mu_i)\,
     F_{a_1 a_2}(x_i',\tvec{y};\mu_i)
\end{align}
for collinear factorisation.  The combinatorial factor $C$ is $1$ if the
observed final states of the hard scatters are different and $2$ if they
are identical.  For simplicity we will consider the case $C=1$ throughout
this paper, unless mentioned otherwise.  As explained in section~2.2.1 of
\cite{Diehl:2011yj}, there are further contributions with DPDs that
describe the interference of different parton species.  They can be
discussed in full analogy to the contributions given in
\eqref{TMD-Xsect-naive} or \eqref{coll-Xsect-naive}, and we do not treat
them explicitly in the present work for ease of notation.

The hard scattering cross sections $\hat{\sigma}_i$ in
\eqref{TMD-Xsect-naive} are for the exclusive final state $V_i$ with
transverse momentum $\tvec{q}_i$, which must satisfy $q_i \ll Q_i$.
By contrast, $\hat{\sigma}_i$ in \eqref{coll-Xsect-naive} is
integrated over $\tvec{q}_i$ and inclusive for final states $V_i + X$.
At leading order in $\alpha_s$ it contains $\delta$ functions that
enforce $x_i' = x_i^{}$ and $\bar{x}_i' = \bar{x}_i^{}$.  The
subtractions for collinear and soft regions in $\hat{\sigma}_i$ are
different in the two factorisation frameworks, but in both cases they
lead to a dependence on the factorisation scale $\mu_i$ that cancels
against the $\mu_i$ dependence of the DPDs, up to powers in $\alpha_s$
beyond the accuracy of the calculation.  This happens separately for
the two hard scatters ($i=1,2$) and by construction works exactly as
in the case of single hard scattering.

As was pointed out in \cite{Diehl:2011yj}, the framework discussed so far
suffers from problems in the region of small transverse distances between
the two partons in a DPD.  The leading behaviour of the collinear
distributions $F(x_i, \tvec{y})$ at small $y$ can be computed from the
perturbative splitting of one parton into two and gives a behaviour like
$y^{-2}$ up to logarithmic corrections.  When inserted in the
factorisation formula \eqref{coll-Xsect-naive} this gives a quadratically
divergent integral at small $y$, which clearly signals an inappropriate
treatment of the ultraviolet region.  As we will review in
section~\ref{sec:tmd-split}, the short-distance behaviour of the
distributions $F(x_i, \tvec{z}_i, \tvec{y})$ is less singular but still
leads to logarithmic divergences in the TMD factorisation formula
\eqref{TMD-Xsect-naive}.

Instead of using DPDs depending on transverse positions, one may
Fourier transform them to transverse momentum space, integrating
\begin{align}
 &  \frac{1}{(2\pi)^{4-4\epsilon}} 
  \int d^{2-2\epsilon} \tvec{z}_1\, d^{2-2\epsilon} \tvec{z}_2\;
    e^{-i (\tvec{z}_1 \tvec{k}_1 + \tvec{z}_2 \tvec{k}_2)}
\intertext{for TMDs and}
   \label{Delta-FT}
 & \int d^{2-2\epsilon}\tvec{y}\; e^{i\tvec{y}\tvec{\Delta}}
\end{align}
for TMDs and collinear distributions.  For collinear distributions, this
transformation must be made before subtracting UV divergences and
setting $\epsilon\to 0$: with $F(x_i, \tvec{y}) \sim y^{-2}$ in $D=4$
dimensions, the Fourier integral \eqref{Delta-FT} would be logarithmically
divergent at $y=0$.  In $D= 4-2\epsilon$ dimensions this singularity turns
into a pole in $1/\epsilon$, which requires an additional subtraction as
we will review in section~\ref{sec:mom-space-dpds}.  Rather than being
associated with the individual operators $\mathcal{O}_1$ and
$\mathcal{O}_2$, this subtraction is related to the singularity arising
when the transverse distance $y$ between the two operators goes to zero.
It leads to an \emph{inhomogeneous} evolution equation for the DPD $F(x_i,
\tvec{\Delta})$ in momentum space, which has been extensively discussed in
the literature \cite{Kirschner:1979im,Shelest:1982dg,Snigirev:2003cq,
  Gaunt:2009re,Ceccopieri:2010kg} for the case $\tvec{\Delta} = \tvec{0}$.
Notice, however, that this extra $\mu$ dependence does \emph{not} cancel
in the cross section when \eqref{coll-Xsect-naive} is rewritten in
transverse momentum space.  Moreover, the additional UV renormalisation of
$F(x_i, \tvec{\Delta})$ does not remove all UV divergences at the cross
section level.  The singularity of $F(x_i, \tvec{y})$ at small $y$
translates into a behaviour $F(x_i, \tvec{\Delta})\sim
\log(\mu^2/\Delta^2)$ at large $\Delta$ (see
section~\ref{sec:mom-space-dpds}), which gives a quadratic divergence for
the $\tvec{\Delta}$ integration in the DPS cross section.

It is easy to identify the origin of the UV divergences just discussed.
Both in the $\tvec{y}$ and $\tvec{\Delta}$ representations, one has
integrated over the full range of the integration variable and thus left
the region in which the approximations leading to the DPS cross section
formulae are valid, namely the region where $\Delta \ll Q_i$ or,
equivalently, $y \gg 1/Q_i$ (see section~2.1.2 in \cite{Diehl:2011yj}).
Outside this region, the DPS approximations are not only unjustified, but
they give divergent integrals in the cross section.

\begin{figure}
\begin{center}
\subfigure[]{\includegraphics[width=0.32\textwidth]{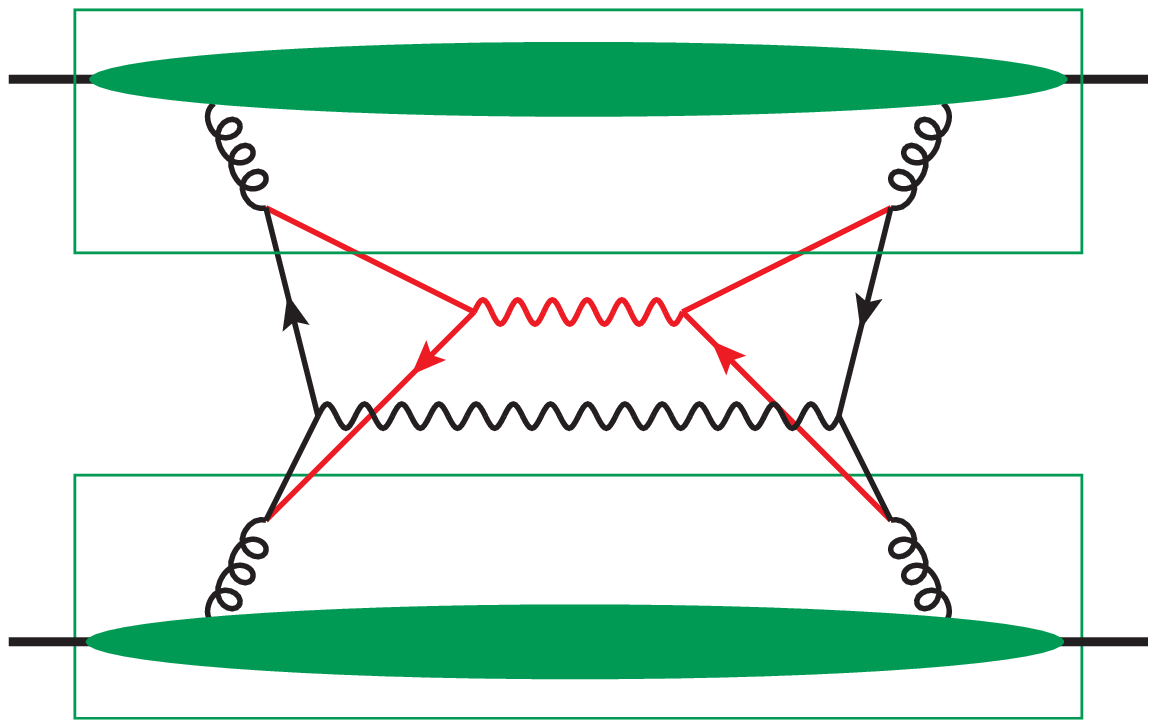}}
\hspace{0.1em}
\subfigure[]{\includegraphics[width=0.32\textwidth]{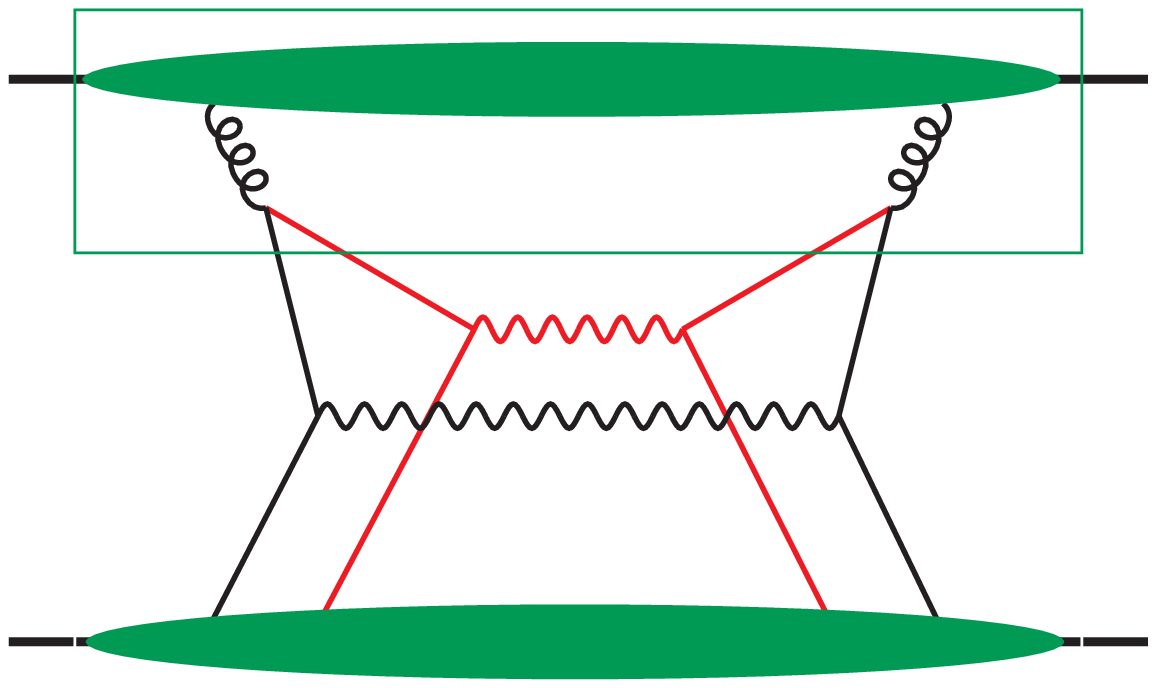}}
\hspace{0.1em}
\subfigure[]{\includegraphics[width=0.32\textwidth]{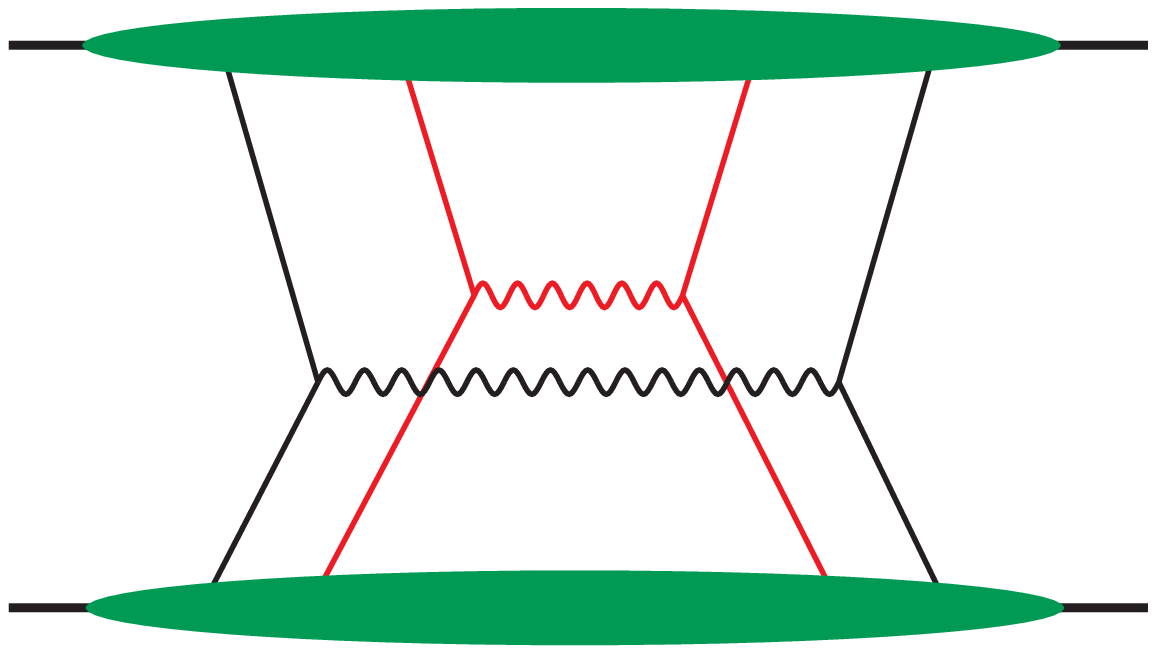}}
\caption{\label{fig:boxed-graphs} (a) A 1v1 contribution to DPS, with
  perturbative splitting DPDs for each proton (indicated by the boxes).
  (b) A 2v1 contribution to DPS, with a perturbative splitting DPD in only
  one proton (indicated by the box).  (c) A 2v2 contribution to DPS.  Here
  and in the following it is understood that partons emerging from the
  oval blobs are approximately collinear to their parent proton.  A line
  for the final state cut across the blobs and the produced vector bosons
  (wavy lines) is not shown for simplicity.}
\end{center}
\end{figure}

This points to another problem, namely that of double counting
contributions between SPS and DPS.  To see this, let us analyse the graph
in figure~\ref{fig:boxed-graphs}a.  Since the transverse boson momenta
$\tvec{q}_1$ and $\tvec{q}_2$ are approximately back to back (up to
effects from the transverse momenta of the incoming gluons) it is
convenient to introduce the combination
\begin{align}
  \label{q-def}
\tvec{q} &= \half (\tvec{q}_1 - \tvec{q}_2) \,.
\end{align}
For $q \ll Q_i$ the graph gives a leading contribution to
$d\sigma/d{q}^2$ if the transverse quark momenta in the loops are all
of order $Q_i$.  This carries the quark lines far off shell, so that this
contribution is naturally associated with SPS, with $g+g \to V_1 V_2$ as
the hard subprocess.  A leading contribution is also obtained from the
region where all transverse quark momenta are much smaller than $Q_i$.
This region is naturally described as DPS, with two disconnected hard
scattering processes $q\bar{q}\to V_1$ and $q\bar{q}\to V_2$ and double
parton distributions with perturbative $g\to q\bar{q}$ splittings,
indicated by the boxes in the graph.  We denote this as a 1v1 (1 versus 1)
contribution to DPS, emphasising its close relation to SPS.

A double counting problem for this graph obviously arises if one takes
the loop integrals in the SPS cross section over all transverse quark
momenta (including the DPS region), and likewise if one integrates the
DPDs cross section over the full range of transverse positions, which
is equivalent to integrating over all transverse momenta in the quark
loops.  The problem persists if one integrates the cross section over
$\tvec{q}$.

Let us now turn to the graph in figure~\ref{fig:boxed-graphs}b and
consider the cross section integrated over $\tvec{q}_1$ and
$\tvec{q}_2$.  In the region of large $q \sim Q_i$, the quark lines at
the top of the graph have transverse momenta of order $q$ and are far
off shell.  The proper description of this region is in terms of a
hard scattering $q\bar{q} + g \to V_1 V_2$, convoluted with a
collinear twist-two distribution (i.e.\ an ordinary PDF) at the top
and a collinear twist-four distribution at the bottom of the graph.
For brevity we refer to this as the ``twist-four contribution''
henceforth.  In the region $q \ll Q_i$, the $g\to q\bar{q}$ splittings
are near collinear and the approximations for DPS are appropriate.  We
call this the 2v1 contribution to DPS, recalling that there is a
$q\bar{q} + g$ subprocess in the graph.  Both small and large $q$ give
leading contributions to the integrated cross section, and in a naive
calculation adding up the twist-four term and the DPS term has again a
double counting problem, as well as divergences in each contribution.
The naive DPS cross section has a logarithmic divergence at small $y$,
which is seen by inserting the $1/y^2$ splitting behaviour of only one
DPD in \eqref{coll-Xsect-naive}.  In turn, the hard scattering cross
section in the twist-four term contains a collinear divergence in the
form of an integral behaving like $dq^2/q^2$ at $q\to 0$, as we will
show in section~\ref{sec:2v1}.

Clearly, one needs a consistent scheme for computing the overall cross
section, without double counting and without divergences in individual
terms.  An intuitive approach for evaluating DPS is to separate the
``perturbative splitting'' part of a DPD from its ``intrinsic''
nonperturbative part.\footnote{\rev{The intrinsic part of a DPD may be
    studied using quark models \cite{Chang:2012nw,Rinaldi:2013vpa,
      Rinaldi:2014ddl,Rinaldi:2016jvu,Rinaldi:2016mlk,Broniowski:2013xba,
      Broniowski:2016trx,Kasemets:2016nio}, at least in the valence
    region, or it may be related to the product of two PDFs if
    correlations between the two partons are neglected.}}
This has been pursued independently by Blok et
al.~\cite{Blok:2011bu,Blok:2013bpa} and by Ryskin and
Snigirev~\cite{Ryskin:2011kk,Ryskin:2012qx}.  Taking the intrinsic DPD for
each proton, one obtains the 2v2 part of DPS, which does not contain any
perturbative splitting and is shown in figure~\ref{fig:boxed-graphs}c.
The splitting part of the DPD is explicitly computed in terms of a single
parton distribution function (PDF) and a perturbative kernel.  This is
multiplied with an intrinsic DPD to compute the 2v1 term.  Finally, the
product of two splitting DPDs is used to compute the 1v1 contribution in
the approach of \cite{Ryskin:2011kk,Ryskin:2012qx}, where an ultraviolet
cutoff must be imposed to regulate the quadratic divergence we mentioned
earlier.  By contrast, the authors of \cite{Blok:2011bu,Blok:2013bpa}
advocate to omit this term and replace it entirely with the SPS
contribution to the cross section.

We are, however, not able to give a field theoretic definition of the
``intrinsic'' or ``nonperturbative'' part of a DPD.  The consideration
of Feynman graphs in the preceding arguments is instructive, but a
satisfactory definition should only appeal to perturbation theory in
regions where it is applicable.  We regard a nonperturbative
definition of DPDs as indispensable for a systematic theory approach,
for instance for deriving evolution equations and other general
properties.  

The setup we propose in this work defines DPDs as operator matrix elements
as described above, containing both splitting and intrinsic contributions.
UV divergences in the DPS cross section are avoided by introducing (smooth
or hard) cutoffs in the integrations over transverse distances.  The
double counting problems are treated within the subtraction formalism used
in standard factorisation theorems, described in detail in sections~10.1
and 10.7 of \cite{Collins:2011zzd} and briefly recalled in
section~\ref{sec:scheme} here.  The subtraction terms that avoid double
counting also remove the above mentioned collinear divergence in the
twist-four term.  A distinction between ``splitting'' and ``intrinsic''
contributions to a DPD will be made in the limit of small transverse
distances, where it can be formulated in terms of an operator product
expansion (see section~\ref{sec:coll-all}), and when making a model ansatz
for DPDs at large distances, which is of course necessary for
phenomenology.


\subsection{Contributions to the cross section: power behaviour and
logarithms}
\label{sec:power}

In preparation for later sections, we now recall some results for the
power behaviour of different contributions to the cross section, referring
to section~2.4 of \cite{Diehl:2011yj} for a derivation.  We also recall
which logarithms appear in the lowest order 1v1 and 1v2 graphs.  As
already stated, we take the process $pp\to V_1 + V_2 + X$ as a concrete
example, but the discussion readily generalises to other cases.

\begin{figure}
\begin{center}
\subfigure[]{\includegraphics[width=0.32\textwidth]{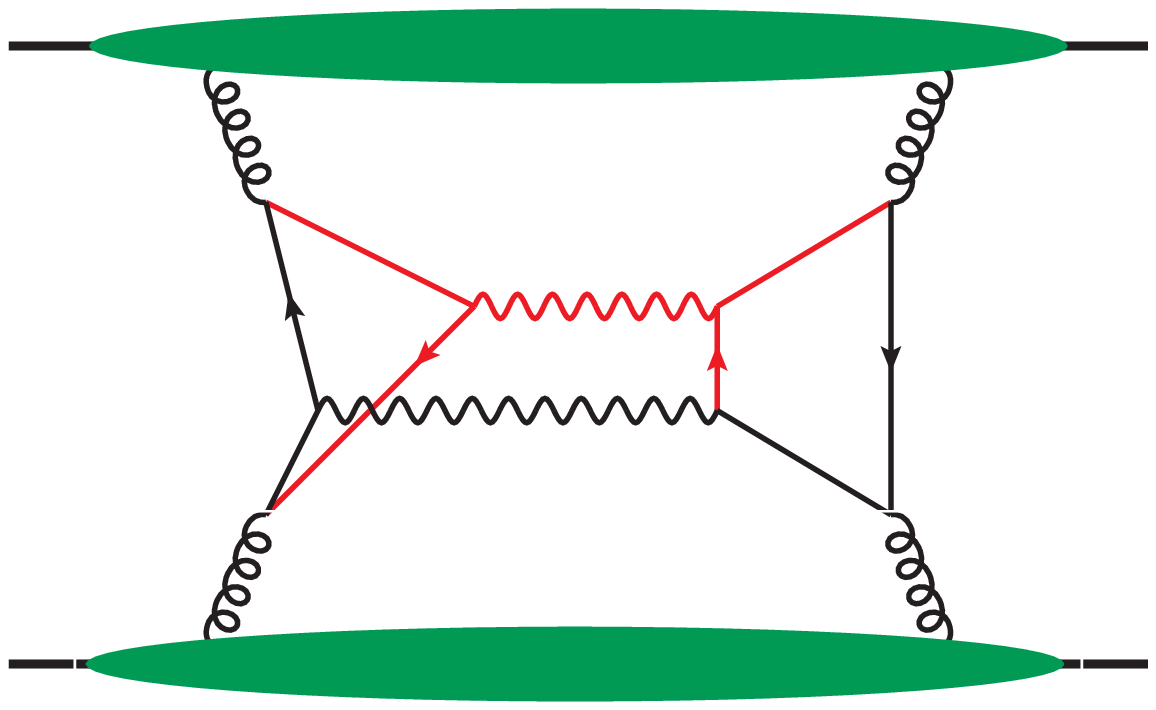}}
\hspace{0.1em}
\subfigure[]{\includegraphics[width=0.32\textwidth]{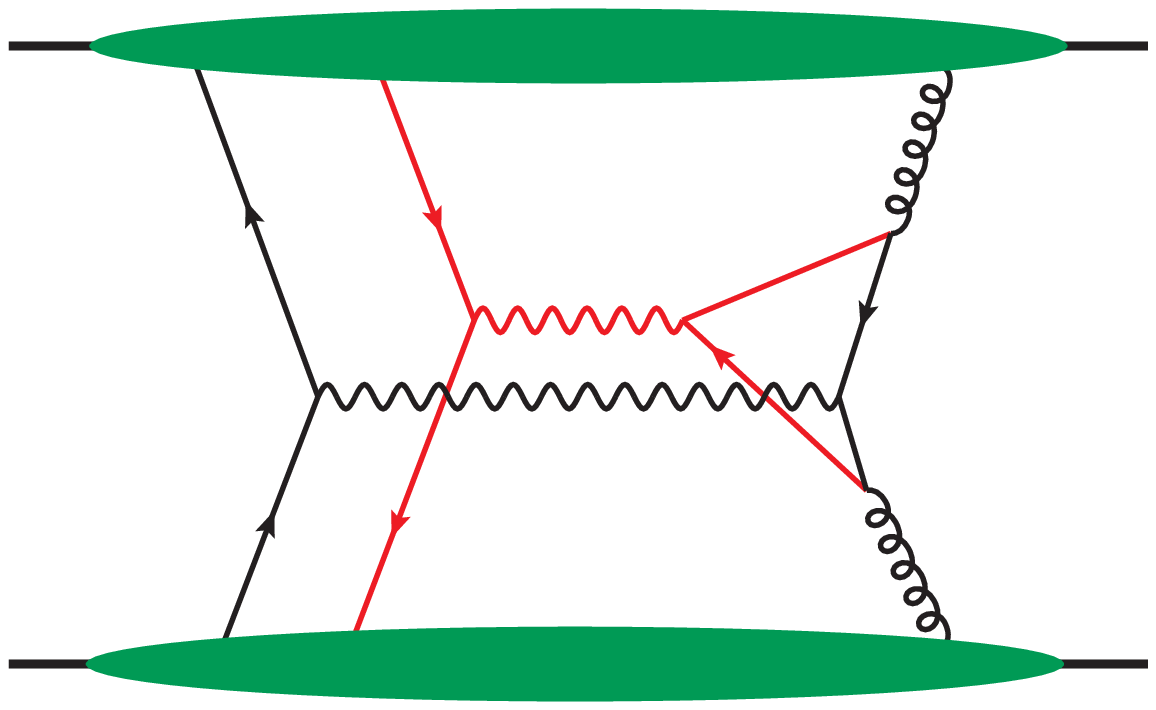}}
\hspace{0.1em}
\subfigure[]{\includegraphics[width=0.32\textwidth]{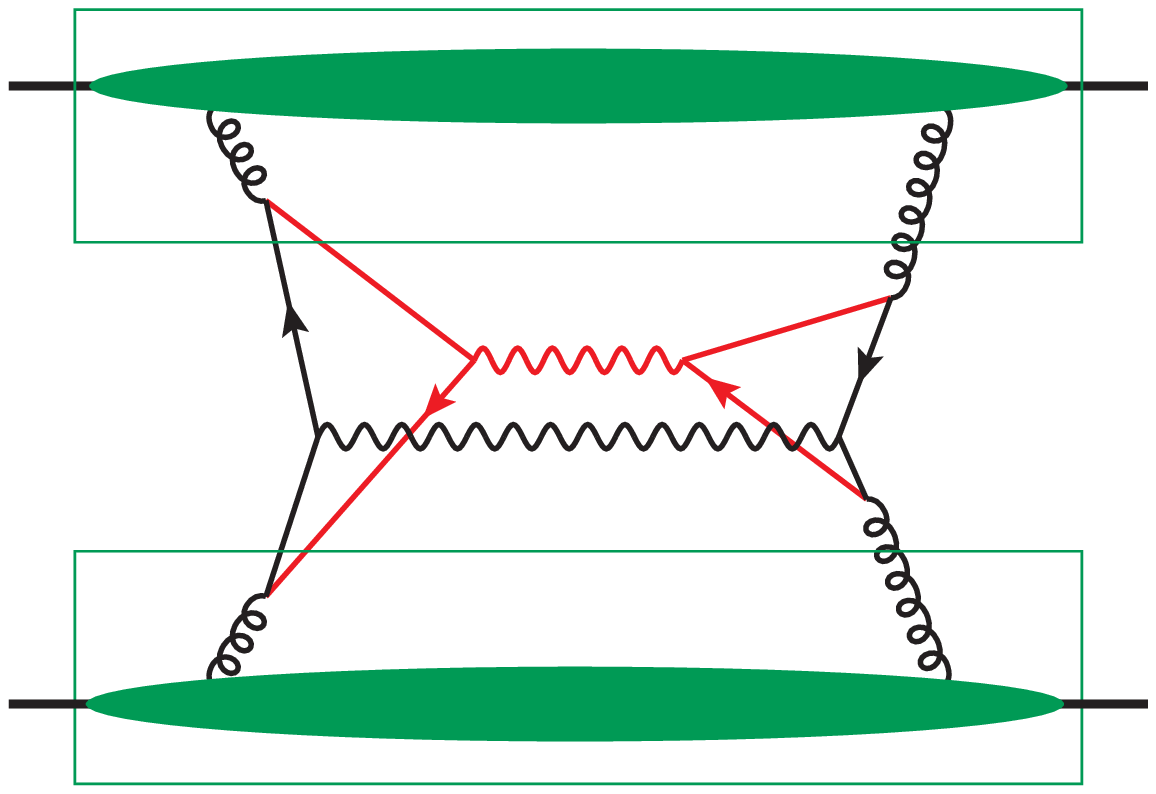}}
\caption{\label{fig:sps-dps-int} (a) A higher-order graph for SPS.  The
  loop on the l.h.s.\ has a momentum region associated with DPS but the
  loop on the r.h.s.\ does not. (b) Interference between DPS in the
  amplitude and SPS in the complex conjugate amplitude. (c) 1v1 graph in
  the region where the quark loop on the left is collinear whilst the one
  on the right is hard.  This corresponds to the region of SPS/DPS
  interference, with the boxes indicating DPDs with perturbative splitting
  for the parton pair in the amplitude.}
\end{center}
\end{figure}

The differential cross section $d\sigma / (d^2\tvec{q}_1\ms
d^2\tvec{q}_2)$ in the region $q_i \ll Q_i$ can be computed using TMD
factorisation.  Here and in the following we write $Q_i$ to denote the
generic size of $Q_1$ and $Q_2$, and likewise for $q_i$.  The transverse
momenta $q_i$ may be of nonperturbative size $\Lambda$ or much larger.  In
the latter case, further simplifications are possible by expressing
transverse momentum dependent distributions in terms of collinear ones
\cite{Buffing:2016wip}, but we shall not discuss this here.  The leading
power behaviour of the cross section is
\begin{align}
  \label{TMD-Xsect-power}
\frac{d\sigma}{d^2\tvec{q}_1\ms d^2\tvec{q}_2} \sim
     \frac{1}{Q_i^4\, q_i^2} \,.
\end{align}
When $q_i$ goes to zero, it should be replaced by $\Lambda$.  Three types
of mechanisms contribute to the leading behaviour, namely DPS, SPS, and
the interference between SPS and DPS.  Corresponding graphs are shown in
figures~\ref{fig:boxed-graphs}c, \ref{fig:sps-dps-int}a and
\ref{fig:sps-dps-int}b, respectively.

As discussed in the previous section, certain graphs contribute both to
DPS and to SPS, depending on the kinematics of their internal lines.  The
1v1 graph in figure~\ref{fig:boxed-graphs} also has leading regions in
which one of the loops is hard and the other is collinear.  These regions
contribute to the SPS/DPS interference, as shown in
figure~\ref{fig:sps-dps-int}c.  The double counting problem thus concerns
both SPS, DPS and their interference.  Note that the SPS graph in
figure~\ref{fig:sps-dps-int}a contributes to the SPS/DPS interference but
not to DPS.

Both the amplitude and its conjugate in the 1v1 graph contains a loop
integral that behaves like $d^2\tvec{k} /\tvec{k}^2$ in the region
$\Lambda, q_i \ll k \ll Q_i$.  When integrated over the full phase space,
each loop thus builds up a so-called DPS logarithm, namely $\log(Q_i/q_i)$
when $q_i \gsim \Lambda$ and $\log(Q_i/\Lambda)$ when $q_i \lsim \Lambda$.
Whether these logarithms reside in SPS, DPS or their interference depends
on how exactly one handles the double counting problem.  We come back to
this issue in section~\ref{sec:tmd-subtr}.

\begin{figure}
\begin{center}
\subfigure[]{\includegraphics[width=0.32\textwidth]{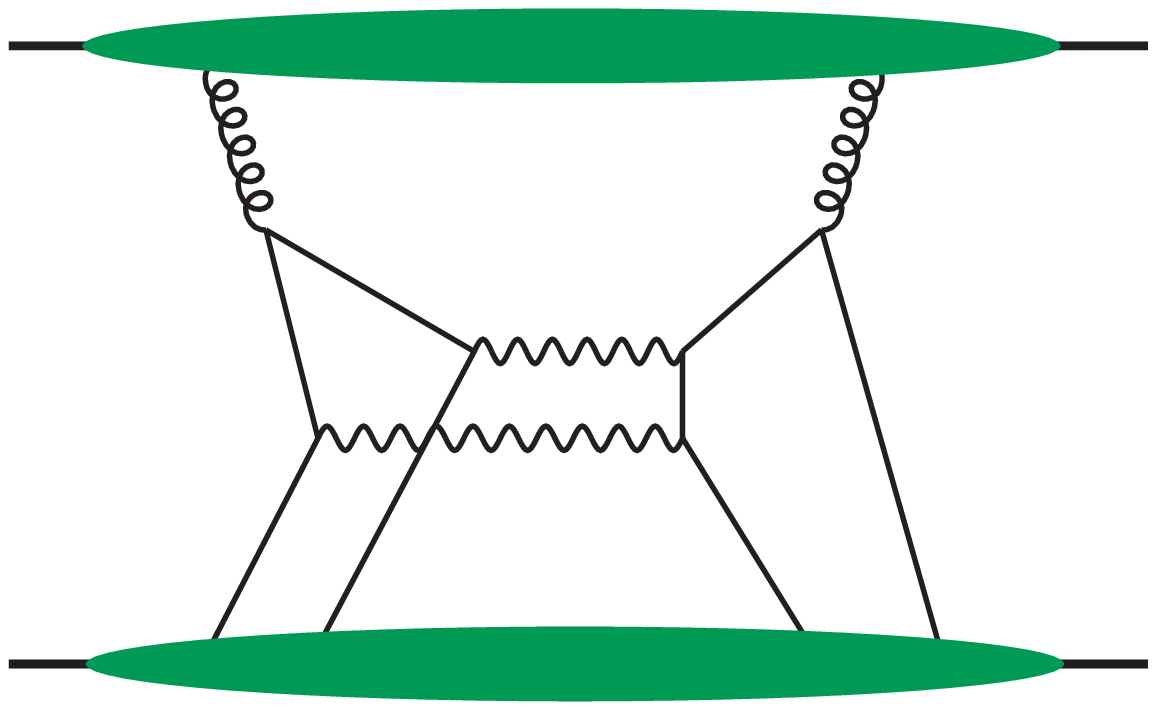}}
\hspace{1em}
\subfigure[]{\includegraphics[width=0.32\textwidth]{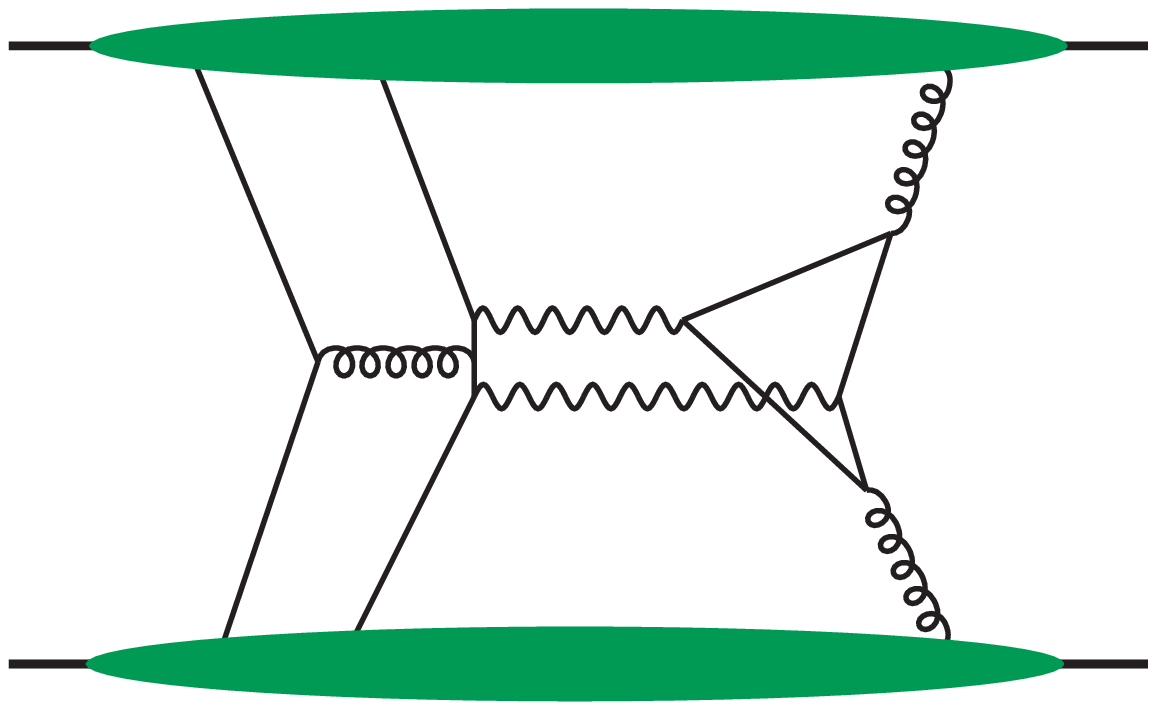}}
\caption{\label{fig:twist-graphs} (a) A graph with a twist-four
  distribution for one proton and a twist-two distribution for the other.
  (b) A graph with twist-three distributions for both protons.}
\end{center}
\end{figure}

Let us now turn to the cross section integrated over $\tvec{q}_1$ and
$\tvec{q}_2$, which can be described using collinear factorisation.  The
leading power behaviour of the cross section, $\sigma \sim 1/Q_i^2$, is
given by the SPS mechanism alone.  Several mechanisms contribute at the
power suppressed level $\sigma \sim \Lambda^2/Q_i^4$.  These are
\begin{enumerate}
\item DPS, which is suppressed because it can only populate the region
  $q_i \ll Q_i$ rather than the full phase space up to $q_i \sim Q_i$,
\item the interference between SPS and DPS, which is suppressed for the
  same reason,
\item hard scattering with a twist-four distribution for one proton and a
  twist-two distribution for the other.  Example graphs are
  figure~\ref{fig:twist-graphs}a, as well as
  figure~\ref{fig:boxed-graphs}b with the box removed.
\item hard scattering with twist-three distributions for both protons.  An
  example graph is figure~\ref{fig:twist-graphs}b.
\end{enumerate}
The rationale for considering such contributions is that -- whilst being
power suppressed compared with SPS -- they may be enhanced by higher
parton luminosities at small momentum fractions $x$, or by coupling
constants in the relevant hard scattering subprocesses.  Let us emphasise
that a complete calculation of the cross section at the level of
$\Lambda^2/Q_i^2$ corrections would be a formidable task, and it is not
even established whether factorisation (in particular the cancellation of
Glauber gluons) holds at that level.

Notice that in collinear factorisation, the SPS/DPS interference term
involves collinear twist three distributions for both protons, because the
SPS mechanism forces the two partons in the interfering DPS amplitude to
be at same transverse position (see section~2.4.1 in \cite{Diehl:2011yj}).
In this sense, mechanism 2 in the above list may be regarded as a special
case of mechanism 4, with a disconnected hard scattering in the amplitude
or its conjugate (see figure~\ref{fig:sps-dps-int}b).

A full treatment of contributions with twist-three or twist-four
distributions is beyond the scope of this paper.  We remark however that
twist-$n$ operators contain $n$ or less than $n$ parton fields, and that
different operators are related by the equations of motion.  For a
detailed discussion we refer to
\cite{Braun:2008ia,Braun:2009vc,Ji:2014eta}.  Twist-$n$ operators with $n$
parton fields were called ```quasipartonic'' in \cite{Bukhvostov:1985rn}
and involve only the ``good'' parton fields in the parlance of light-cone
quantisation \cite{Jaffe:1996zw}.  These are exactly the fields appearing
in the definitions of multiparton distributions, so that graphs with a
double counting issue between higher-twist hard scattering and DPS (or the
SPS/DPS interference) involve only quasipartonic operators.

The matrix elements of quasipartonic twist-three operators in an
unpolarised target satisfy the important selection rule that the
helicities carried by the parton lines must balance .  This excludes
three-gluon operators since three helicities $\pm 1$ cannot add up to
zero.  For quark-antiquark-gluon operators it forces the quark and
antiquark fields to have opposite chirality, i.e.\ one only has the
operator combination $\bar{q} \sigma^{+j} q$, where the transverse index
$j$ is contracted with the polarisation index of the gluon.  As for
non-quasipartonic twist-three distributions in an unpolarised target, one
finds that they are absent in the pure gluon sector \cite{Mulders:2000sh},
whereas the corresponding quark-antiquark distributions are again
chiral-odd \cite{Jaffe:1991ra}.  Since chiral-odd distributions cannot be
generated by gluon ladder graphs, they lack the small $x$ enhancement that
is one of the motivations to keep higher twist contributions in the cross
section.  We will therefore not discuss them further in this work.  Note
that corresponding selection rules do \emph{not} hold for TMD correlators,
where an imbalance in the helicities of the parton fields can be
compensated by orbital angular momentum.

Let us finally recall the appearance of DPS logarithms in collinear
factorisation.  The 2v1 graph (figure~\ref{fig:boxed-graphs}b) has a
behaviour $d\sigma/dq^2 \sim 1/q^2$ in the region $\Lambda \ll q \ll Q_i$,
which gives a $\log(Q_i/\Lambda)$ when integrated over the full phase
space.  Depending on how the double counting between DPS and the
twist-four mechanism is resolved, this logarithm can appear in different
contributions to the cross section.  We will discuss this in
section~\ref{sec:2v1}.

\section{Short-distance limit of DPDs}
\label{sec:short-dist}

In this section, we analyse the behaviour of DPDs in the limit where the
transverse distance between partons becomes small compared with the scale
of nonperturbative interactions.  In this region, the splitting of one
parton into two becomes dominant.  Generalising results in
\cite{Diehl:2011yj} we give expressions in $D= 4-2\epsilon$ dimensions,
which are necessary in intermediate steps when constructing a
factorisation formula for the cross section.


\subsection{TMDs}
\label{sec:tmd-split}

\begin{figure}
\begin{center}
\includegraphics[width=0.5\textwidth]{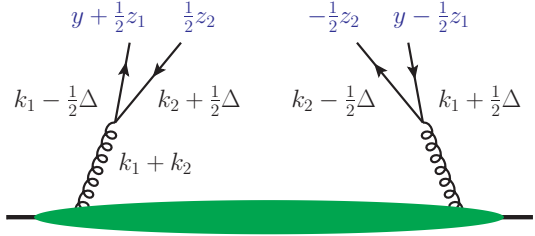}
\caption{\label{fig:split-mom} The perturbative splitting mechanism
  for a DPD, with momentum and position assignments.  Here and in the
  following, the line for the final-state cut of the spectator partons
  is not shown for simplicity.}
\end{center}
\end{figure}

A useful choice of position variables for describing the parton splitting
mechanism is
\begin{align}
\tvec{y}_\pm  &= \tvec{y} \pm \half( \tvec{z}_1 - \tvec{z}_2 ) \,,
&
\tvec{Z} &= \half( \tvec{z}_1 + \tvec{z}_2 )
\end{align}
with Fourier conjugate momenta\,\footnote{The momentum
  $\tvec{\Delta}$ is called $\tvec{r}$ in
  \protect\cite{Diehl:2011tt,Diehl:2011yj}.}
\begin{align}
\tvec{k}_\pm &= \half( \tvec{k}_1 - \tvec{k}_2 \pm \tvec{\Delta} ) \,,
&
\tvec{K} &= \tvec{k}_1 + \tvec{k}_2 \,.
\end{align}
The relation between DPDs in position and momentum space reads
\begin{align}
  \label{split-ft}
& F(x_i,\tvec{y}_\pm,\tvec{Z}) 
\nonumber \\
 &\qquad = \frac{1}{(2\pi)^{2-2\epsilon}}
  \int d^{2-2\epsilon} \tvec{K}\,
       d^{2-2\epsilon} \tvec{k}_+\, d^{2-2\epsilon} \tvec{k}_-\; 
  e^{i \tvec{Z}\tvec{K}
   + i (\tvec{y}_+ \tvec{k}_- - \tvec{y}_- \tvec{k}_+)}\,
  F(x_i,\tvec{k}_\pm,\tvec{K})
\end{align}
in $D= 4-2\epsilon$ dimensions.  As seen in
figure~\ref{fig:split-mom}, one can identify $\tvec{y}_+$
($\tvec{y}_-$) as the transverse distance between the two partons on
the left (right) hand side of the final state cut in the DPD.
Correspondingly, the transverse momentum difference between the
partons on the left (right) hand side of the cut is $\tvec{k}_-$
($\tvec{k}_+$).  The splitting singularities of the DPDs thus occur at
$y_\pm \to 0$ or $k_\pm \to \infty$.

The perturbative splitting contribution \rev{$F_{\text{spl,pt}}$} to
transverse-momentum
dependent DPDs in momentum space has been calculated at leading order
in section 5.2.2 of \cite{Diehl:2011yj}.  Generalising these results
to $D=4-2\epsilon$ dimensions, we have
\begin{align}
  \label{TMD-split}
F_{a_1 a_2,\ms \text{spl,pt}}(x_i,\tvec{k}_\pm,\tvec{K}) &=
   \frac{\tvec{k}_-^j}{k_-^2}\, \frac{\tvec{k}_+^{j'}}{k_+^2}\,
   \frac{(2 \mu)^{2\epsilon}}{\pi^{1-2\epsilon}}\, 
\nonumber \\[0.2em]
 & \quad \times
         \biggl[ \frac{f_{a_0}(x_1+x_2,\tvec{K})}{x_1+x_2} \,
         \frac{\alpha_s}{2\pi}\, T^{jj'}_{a_0\to a_1 a_2}\biggl(
            \frac{x_1}{x_1+x_2}, \epsilon \biggr)  + \ldots \,\biggr] \,,
\end{align}
where $j$, $j'$ are transverse Lorentz indices and
$f_{a_0}(x_1+x_2,\tvec{K})$ is an unpolarised single-parton
TMD.\footnote{Compared with section 5.2 of \protect\cite{Diehl:2011yj}, the
  kernel $T^{jj'}$ used here has the opposite order of indices $jj'$ and
  includes a colour factor, e.g.\ $T_F=1/2$ for the colour singlet
  distribution ${}^{1}F_{g\to q\bar{q}}$.}
The ellipsis denotes a term that involves a TMD for polarised partons
in an unpolarised proton and depends on $\tvec{K}$ but not on
$\tvec{k}_\pm$.  In position space we then get
\begin{align}
  \label{TMD-split-y}
F_{a_1 a_2,\ms \text{spl,pt}}(x_i,\tvec{y}_\pm,\tvec{Z}) &=
   \frac{\tvec{y}_+^{j}}{y_+^{2-2\epsilon}}\,
   \frac{\tvec{y}_-^{j'}}{y_-^{2-2\epsilon}}\,
   \mu^{2\epsilon}\, \frac{\Gamma^2(1-\epsilon)}{\pi^{1-2\epsilon}}\,
\nonumber \\[0.2em]
 & \quad \times
      \biggl[ \frac{f_{a_0}(x_1+x_2,\tvec{Z})}{x_1+x_2} \,
         \frac{\alpha_s}{2\pi}\, T^{jj'}_{a_0\to a_1 a_2}\biggl(
            \frac{x_1}{x_1+x_2}, \epsilon \biggr) + \ldots \,\biggr]
\end{align}
using the Fourier integral \eqref{vector-FT}, where the term denoted by an
ellipsis depends on $\tvec{Z}$ but not on $\tvec{y}_\pm$.  It is
understood that for transverse quark or linear gluon polarisation, both
$F_{a_1 a_2}$ and the kernel $T$ carry additional transverse Lorentz
indices.  $f_{a_0}(x_1+x_2,\tvec{Z})$ is the Fourier transform of
$f_{a_0}(x_1+x_2,\tvec{K})$.  The form \eqref{TMD-split} gives the leading
behaviour of the DPD for large $k_\pm \gg \Lambda$, and correspondingly
\eqref{TMD-split-y} gives the leading behaviour for $y_\pm \ll 1/\Lambda$.

If one inserts these results into the cross section formula and sets
$D=4$, logarithmic divergences appear at $y_+=0$ and $y_-=0$.  To make
them explicit we transform variables to
\begin{align}
  \label{TMD-int-vars}
\int d^2\tvec{y}\, d^2\tvec{z}_1\, d^2\tvec{z}_2\;
  e^{-i (\tvec{q}_1^{} \tvec{z}_1^{} + \tvec{q}_2^{} \tvec{z}_2^{})}
&= \int d^2\tvec{Z}\, d^2\tvec{y}_+\, d^2\tvec{y}_-\;
       e^{-i (\tvec{q}_1 + \tvec{q}_2) \tvec{Z}}
       e^{-i \tvec{q} (\tvec{y}_+ - \tvec{y}_-)}
\end{align}
with $\tvec{q}$ defined in \eqref{q-def}.  Performing the angular
integration in
\begin{align}
  \label{TMD-angular-int}
\frac{1}{\pi} \int d^2\tvec{y}\; e^{\ms\pm i\tvec{q} \tvec{y}}\,
              \frac{\tvec{y}^j \tvec{y}^{j'}}{y^4}
&= \delta^{jj'} \! \int \frac{dy}{y}\, J_0(y q)
     + \biggl( \delta^{jj'} - \frac{2\tvec{q}^j \tvec{q}^{j'}}{q^2} \biggr)
   \int \frac{dy}{y}\, J_2(y q) \,,
\end{align}
where $\tvec{y}$ stands for $\tvec{y}_+$ or $\tvec{y}_-$, we see that the
integral with $J_0$ is divergent at $y=0$.  Given the range of validity of
\eqref{TMD-split-y} one should impose $y \ll 1/\Lambda$ in
\eqref{TMD-angular-int}, although the integrals are finite for $y \to
\infty$ due to the oscillations of the Bessel functions.

\begin{figure}
\begin{center}
\subfigure[]{\includegraphics[width=0.32\textwidth]{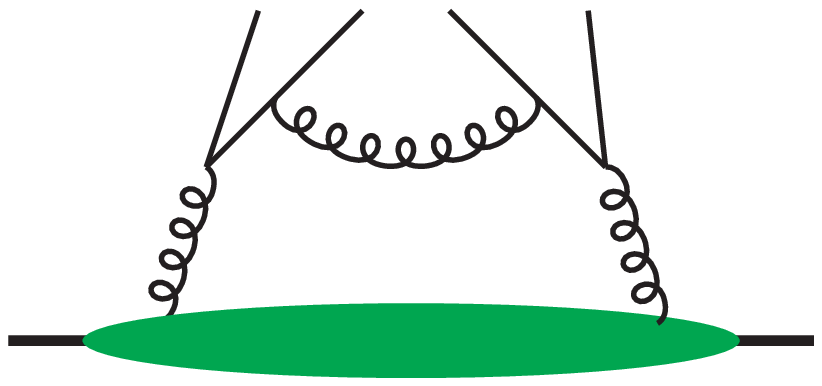}}
\hspace{0.1em}
\subfigure[]{\includegraphics[width=0.32\textwidth]{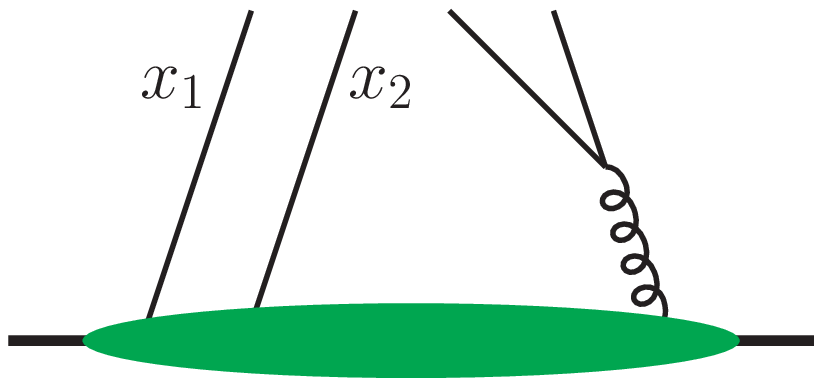}}
\hspace{0.1em}
\subfigure[]{\includegraphics[width=0.32\textwidth]{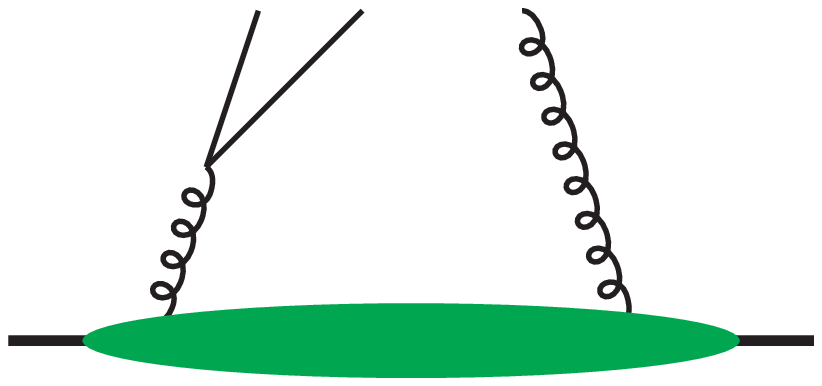}}
\caption{\label{fig:more-split} (a) A higher order contribution to the
  splitting part of a DPD.  (b) Graph for a DPD with perturbative
  splitting only to the right of the final-state cut.  The blob denotes a
  distribution $D_{\alpha_1 \alpha_2 | \alpha_0}$.  (c) Graph for
  perturbative splitting in the distribution $D_{\alpha_1 \alpha_2 |
    \alpha_0}$.}
\end{center}
\end{figure}

The perturbative splitting contribution to DPDs at higher order in
$\alpha_s$ involves graphs with additional partons radiated into the final
state as shown in figure~\ref{fig:more-split}a, as well as virtual
corrections.  It is natural to expect that it will again be singular at
$y_\pm = 0$.  A calculation of this contribution is outside the scope of
the present work, so that we will limit our analysis of TMD factorisation
to perturbative splitting at LO.

To compute the DPD cross section, we must also consider the case where
only one of the distances $y_+$ or $y_-$ is small, whereas the other
one remains large.  In this case, one has a perturbative splitting only on
one side of
the final-state cut, as illustrated in figure~\ref{fig:more-split}b.  We
will not discuss the detailed expression of the DPD in this regime, but
give its general structure.  Setting $D=4$ for simplicity, we have
\begin{align}
  \label{TMD-one-split}
F_{\alpha_1 \alpha_2,\ms
 y_- \to 0}(x_i,\tvec{y}_\pm,\tvec{Z}) &=
   \frac{\tvec{y}_-^{j'}}{y_-^{2}}\,
   \bigl[ U^{j'}_{\alpha_0\to \alpha_1 \alpha_2}(x_i) \bigr]^* \,
   D_{\alpha_1 \alpha_2| \alpha_0}(x_i, \tvec{y}_+, \tvec{Z}) \,,
\end{align}
were $U_{\alpha_0\to \alpha_1 \alpha_2}$ is a kernel
for the splitting $\alpha_0 \to \alpha_1 \alpha_2$ in the amplitude
(hence its complex conjugate appears in \eqref{TMD-one-split}).
$D_{\alpha_1 \alpha_2| \alpha_0}$ is the position space version of a
transverse-momentum dependent twist-three distribution, constructed
from the hadronic matrix element
\begin{align}
  \label{tw3-mat-el}
& \bigl\langle p \ms\big| \phi_0\bigl( -\half Z \bigr)\,
   \phi_2\bigl( \half (Z - y_+) \bigr)\,
   \phi_1\bigl( \half (Z + y_+) \bigr) \big|\ms p \bigr\rangle
\nonumber \\
 &\qquad\quad
 = \bigl\langle p \ms\big| \phi_0\bigl( \half (y - Z) \bigr)\,
   \phi_2\bigl( \half z_2 \bigr)\,
   \phi_1\bigl( y + \half z_1 \bigr) \big|\ms p \bigr\rangle \,,
\end{align}
where $\phi_i$ is a ``good'' field for parton $\alpha_i$
(cf.~section~\ref{sec:power}).  Distributions $D_{\alpha_0| \alpha_1
  \alpha_2}(x_i, \tvec{y}_-, \tvec{Z})$ where $\alpha_{0}$ belongs to the
amplitude and $\alpha_1, \alpha_2$ to the complex conjugate amplitude are
defined in analogy.  In the second step of \eqref{tw3-mat-el} we have used
translation invariance and shifted the parton fields to the same position as
in the corresponding DPD (see figure~\ref{fig:split-mom}).  \rev{The labels
  $\alpha_i$ denote the parton species; it is understood that in
  \eqref{TMD-one-split} and the following equations parton helicities are
  taken fixed on the l.h.s.\ and must be appropriately summed over on the
  r.h.s.  Note the difference between this notation and the labels $a_i$,
  which denote parton species and polarisation (none, longitudinal,
  transverse or linear) and thus refer to a \emph{pair} of parton legs.  The
  notation with $a_i$ is hence not suitable for distributions with three
  parton fields.}

If $y_+$ is small, then $D_{\alpha_1 \alpha_2| \alpha_0}(x_i, \tvec{y}_+,
\tvec{Z})$ itself can be generated by perturbative splitting, as shown in
figure~\ref{fig:more-split}c.  We have
\begin{align}
  \label{tw3-split}
D_{\alpha_1 \alpha_2| \alpha_0,\ms y_+ \to 0}(x_i, \tvec{y}_+, \tvec{Z})
 &= \frac{\tvec{y}_+^{j}}{y_+^{2}}\,
       U^{j}_{\alpha_0\to \alpha_1 \alpha_2}(x_i)\,
    f_{\alpha_0}(x_1+x_2, \tvec{Z}) \,.
\end{align}
Notice that a quark and antiquark produced by perturbative splitting
have opposite helicities, so that the corresponding quark-antiquark
operator $\phi_2\, \phi_1$ in $D_{q\bar{q}\ms| g}$ must be chiral-even.
Inserting \eqref{tw3-split} into \eqref{TMD-one-split} we obtain
\begin{align}
  \label{TMD-two-splits}
& F_{\alpha_1 \alpha_2,\ms
    y_\pm \to 0}(x_i,\tvec{y}_\pm,\tvec{Z})
\nonumber \\
&\quad =
   \frac{\tvec{y}_+^{j}}{y_+^{2}}\, \frac{\tvec{y}_-^{j'}}{y_-^{2}}\,
   U^{j}_{\alpha_0\to \alpha_1 \alpha_2}(x_i)\,
   \bigl[ U^{j'}_{\alpha_0\to \alpha_1 \alpha_2}(x_i) \bigr]^* \,
   f_{\alpha_0}(x_1+x_2, \tvec{Z}) \,.
\end{align}
Taking appropriate linear combinations of parton helicities, we recover
the form of $F_{a_1 a_2,\ms \text{spl,pt}}$ in \eqref{TMD-split-y} at
$\epsilon=0$.


\subsection{Collinear DPDs: splitting contribution}
\label{sec:coll-split}

We now turn to collinear DPDs, i.e.\ to the case where
$\tvec{z}_1 = \tvec{z}_2 = \tvec{0}$.  Let us first consider
distributions for two unpolarised or two longitudinally polarised
partons, so that the DPDs do not carry any transverse Lorentz indices.
The lowest order splitting has been computed in \cite{Diehl:2011yj}.
For $4-2\epsilon$ dimensions, one finds the general form
\begin{align}
  \label{split-lo}
& F_{a_1 a_2,\ms \text{spl,pt}}(x_1,x_2,\tvec{y};\mu)
\nonumber \\ 
 &\quad =
  \frac{\mu^{2\epsilon}}{y^{2-4\epsilon}}\,
  \frac{\Gamma^2(1-\epsilon)}{\pi^{1-2\epsilon}}\,
  \frac{f_{a_0}(x_1+x_2;\mu)}{x_1+x_2}\, 
  \frac{\alpha_s(\mu)}{2\pi}\,
     P_{a_0\to a_1 a_2}\biggl( \frac{x_1}{x_1+x_2}, \epsilon \biggr) \,.
\end{align}
The kernel for the splitting $g\to q\bar{q}$ reads for instance
\begin{align}
  \label{split-gqq}
P_{g\to q\bar{q}\,}(u,\epsilon) &= \frac{f}{2}\,
  \frac{u^2 + (1-u)^2 - \epsilon}{1-\epsilon}
\end{align}
with a factor $f = 1$ for the colour singlet and
$f = - 1/\sqrt{N^2-1}$ for the colour octet DPD.  In terms of the
kernel in \eqref{TMD-split} we have
$T^{jj'}_{g\to q\bar{q}} = \delta^{jj'}_{\phantom{q}} P^{}_{g\to
  q\bar{q}}$.
We recognise in $P_{g\to q\bar{q}}(u,0)$ the usual DGLAP splitting
kernel without the terms proportional to $\delta(1-u)$.

Going beyond leading order, one can deduce the general form of the
perturbative splitting contribution using dimensional analysis and
boost invariance.  For colour singlet distributions one finds
\begin{align}
  \label{split-ho}
& {}^{1}F_{a_1 a_2,\ms \text{spl,pt}}(x_1,x_2,\tvec{y};\mu)
\nonumber \\
 & \qquad =
  \frac{1}{\pi^{1-\epsilon}\, y^{2-2\epsilon}}\, \sum_{a_0}
  \int\limits_{x_1+x_2}^1 \frac{dv}{v}\, \frac{f_{a_0}(v;\mu)}{v}\,
  V_{a_0\to a_1 a_2}\biggl( \frac{x_1}{v}, \frac{x_2}{v}, 
       \alpha_s(\mu), y \mu, \epsilon \biggr)
\end{align}
in $D=4-2\epsilon$ dimensions.  The convolution integral over $v$ is
familiar from factorisation formulae for hard scattering processes.  Both
$f$ and $V$ on the right-hand side are understood to include all necessary
subtractions, so that they are finite at $\epsilon=0$.  The splitting
kernel $V$ is a double series
\begin{align}
  \label{V-series}
V_{a_0\to a_1 a_2}(v_1, v_2, \alpha_s, y \mu, \epsilon)
 &= \sum_{n=1}^\infty \alpha_s^n \sum_{m=1}^n (y \mu)^{2\epsilon m}\,
    V^{(n,m)}_{a_0\to a_1 a_2}(v_1, v_2, \epsilon) \,.
\end{align}
The $\mu$ (and thus on dimensional grounds the $y$) dependence of $V$
follows from the fact that the mass parameter of dimensional
regularisation appears in graphs only via $\mu^{2\epsilon} \alpha_s(\mu)$;
terms with $n > m$ in \eqref{V-series} are due to the subtractions of
ultraviolet or collinear divergences.  At lowest order, the hard splitting
graphs are disconnected (with no partons across the final state cut), so
that
$V^{(1,1)}(v_1,v_2,\epsilon) = \delta(1-v_1-v_2)\, V^{(1)}(v_1,\epsilon)$.
Inserted into \eqref{split-ho} this gives a form consistent with
\eqref{split-lo}.  Using that at order $\alpha_s^n$ the poles of highest
order are $1/\epsilon^{n-1}$, we find
\begin{align}
  \label{V-series-zero}
V_{a_0\to a_1 a_2}(v_1, v_2, \alpha_s, y \mu, 0)
 &= \sum_{n=1}^\infty \alpha_s^n \sum_{m=0}^{n-1} \log^m(y \mu)\,
    V^{[n,m]}_{a_0\to a_1 a_2}(v_1, v_2) \,.
\end{align}
in the physical limit $\epsilon=0$.

For colour nonsinglet DPDs one must regulate rapidity divergences,
which complicates the preceding result.  Taking e.g.\ Wilson lines
along non-lightlike paths introduces additional vectors and changes
the analysis of boost properties of the kernel.  We will not pursue
this issue here.

DPDs with transverse quark or linear gluon polarisation carry
transverse Lorentz indices.  Their perturbative splitting expressions
thus have a tensor structure containing additional factors of
$\tvec{y}^j/y$ compared with the formulae above.  At leading order one
readily finds from \eqref{TMD-split-y} and the appropriate
splitting kernels that the factor $1/ y^{2-4\epsilon}$ in
\eqref{split-lo} is to be replaced with
$\tvec{y}^j \tvec{y}^{j'} / y^{4-4\epsilon}$ times a tensor
constructed only from Kronecker deltas.


\subsection{Collinear DPDs: all contributions}
\label{sec:coll-all}

Let us now study the small $y$ behaviour of collinear DPDs in more
general terms.  We start by writing the relation between
unrenormalised DPDs in position and momentum space as
\begin{align}
  \label{match-master}
  (2\pi)^{D-2} F(\tvec{y})
&= \int d^{D-2} \tvec{\Delta}\;  F(\tvec{\Delta})
       + \int d^{D-2} \tvec{\Delta}\; [e^{-i\tvec{y} \tvec{\Delta}} - 1]\,
          F(\tvec{\Delta}) \,,
\end{align}
omitting all arguments other than $\tvec{y}$ and $\tvec{\Delta}$.  The
first term on the r.h.s.\ is a collinear twist-four distribution,
independent of any transverse variable.  For small $y$, the second
term is dominated by large $\tvec{\Delta}$, so that one can replace
$F(\tvec{\Delta})$ by its approximation for large $\Delta$, following
the power counting analysis of section 5.2 in \cite{Diehl:2011yj}.
This leads us to write
\begin{align}
  \label{small-y-dec}
F_{y \to 0}(\tvec{y}) &=
   F_{\text{spl,pt}}(\tvec{y})
   + F_{\text{tw3,pt}}(\tvec{y}) + F_{\text{int,pt}}(\tvec{y}) \,,
\end{align}
where the three terms on the r.h.s.\ will be described shortly.  In $D=4$
dimensions, they respectively go like $y^{-2}$, $y^{-1}$ and $y^{0}$, up
to logarithmic corrections.  Further terms from the perturbative expansion
of $F(\tvec{\Delta})$ give contributions to $F(\tvec{y})$ that vanish like
$y$ or faster.

One may also derive the expansion \eqref{small-y-dec} from the
operator product expansion, without taking recourse to the transverse
momentum representation \eqref{match-master}.  In the definition of
collinear DPDs one has a product
$\mathcal{O}_2^{}(0, z_2)\, \mathcal{O}_1^{}(y, z_1)$ of operators
with $\tvec{z}_1 = \tvec{z}_2 = \tvec{0}$ but nonzero $\tvec{y}$.
This can be expanded around $\tvec{y} = \tvec{0}$ in terms of
light-ray operators where all fields are at transverse
position~$\tvec{0}$.  These operators have twist 2, 3, 4 for the
first, second and third term in \eqref{small-y-dec}, respectively.

The spitting contribution $F_{\text{spl,pt}}$ is given by graphs as in
figures~\ref{fig:split-mom} and \ref{fig:more-split}a and has already
been discussed in the previous subsection.  The term
$F_{\text{tw3,pt}}$ originates from two types of graphs.  The first
type involves a single perturbative splitting and a quasipartonic
collinear twist-three distribution as shown in
figure~\ref{fig:more-split}b.  The second type has two splittings as
in figure~\ref{fig:split-mom} and a twist-three distribution with one
``good'' and one ``bad'' parton field.  Given the helicity constraints
discussed in section \ref{sec:power}, collinear twist-three
distributions in an unpolarised proton involve a quark and antiquark
with opposite chirality (and possibly an extra gluon).  As announced
earlier, we discard twist-three terms in the following, since they are
expected to become unimportant at small momentum fractions $x_1, x_2$.

\begin{figure}
\begin{center}
\subfigure[]{\includegraphics[width=0.32\textwidth]{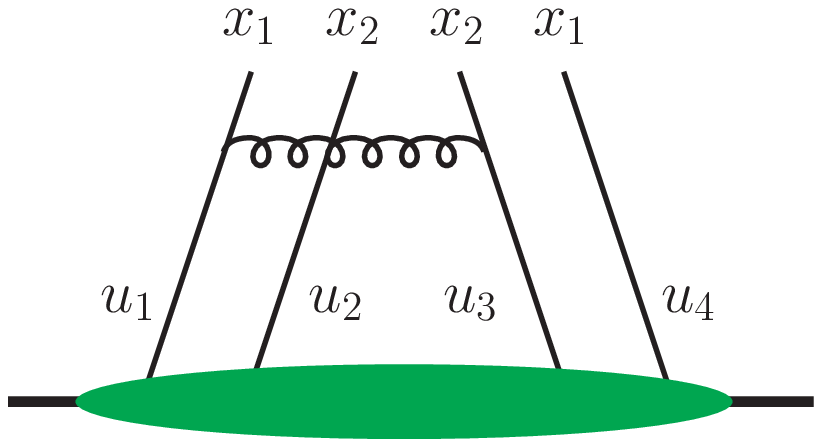}
\hspace{0.1em}}
\subfigure[]{\includegraphics[width=0.32\textwidth]{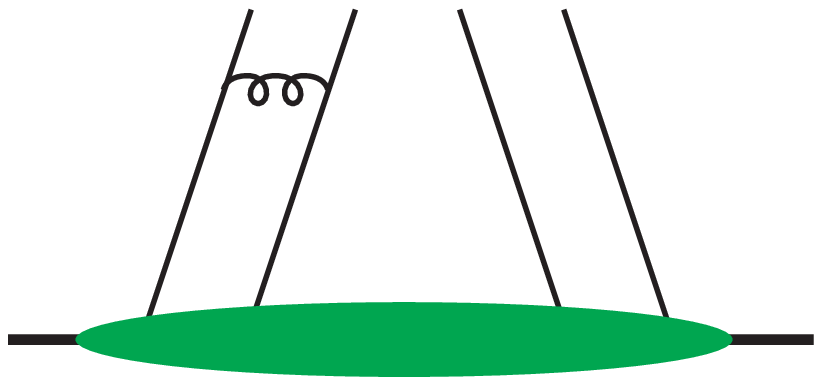}}
\hspace{0.1em}
\subfigure[]{\includegraphics[width=0.32\textwidth]{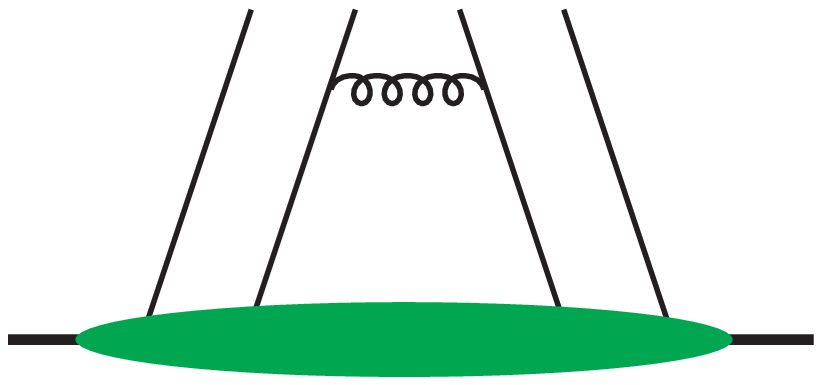}}
\caption{\label{fig:intr-nlo} Graphs for the short-distance behaviour of a
  DPD that involve a twist-four distribution. $x_i$ and $u_i$ denote
  longitudinal momentum fractions.}
\end{center}
\end{figure}

Finally, the term $F_{\text{int,pt}}$ contains contributions without any
perturbative splitting; we hence refer to it as the ``intrinsic'' part of
the DPD.  It can be written as
\begin{align}
  \label{match-tw4}
F_{\text{int,pt}}(x_1,x_2,\tvec{y};\mu) &= 
  G(x_1,x_2,x_2,x_1;\mu)
    + C(\cdots\!,\tvec{y}; \mu) \otimes G(\cdots\bs; \mu)
    + \ldots
\end{align}
where $G$ is a quasipartonic collinear twist-four distribution and $C$
a perturbative splitting kernel, corresponding to graphs as in
figure~\ref{fig:intr-nlo}.  The convolution $\otimes$ is in the
longitudinal momentum fractions indicated by $\cdots$
(cf.~figure~\ref{fig:intr-nlo}a).  The first term in \eqref{match-tw4}
corresponds to the first term in \eqref{match-master} and is the only
contribution that does not involve a hard splitting at all.  The
ellipsis denotes terms with non-quasipartonic twist-four distributions
containing three or two parton fields, together with one or two parton
splittings.  While having the same power behaviour in $y$, one may
expect that at small $x_1, x_2$ these terms become less important than
the terms with quasipartonic twist-four distributions, which should
roughly grow as fast as the square of two parton densities.

We now take a closer look at the second term in \eqref{match-tw4}.  The
kernel $C$ can be determined by computing both sides of \eqref{match-tw4}
for a given graph.  At order $\alpha_s$ only ``non-diagonal''
interactions, i.e.\ interactions connecting partons 1 and 2 as in
figure~\ref{fig:intr-nlo}a and b, contribute to $C$.  The ladder graph in
figure~\ref{fig:intr-nlo}c is independent of $y$ and thus gives identical
contributions to the matrix elements $F_{\text{int,pt}}$ and $G$.  As a
consequence it does not contribute to $C$.

At this point we can comment on the scale evolution of the different
terms in \eqref{match-tw4}.  The l.h.s.\ evolves according to the
homogeneous double DGLAP equation for DPDs, which describes
``diagonal'' interactions, either between the partons with final
momentum fraction $x_1$ or between those with final momentum fraction
$x_2$.  By contrast, the evolution of $G(x_1,x_2,x_2,x_1;\mu)$
contains both diagonal and non-diagonal ladder interactions
\cite{Bukhvostov:1985rn}.  The non-diagonal interactions in the
evolution must thus be cancelled by the $\mu$ dependence of the term
$C \otimes G$.  At leading order in $\alpha_s$, this dependence comes
only from the coefficient function $C$, which indeed contains just
non-diagonal interactions as just discussed.

We finally emphasise that an unambiguous decomposition of $F(\tvec{y})$
into splitting, intrinsic and twist-three parts is only possible in the
limit of small $y$.  If $y$ is of hadronic size, neither the operator
product expansion nor the notion of perturbative parton splitting make
sense.  One may however use the short-distance decomposition
\eqref{small-y-dec} as a starting point for a model parameterisation of
DPDs in the full $y$ range.  We describe a simple version of this strategy
in section~\ref{sec:numerics}.

\section{A scheme to regulate DPS and avoid double counting}
\label{sec:scheme}

In this section, we present a scheme that regulates the DPS cross
section and solves the double counting problem between DPS and SPS, as
well as between DPS and the twist-four contribution
(figure~\ref{fig:boxed-graphs}b).  Before doing so, we discuss some
general considerations that motivate our scheme.

The following properties are in our opinion desirable for any
theoretical setup describing double parton scattering.
\begin{enumerate}
\item \label{prop:non-pert} It should permit a field theoretical
  definition of DPDs, without recourse to perturbation theory.  This
  is the same standard as for the ordinary parton distributions in SPS
  processes.  In particular, it allows one to derive general
  properties and to investigate these functions
  using nonperturbative methods, for instance lattice calculations.

  One may object that so far not even ordinary PDFs can be computed to
  a precision sufficient for phenomenology.  However, important
  progress has been made in the area of lattice computations, and more
  can be expected for the future.  Furthermore, whereas ordinary PDFs
  are being extracted with increasing precision from experiment, it is
  hard to imagine a similar scenario for DPDs, because of their sheer
  number and because DPS processes are much harder to measure and
  analyse than most processes from which PDFs are extracted.  In such
  a situation, even semi-quantitative guidance from nonperturbative
  calculations (such as the relevance of correlations of different
  types) is highly valuable.

  As already discussed in section~\ref{sec:scene}, the requirement of
  a nonperturbative definition prevents us from separating the
  ``perturbative splitting'' contribution of a DPD in a controlled way
  for all distances $\tvec{y}$ (or equivalently for all conjugate
  momenta $\tvec{\Delta}$).
\item \label{prop:high-ord} To pave the way for increased theoretical
  precision, the scheme should permit a formulation at higher orders
  in perturbation theory.  Furthermore, the complexity of the required
  higher order calculations should be manageable in practice.
\item \label{prop:coll} For collinear factorisation, one would like to
  use as much as possible existing higher-order results for SPS
  processes, namely partonic cross sections and splitting functions.
  This means that the scheme should not modify the collinear
  subtractions to be made in hard scattering kernels, nor the validity
  of standard DGLAP evolution for DPDs in the colour singlet channel.
\item \label{prop:tmd} For TMD factorisation, it is desirable not to
  modify Collins-Soper evolution and the handling of rapidity
  divergences.  This again allows one to re-use calculations done for
  SPS, although rapidity evolution for DPS is necessarily more
  complicated due to the complexities caused by colour
  \cite{Diehl:2011yj,Buffing:2016wip}.
\item \label{prop:match} One would like to keep procedures as similar
  as possible for collinear and TMD factorisation.  This will in
  particular facilitate the computation of DPS processes at
  perturbatively large transverse momenta in terms of collinear DPDs
  \cite{Buffing:2016wip}, adapting the well known procedure for single
  Drell-Yan production \cite{Collins:1984kg}.
\end{enumerate}

In principle one can use dimensional regularisation to handle the UV
divergences that are induced in the DPS cross section by the
perturbative splitting mechanism, as is done with the UV divergences
that arise in simpler situations such as single hard scattering.
However, contrary to that case, the UV divergences discussed in
section~\ref{sec:scene} arise not at the level of individual DPDs but
only when two DPDs are multiplied together and integrated over
$\tvec{y}$.  This means that if one treats these divergences in
dimensional regularisation, only the product of two DPDs is defined in
$D=4$ dimensions but not the DPDs separately.  This possibility was
explored in \cite{Manohar:2012pe}.  However, DPDs and their products
remain nonperturbative functions at large $\tvec{y}$, which according
to present knowledge cannot be reduced to simpler quantities in a
model independent way.  \rev{In practice, one therefore needs to model or
  parameterise them at some starting scale.  This is more involved for the
  product of DPDs than for DPDs themselves, as is the practical
  implementation of scale evolution.  We will come back to this scheme in
  section~\ref{sec:compare}.}

\begin{description}
\item[Ultraviolet regularisation.] We define the regularised DPS cross
  section by multiplying the integrand in the DPS formula
  \eqref{TMD-Xsect-naive} for measured transverse momenta with
  $\Phi(y_+ \nu) \, \Phi(y_- \nu)$ and the integrand in the collinear
  DPS formula \eqref{coll-Xsect-naive} with $\Phi^2(y\nu)$.  The
  function $\Phi(u)$ goes to $1$ for $u \gg 1$ and to $0$ for
  $u \to 0$, and we can restrict ourselves to the case where
  $0\le \Phi(u) \le 1$ for all $u$.  More specific requirements are
  given below.

  Collinear and transverse-momentum dependent DPDs are defined as
  specified in section~\ref{sec:scene}, without any modifications.
  Constructed from operator matrix elements, they contain both splitting
  and non-splitting contributions.  They quantify specific properties of
  the proton and have a simple physical interpretation, with the same
  limitations as single parton densities.  (We recall that a literal
  density interpretation of PDFs and TMDs is hindered by the presence of
  Wilson lines and of ultraviolet renormalisation.)
\item[Double counting subtraction.] To treat the double counting
  between DPS and other contributions, we adapt the recursive subtraction
  formalism of Collins, which we briefly sketch now (details are given
  in sections~10.1 and 10.7 of \cite{Collins:2011zzd}).  Consider a
  graph (or sum of graphs) $\Gamma$ that receives leading
  contributions from a set of loop momentum regions $R$.  An
  approximation for $\Gamma$ is then given by
  \begin{align}
    \label{subtr-def}
    \Gamma &\,\approx\, \sum_R C_R\ms \Gamma
    & \text{with~~}
      C_R\ms\Gamma &= T_R\ms\Gamma - \sum_{R'<R} T_R\ms C_{R'}\ms \Gamma \,.
  \end{align}
  In each term one integrates over \emph{all} loop momenta.  The operator
  $T_R$ applies approximations designed to work in momentum region $R$.
  Subtraction terms avoid double counting the contributions from smaller
  regions $R'$ (regions that are contained in $R$).  In these subtraction
  terms one applies the approximations designed for $R$ and those designed
  for the smaller regions.  One can show \cite{Collins:2011zzd} that
  $C_R\ms \Gamma$ then provides a valid approximation in the region $R$
  and in all smaller regions, and $\sum_R C_R\ms \Gamma$ gives a valid
  approximation of the graph in all relevant regions.  All approximations
  discussed here are valid up to power corrections.

  In our context, we have graphs in which a set of collinear partons
  split into partons that can be either collinear (as in DPS) or hard
  (as in SPS).  A slight adaptation of the above formalism is required
  since we compute DPS using DPDs and a regulating function $\Phi$
  that depend on transverse distances $\tvec{y}$ rather than
  transverse momenta.  A collinear splitting region $R'$ then
  corresponds to large $y$ and the corresponding hard region $R$ to
  small $y$, but we keep the ordering of regions $R'<R$ from momentum
  space when implementing \eqref{subtr-def}.  We will show in
  section~\ref{sec:mom-subtr} that our use of subtractions in position
  space is equivalent to the one in momentum space up to power
  suppressed effects.

  The subtraction terms for the DPS region turn out to have a very
  simple form.  They can be obtained by replacing the DPDs in the UV
  regularised DPS cross section with their appropriate limits for
  small ${y}_{\pm}$ in TMD factorisation and for small $y$ in collinear
  factorisation.  Details will be given in
  sections~\ref{sec:coll-subtr}, \ref{sec:tmd-subtr} and
  \ref{sec:higher-order}.
\end{description}
Criteria \ref{prop:non-pert} and \ref{prop:match} above are obviously
satisfied in this scheme.  Regarding criterion~\ref{prop:high-ord}, we
note that the higher-order calculations required for the double
counting subtraction terms are for the short-distance limit of DPDs,
which involve much simpler Feynman graphs than the full scattering
process.

The introduction of a function $\Phi$ in the DPS cross section avoids
an explicit modification of the definition of DPDs and thus respects
criteria \ref{prop:coll} and \ref{prop:tmd}.  In particular, the
collinear DPDs $F(x_i,\tvec{y})$ in transverse position space follow
the homogeneous DGLAP evolution equation \eqref{DGLAP-mu1}.  Since for
colour singlet DPDs, the evolution kernels are the familiar DGLAP
kernels, the associated scale dependence in the cross section cancels
by construction against the one of the hard cross sections computed
with the same collinear subtraction as for SPS.

In our scheme, we have introduced an additional momentum scale $\nu$
to separate DPS from SPS and the twist-four contribution.  In
practical calculations one may take $\nu$ equal to the UV
renormalisation scale $\mu$ in DPDs, but we find it useful to keep it
separate in the general discussion.  As a minimal requirement, $\nu$
must be of perturbative size, so that the double counting subtraction
terms remove all contributions from nonperturbative regions in the
hard kernels of the SPS and twist-four cross sections, making their
calculation consistent.  By construction, the $\nu$ dependence in the
physical cross section cancels between DPS and the double counting
subtraction terms, up to higher orders in $\alpha_s$ that are beyond
the accuracy of the computation.  We will come back to this issue in
section~\ref{sec:dglap-subtr}.


Let us now take a closer look at the properties required for the
function $\Phi$.
\begin{itemize}
\item The introduction of $\Phi$ in the DPS cross section must not
  spoil the correct description of the physics in the region where the
  DPD approximations work.  Specifically, we require that the
  modifications introduced by $\Phi$ in that region should be power
  suppressed in the large scale.  This requires that $\Phi(u)$ must
  approach $1$ sufficiently fast when $u\gg 1$.

  Anticipating the momentum space analysis in
  section~\ref{sec:mom-subtr}, we demand that for collinear
  factorisation the Fourier transformation of $1-\Phi^2$ exists in
  $2-2\epsilon$ dim (for positive and negative $\epsilon$).  Therefore
  the integral
  $\int du\, u^{1 - 2\epsilon} \bigl[ 1 - \Phi^2(u) \bigr]$ must
  converge for $u\to \infty$.  The corresponding criterion for TMD
  factorisation is obtained by replacing $\Phi^2 \to \Phi$.

\item $\Phi(u)$ at small $u$ must regulate the UV divergences of the
  naive DPD cross section in 2 dimensions.  This means that
  $\int du\, u^{-1}\, \Phi(u)$ must be integrable at $u=0$ in TMD
  factorisation.  For collinear factorisation we have the stronger
  requirement that $\int du\, u^{-3}\, \Phi^2(u)$ be integrable at
  $u=0$.  These criteria are satisfied if for $u\to 0$
\begin{align}
  \label{Phi-conditions}
  \Phi(u) &= O(u^\delta)     &&\mbox{for TMD factorisation}
\nonumber \\
  \Phi(u) &= O(u^{1+\delta}) &&\mbox{for collinear factorisation}
\end{align}
with some $\delta > 0$.

\item To compute the double counting subtraction terms for SPS and the
  twist-four contribution, we must perform integrals over $\tvec{y}$ in
  $2$ and $2-2\epsilon$ transverse dimensions, respectively, as we
  shall see in the next sections.  We choose $\Phi$ such that the
  required integrals are known analytically.  This is especially
  important for the twist-four contribution, where we must expand the
  result around $\epsilon=0$.
\end{itemize}

A suitable function for both collinear and TMD factorisation is the
step function
\begin{align}
  \label{step-fct}
\Phi(u) = \Theta(u-b_0) & & \text{with~~} b_0 = 2 e^{-\gamma} \,,
\end{align}
where $\gamma$ is Euler's constant.  This corresponds to a hard lower
cutoff $y > b_0/\nu$ in the $y$ integration, where the constant
$b_0 \approx 1.12$ is taken to simplify certain analytical results.
An alternative choice is
\begin{align}
\Phi(u) = 1 - \exp(-u^2/4) \,.
\end{align}
Important integrals for these functions are given in table~\ref{tab:phi}.
Further possible choices for collinear factorisation are
$\Phi^2(u) = 1 - \exp(- u^p)$ or $\Phi^2(u) = u^p /(1 + u^p)$ with
$p>2$; the corresponding integrals in the third and fourth row of
table~\ref{tab:phi} can be performed and give Euler $\Gamma$
functions.

\begin{table}[th]
\begin{displaymath}
\renewcommand{\arraystretch}{2.2}
\begin{array}{c@{~~~~}c@{~~~~~}c}   \hline \hline
\Phi(u)   & \Theta(u-b_0) & 1 - \exp(-u^2/4) \\ \hline
2 \displaystyle \int_0^\infty \frac{du}{u}\, J_0(u r)\, \Phi(u) &
  \log\dfrac{1}{r^2} + \mathcal{O}(r^2) &
  \log\dfrac{1}{r^2} - \gamma + \mathcal{O}(r^2) \\
2 \displaystyle \int_0^\infty \frac{du}{u^{1-2\epsilon}}\, \Phi^2(u) &
  - \dfrac{(b_0)^{2\epsilon}}{\epsilon}
     = - \dfrac{1}{\epsilon} + \mathcal{O}(\epsilon^0) & 
   2^{\epsilon}\ms \bigl[ 1 - 2^{1+\epsilon} \bigr]\, \Gamma(\epsilon)
     = - \dfrac{1}{\epsilon} + \mathcal{O}(\epsilon^0) \\
2 \displaystyle \int_{0}^\infty \frac{du}{u^{3-2\epsilon}}\, \Phi^2(u) &
  \dfrac{1}{1-\epsilon}\, \Bigl( \dfrac{1}{b_0} \Bigr)^{2-2\epsilon} &
  \dfrac{1- 2^{\epsilon}}{2^{1-\epsilon}}\,
     \Gamma(\epsilon-1) = \dfrac{\log 2}{2} + \mathcal{O}(\epsilon)
\\[0.5em] \hline \hline
\end{array}
\end{displaymath}
\caption{\label{tab:phi} Examples for the cutoff function $\Phi$ and
  relevant integrals.  Exact results for the integral in the second row
  are given in \protect\eqref{bessel-step} and
  \protect\eqref{bessel-exp}.  The integral in the third row converges
  for $\epsilon < 0$.}
\end{table}


\subsection{Leading order analysis: collinear factorisation}
\label{sec:coll-subtr}

\rev{In this section we show how our formalism works in collinear
  factorisation, concentrating on the leading order in $\alpha_s$.
  Following the procedure of all-order factorisation proofs (see for
  instance \cite{Collins:2011zzd}), we rewrite individual Feynman graphs
  in a way consistent with the final factorisation formula.  At the end of
  this procedure, the factors associated with long-distance physics in
  that formula, such as PDFs and DPDs, are expressed in terms of operator
  matrix elements and thus defined in a nonperturbative way.  During our
  Feynman graph analysis we can separately consider splitting and
  intrinsic contributions to a DPD, $F_{\text{spl}}$ and $F_{\text{int}}$.
  Likewise, we can consider separate contributions $\sigma_{\text{1v1}}$,
  $\sigma_{\text{2v1}}$ and $\sigma_{\text{2v2}}$ to the DPS cross
  section, which respectively involve the combinations $F_{\text{spl}} \,
  F_{\text{spl}}$, $F_{\text{int}} \, F_{\text{spl}} + F_{\text{spl}} \,
  F_{\text{int}}$ and $F_{\text{int}} \, F_{\text{int}}$, corresponding to
  graphs a, b and c in figure~\ref{fig:boxed-graphs}.  In the final
  factorisation formula, only the full DPDs $F$ and the full DPS cross
  section $\sigma_{\text{DPS}}$ will appear.  For brevity, we write
  $\sigma$ instead of $d\sigma /(dx_1\ms dx_2\ms d\bar{x}_1\ms
  d\bar{x}_2)$ here and in the following.}


\subsubsection{Squared box graph}
\label{sec:box}

Let us start with the squared box graph in figure~\ref{fig:boxed-graphs}a,
integrated over all transverse momenta in the final state.  As discussed in
section~\ref{sec:scene}, it receives its leading contribution from the
region $q \sim Q$.  \rev{The quark loops in the amplitude and its conjugate
  are then in the hard momentum region.  This gives a part of the SPS cross
  section $\sigma_{\text{SPS}}$ and is computed as usual in terms of PDFs
  and the cross section for $gg$ annihilation into two gauge bosons.  Other
  graphs that contribute to SPS, such as the one in
  figure~\ref{fig:sps-dps-int}a, have no overlap with DPS.  In the present
  subsection we restrict $\sigma_{\text{SPS}}$ and other terms in the cross
  section to refer to the double box graph only.}

The DPS region of the graph has near-collinear splittings in both the
amplitude and its conjugate, which is only possible at $q \ll Q$.
\rev{This gives the 1v1 contribution $\sigma_{\text{1v1}}$ to the DPS
  cross section, which is power suppressed compared with
  $\sigma_{\text{SPS}}$ because we integrate over the transverse boson
  momenta.  Since the double box graph is integrated over all $q$ in
  $\sigma_{\text{SPS}}$, we subtract the contribution of the DPS region in
  order to prevent double counting.  (At leading power accuracy this is
  not strictly necessary, but we will shortly see that it is useful).}
According to \eqref{subtr-def} this contribution is obtained by applying
to the graph the approximations for DPS as well as those for SPS.  This
gives the DPS expression with the $g\to q\bar{q}$ splittings computed
perturbatively, the incoming gluons treated as collinear on-shell partons,
and quark masses in the loop set to zero.  These approximations are just
what goes into the \rev{perturbative splitting approximation
  $F_{\text{spl,pt}}$ of a DPD, which is given in \eqref{split-lo}.  The
  subtraction term thus reads}
\begin{align}
  \label{1v1-subtr-lo}
\frac{d\sigma_{\text{1v1,pt}}}{dx_1\, dx_2\,
  d\bar{x}_1\, d\bar{x}_2}  &= \sum_{a_1 a_2 b_1 b_2}
    \hat{\sigma}_{a_1 b_1}\ms \hat{\sigma}_{a_2 b_2}\ms
  \int d^2\tvec{y}\, \Phi^2(y \nu)\,
     F_{b_1 b_2,\ms \text{spl,pt}}(\bar{x}_i, \tvec{y})\,
     F_{a_1 a_2,\ms \text{spl,pt}}(x_i, \tvec{y}) \,.
\end{align}
\rev{Note that in $\sigma_{\text{1v1,pt}}$ we use the short-distance
  approximation $F_{\text{spl,pt}}$ (even for large $\tvec{y}$, where it
  is not valid), whereas in $\sigma_{\text{1v1}}$ we use the
  un-approximated splitting part $F_{\text{spl}}$ of each DPD.}  To ensure
that $\sigma_{\text{1v1}}$ and $\sigma_{\text{1v1,pt}}$ receive
contributions only from the DPS region, one should take $\nu \ll Q$.

To compute \eqref{1v1-subtr-lo} we can use the known form
\eqref{split-lo} for the splitting DPDs, with the kernels collected in
\cite{Diehl:2014vaa}.  With $\sigma_{\text{SPS}}$ being computed in
fixed order perturbation theory, we make the same approximations in
$\sigma_{\text{1v1,pt}}$, taking the renormalisation and
factorisation scales in the splitting kernel and in the collinear PDF
fixed.  In $D=2$ dimensions the splitting DPDs then depend on $y$ like
$1/y^2$.  The $y$ integration in \eqref{1v1-subtr-lo} is thus readily
performed,
\begin{align}
  \label{coll-subt-int}
\int \frac{d^2\tvec{y}}{y^4}\, \Phi^2(y\nu) &= 
  \nu^2\! \int \frac{d^2\tvec{u}}{u^4}\, \Phi^2(u) \,.
\end{align}
This is proportional to $\nu^2$, which is expected since the
unregulated integral has a quadratic divergence in $y$ and $\nu$ is
the scale regulating this divergence.  The relevant integral for
different $\Phi(u)$ is obtained from the last row of
table~\ref{tab:phi} by setting $\epsilon=0$.

We anticipate that the subtraction term at higher orders in $\alpha_s$
involves an integral as in \eqref{1v1-subtr-lo}, with splitting DPDs
given by \eqref{split-ho} and \eqref{V-series-zero}.  Compared with
\eqref{coll-subt-int} the integrand then has additional powers of
$\log (y \mu)$.  The integrals can be obtained from those in the last
row of table~\ref{tab:phi} by Taylor expanding around $\epsilon=0$.

The complete contribution of the double box graph to the cross section is
\begin{align}
  \label{1v1-combined}
\sigma_{\text{SPS}} - \sigma_{\text{1v1,pt}} + \sigma_{\text{1v1}} \,.
\end{align}
As is well known, an
appropriate choice for the factorisation scale $\mu$ in collinear
factorisation is $\mu \sim Q$.  In the same spirit we take
$\nu \sim Q$ rather than $\nu \ll Q$, although this extends the $y$
integrals in $\sigma_{\text{1v1}}$ and $\sigma_{\text{1v1,pt}}$ to
values $y \sim 1/Q$ where the DPS approximation does not work.  Both
$\sigma_{\text{1v1}}$ and $\sigma_{\text{1v1,pt}}$ are then of the
same order as $\sigma_{\text{SPS}}$ rather than power suppressed.  Let
us see that this still gives the correct approximation of the overall
cross section.

Although $\sigma_{\text{SPS}}$ is naturally computed in momentum
space, one can perform a Fourier transform w.r.t.\ the transverse
momentum difference $\tvec{\Delta}$ (defined as in
figure~\ref{fig:split-mom} but for the graph of the cross section).
Then $\sigma_{\text{SPS}}$ is a loop integral over the conjugate
distance $\tvec{y}$, just as the two other terms in
\eqref{1v1-combined}.

We now show how the subtraction formalism works in this case.  In the
region of small $y \sim 1/Q$, one has
$\sigma_{\text{1v1}} \approx \sigma_{\text{1v1,pt}}$, because the
perturbative approximation of the DPD is designed to work at short
distances.  The dependence on the cutoff function $\Phi(y \nu)$
cancels between these two terms.  One is therefore left with
$\sigma_{\text{SPS}}$, which gives the appropriate description of the
graph for $y \sim 1/Q$.  We note that the DPDs in
$\sigma_{\text{1v1,pt}}$ are computed at fixed order in $\alpha_s$,
whereas in $\sigma_{\text{1v1}}$ they should be resummed using the
double DGLAP equations.  The cancellation between the two terms is
therefore only up to higher orders in $\alpha_s$, which are beyond the
accuracy of the result.  We shall investigate this in detail in
section~\ref{sec:dglap-subtr}.

Turning to the region of large $y \gg 1/Q$, one finds that
$\sigma_{\text{SPS}} \approx \sigma_{\text{1v1,pt}}$, because
precisely in that region the DPS approximation is designed to work for
the graph.  We are therefore left with the DPS term
$\sigma_{\text{1v1}}$, which is exactly what we want to describe this
region.  The cutoff function $\Phi(y \nu)$ has no effect here, since
we take $\nu \sim Q$ and require $\Phi(u) \approx 1$ for $u \gg 1$.
The combination \eqref{1v1-combined} thus gives the correct
approximation for both small and large $y$.


\subsubsection{2v1 graph}
\label{sec:2v1}

We now derive a description for the 2v1 graph in
figure~\ref{fig:boxed-graphs}b.  Integrated over transverse momenta in
the final state, the graph receives leading contributions from small
and large transverse momenta of the quarks on the splitting side
(i.e.\ the top of the graph), corresponding to small or large
transverse momenta of the produced gauge bosons.  The region of small
transverse momenta corresponds to DPS, whilst the contribution from
large transverse momenta is correctly described by the twist-four
mechanism introduced in section~\ref{sec:scene}.

\begin{figure}
\begin{center}
\includegraphics[width=0.44\textwidth]{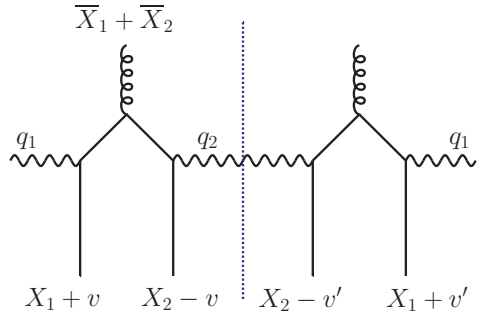}
\caption{\label{fig:2v1-labels} Lowest order graph for the 2v1 mechanism,
  with the blobs connecting partons with hadrons being omitted at the top
  and the bottom.  The vector boson with momentum $q_1$ is understood to
  cross the final state cut.  The four quark lines at the bottom are right
  moving, the two gluons entering at the top left moving.}
\end{center}
\end{figure}

Whereas the SPS contribution in the previous subsection gives a finite
result when computed as a perturbative two-loop graph, the twist-four
contribution has a collinear divergence when integrated over the full
region of phase space.  To exhibit this divergence, we use dimensional
regularisation and write
\begin{align}
  \label{tw4-start}
\frac{d\sigma_{\text{tw4}}}{dx_1\, dx_2\,
  d\bar{x}_1\, d\bar{x}_2}
 &= \int d^{D-2}\tvec{q}\, \int_{-X_1}^{X_2} dv \int_{-X_1}^{X_2} dv'\,
     \sum_{a_1 a_2} G_{a_1 a_2}(X_1 + v, X_2 - v, X_2 - v', X_1 + v')\,
\nonumber \\
  &\quad \times H_{a_1 a_2,g}(\tvec{q}, X_i, v, v', \widebar{X}_i, s)\,
         f_g(\widebar{X}_1 + \widebar{X}_2)\,.
\end{align}
As in the previous subsection, we restrict the meaning of
$\sigma_{\text{tw4}}$ and other terms in the cross section to the
contribution from a single graph.  The labels $a_1$ and $a_2$ indicate
parton species and polarisation, and $G$ is a collinear twist-four
distribution.  $H$ includes overall factors and the squared amplitude
of the hard scattering process $q\bar{q} + g \to V_1 V_2$.  Note that
in the hard scattering graph one has
$\tvec{q}_1 = - \tvec{q}_2 = \tvec{q}$.  Momentum fractions are as
shown in figure~\ref{fig:2v1-labels}. The plus- and minus-momentum
fractions of the produced bosons are respectively given by
\begin{align}
  \label{big-X-def}
X_i &= \frac{q_i^+}{p^+}
     = \sqrt{\frac{Q_i^2 + {q}_i^2}{Q_i^2}}\, x_i \,,
&
\widebar{X}_i &= \frac{q_i^-}{\bar{p}^{\, -}}
           = \sqrt{\frac{Q_i^2 + {q}_i^2}{Q_i^2}}\, \bar{x}_i
\end{align}
with $x_i$ and $\bar{x}_i$ defined in \eqref{little-x-def}.  For the
momenta of internal lines in the hard scattering subprocess, minus and
transverse components are fixed by the final state kinematics, so that
there is only a loop integration over the plus-momentum fractions $v$
and $v'$.  Note that by virtue of \eqref{big-X-def} both $X_i$ and
$\widebar{X}_i$ have an implicit dependence on $q^2$ at given
$x_i, \bar{x}_i$.  An upper limit on $q^2$ follows from the
requirements that $X_1+X_2 \le 1$ and
$\widebar{X}_1 + \widebar{X}_2 \le 1$.

The expression \eqref{tw4-start} includes a contribution from DPS
region, characterised by $q_i \ll Q_i$ and $v, v' \ll 1$.  Let us
approximate the integrand in that region.  We first perform a Fierz
transformation for the Dirac indices of the upper quark lines at the
gauge boson vertices and retain only the leading Dirac structures.
After this, the cross sections $\hat{\sigma}_i$ for $q\bar{q}$
annihilation appear in the expression.  Furthermore we set $v=v'=0$ in
$G$ and extend the integrations over $v$ and $v'$ to the full real
axis (after which they are easily done using Cauchy's theorem).
Approximating $X_i \approx x_i$ and $\widebar{X}_i \approx \bar{x}_i$
and collecting all factors, we obtain
\begin{align}
  \label{tw4-final}
& \frac{d\sigma_{\text{tw4}}}{dx_1\, dx_2\,
  d\bar{x}_1\, d\bar{x}_2} \bigg|_{q \ll Q_i \text{~and~} v,v' \ll 1}
 =  \sum_{a_1 a_2 b_1 b_2} 
    \hat{\sigma}_{a_1 b_1}\ms \hat{\sigma}_{a_2 b_2}\ms
    G_{a_1 a_2}(x_1, x_2, x_2, x_1)\,
\nonumber \\
 & \qquad\quad \times \int\limits_{q \ll Q_i} d^{D-2}\tvec{q}\;
     \frac{\tvec{q}^{j} \tvec{q}^{j'}}{q^4}\,
     \frac{(2 \mu)^{4-D}}{\pi^{D-3}}\,
            \frac{\alpha_s}{2\pi}\; T^{jj'}_{g\to b_1 b_2}\biggl(
            \frac{\bar{x}_1}{\bar{x}_1+\bar{x}_2} \biggr)\,
     \frac{f_g(\bar{x}_1+\bar{x}_2)}{\bar{x}_1+\bar{x}_2}
\end{align}
with the splitting kernel $T$ from \eqref{TMD-split}.  If we drop the
restriction $q \ll Q_i$, then the expression in the second line of
\eqref{tw4-final} becomes the perturbative splitting approximation
$F_{\text{spl,pt}}(\bar{x}_i, \tvec{\Delta}=\tvec{0})$ of the
transverse-momentum integrated DPD.  This DPD is evaluated at
$\tvec{\Delta} = \tvec{0}$ in order to fulfil momentum conservation at
the gauge boson vertices, because the transverse momenta of the right
moving partons are set to zero in the overall hard scattering kernel~$H$.

The $\tvec{q}$ integral of \eqref{tw4-final} has an infrared divergence
in $D=4$ dimensions, which must be cancelled by the subtraction term for
the DPS region.  That term is obtained by applying the DPS approximations
to the graph, in addition to those for the twist-four region.  This gives
again the form \eqref{tw4-final}, but with an unrestricted integration
over $\tvec{q}$, and a regulator of the corresponding ultraviolet
divergence.  To implement the regulator, we Fourier transform the
collinear DPD, $F_{\text{spl,pt}}(\bar{x}_i, \tvec{\Delta}=\tvec{0}) =
\int d^{D-2}\tvec{y}\; F_{\text{spl,pt}}(\bar{x}_i, \tvec{y})$, and then
multiply with $\Phi^2(y \nu)$ as prescribed by our formalism.  This gives
\begin{align}
  \label{tw4-sub}
& \frac{d\sigma_{\text{2v1,pt}}}{dx_1\, dx_2\, d\bar{x}_1\, d\bar{x}_2}
\nonumber \\
 &\quad = \sum_{a_1 a_2 b_1 b_2}
    \hat{\sigma}_{a_1 b_1}\ms \hat{\sigma}_{a_2 b_2}\ms
    G_{a_1 a_2}(x_1, x_2, x_2, x_1)\,
    \int d^{D-2}\tvec{y}\; \Phi^2(y \nu)\,
         F_{b_1 b_2\, \text{spl,pt}}(\bar{x}_i, \tvec{y}) \,.
\end{align}
Alternatively, one can obtain the subtraction term by starting with
\rev{the contribution $\sigma_{\text{2v1}}$ of graph
  \ref{fig:boxed-graphs}b to the DPS cross section} and applying the
additional approximations adequate for the twist-four region.  Let us
translate the latter from transverse momentum (where the twist-four
approximations are formulated) to transverse position (where our UV
regulator for the DPS cross section is local).  On the l.h.s.\ of the
relation
\begin{align}
\int \frac{d^{D-2}\tvec{\Delta}}{(2\pi)^{D-2}}\;
  F_{\text{int}}(x_i, \tvec{\Delta})\,
  F_{\text{spl}}(\bar{x}_i, -\tvec{\Delta})
 &= \int d^{D-2}\tvec{y}\;
  F_{\text{int}}(x_i, \tvec{y})\, F_{\text{spl}}(\bar{x}_i, \tvec{y})
\end{align}
one neglects $\tvec{\Delta}$ in the upper part of the graph, replacing
$F_{\text{spl}}(\bar{x}_i, \tvec{\Delta})$ by its value at
$\tvec{\Delta} = \tvec{0}$.  The integral over $\tvec{\Delta}$ of
$F_{\text{int}}(x_i, \tvec{\Delta})$ gives the collinear twist-four
distribution $G(x_1,x_2,x_2,x_1)$.  In $\tvec{y}$ space this
corresponds to replacing $F_{\text{int}}(x_i, \tvec{y})$ by its value
at $\tvec{y} = \tvec{0}$ while retaining the $\tvec{y}$ dependence of
$F_{\text{spl}}(\bar{x}_i, \tvec{y})$.  Including the regulator
function $\Phi^2(y \nu)$ under the $\tvec{y}$ integral, one gets
$F_{\text{int}}(x_i, \tvec{y} = \tvec{0}) \int d^{D-2}\tvec{y}\;
\Phi^2(y \nu)\, F_{\text{spl}}(\bar{x}_i, \tvec{y})$.
The twist-four approximation implies computing the $g\to q\bar{q}$
splitting in perturbation theory, with massless quarks and the gluons
taken collinear and on shell.  This corresponds to replacing
$F_{\text{spl}}$ with $F_{\text{spl,pt}}$.  For the graph under
discussion one has
$F_{\text{int}}(x_i, \tvec{y}=\tvec{0}) = G(x_1,x_2,x_2,x_1)$
according to \eqref{match-tw4} and thus obtains \eqref{tw4-sub}.  We
see that, as in the case of the 1v1 graph, the subtraction term is
obtained from the DPS cross section by replacing the DPDs with their
small $y$ approximation at the appropriate order in $\alpha_s$.  In
section~\ref{sec:higher-order} we will find that this also holds at
higher orders, where we have to take into account the second term
in~\eqref{match-tw4}.

Let us finally perform the $\tvec{y}$ integral in the subtraction term
\eqref{tw4-sub}.  Collecting all $\tvec{y}$ and $\epsilon$ dependent
factors in \eqref{split-lo}, we have
\begin{align}
I(\mu,\nu) &=
    \mu^{2\epsilon}\, \frac{\Gamma^2(1-\epsilon)}{\pi^{1-2\epsilon}}\,
    P(\epsilon)
    \int \frac{d^{2-2\epsilon}\tvec{y}}{y^{2-4\epsilon}}\; \Phi^2(y \nu)
\nonumber \\
 &= \biggl( \frac{\mu}{\nu} \biggr)^{2\epsilon}\,
    \frac{\Gamma^2(1-\epsilon)}{\pi^{1-2\epsilon}}\;
    P(\epsilon)\, \Omega_{2-2\epsilon}
    \int_0^\infty \frac{du}{u^{1-2\epsilon}}\; \Phi^2(u) \,,
\end{align}
where $\Omega_{2n} = 2 \pi^{n} / \Gamma(n)$ is the surface of a sphere in
$2n$ dimensions.  For brevity we do not display the momentum fraction in
the splitting kernel $P$.  The $\tvec{y}$ integral is infrared divergent
in $2$ dimensions and converges for $\epsilon < 0$.  Using the third row
of table~\ref{tab:phi}, we get in particular
\begin{align}
  \label{2v1-subtr}
I(\mu,\nu) &= \biggl[ - \frac{1}{\epsilon} - \log(4\pi) + \gamma 
 + \log \frac{\nu^2}{\mu^2} \ms\biggr]\, P(0) - P'(0) + \mathcal{O}(\epsilon)
\end{align}
for $\Phi(u) = \Theta(u-b_0)$, where $P' = \partial P/\partial\epsilon$.

The contribution of the graph to the overall cross section is finally
given by
\begin{align}
  \label{2v1-combined}
\sigma_{\text{tw4}} - \sigma_{\text{2v1,pt}} + \sigma_{\text{2v1}} \,,
\end{align}
in analogy to \eqref{1v1-combined}.  By the same argument as in the
previous subsection, one finds that this combination reproduces the graph
over the full range of $y$ values, with the second term cancelling against
the third term for $y \sim 1/Q$ and against the first term for $y \gg
1/Q$.

As we have just seen, the product of $(\mu/\nu)^{2\epsilon}$ with the
$1/\epsilon$ pole gives a logarithm $\log (\nu/\mu)$ in the
subtraction term $\sigma_{\text{2v1,pt}}$.  In turn, the $\tvec{q}$
integral in the expression \eqref{tw4-final} of $\sigma_{\text{tw4}}$
gives $\log (Q/\mu)$, where $Q$ comes from the upper integration limit
and the $\mu$ dependence from the regulated infrared divergence.  In
the combination $\sigma_{\text{tw4}} - \sigma_{\text{2v1,pt}}$ this
gives $\log (Q/\nu)$.  The $\nu$ dependence is cancelled by a
$\log (\nu/\Lambda)$ from the $\tvec{y}$ integral of the DPS cross
section $\sigma_{\text{2v1}}$, where $\nu$ regulates the ultraviolet
divergence.  Combining all terms, one obtains $\log (Q/\Lambda)$ in
\eqref{2v1-combined}.  With the choice $\nu \sim Q$, the large part of
the logarithm is contained only in the DPS term.


\subsubsection{Combining contributions}
\label{sec:coll-combine}

We can now add all contributions to the physical cross section, including
1v1, 2v1 and 2v2 graphs in DPS.  We include the same regulator $\Phi(y
\nu)$ in all DPD contributions, so that the splitting and intrinsic
contributions to the individual DPDs can be added up to \rev{distributions
  $F = F_{\text{spl}} + F_{\text{int}}$ that are defined by operator
  matrix elements as described in section~\ref{sec:scene}.}

2v2 graphs like the one in figure~\ref{fig:boxed-graphs}c are dominated by
the DPS region, where transverse parton momenta are small, whilst regions
with large transverse momenta are subleading.  In impact parameter space
this means that including the $\Phi$ regulator to suppress the region of
small $y$ does not significantly change \rev{$\sigma_{\text{2v2}}$.}  We
will see in section~\ref{sec:mom-subtr} that the effect of $\Phi$ in the
2v2 term is at the level of power corrections.

The master formula for the cross section with integrated transverse
momenta is then
\begin{align}
  \label{coll-master}
\sigma &= \sigma_{\text{SPS}} - \sigma_{\text{1v1,pt}}
 + \sigma_{\text{tw4}} - \sigma_{\text{2v1,pt}}
 + \sigma_{\text{DPS}} \,,
\end{align}
\rev{where $\sigma_{\text{DPS}} = \sigma_{\text{1v1}} +
  \sigma_{\text{2v1}} + \sigma_{\text{2v2}}$ includes all contributions to
  the DPS cross section and involves the full DPDs $F$.  From the
  discussion in the previous two subsections it follows that the $\nu$
  dependence of $\sigma_{\text{DPS}}$ cancels} against the one of
$\sigma_{\text{1v1,pt}} + \sigma_{\text{2v1,pt}}$, up to terms of higher
order in $\alpha_s$ that are beyond the accuracy of the calculation.  As
discussed in section~\ref{sec:power} we neglect contributions involving
twist-three functions in \eqref{coll-master}.

The terms in \eqref{coll-master} are now meant to include all
contributing graphs, including graphs without a double counting
issue. The hard scattering cross sections needed for the DPS term are
available for many final states, and $\sigma_{\text{1v1,pt}}$ can be
readily obtained from existing results at leading order in $\alpha_s$.
For double gauge boson production, the SPS cross section has been
computed at high perturbative order, including the double box graphs
as well as the first radiative corrections to them
\cite{Caola:2016trd}.  On the other hand, to the best of our
knowledge, no calculations of $\sigma_{\text{tw4}}$ have been
performed so far.  To leading logarithmic accuracy, one can however
omit the terms $\sigma_{\text{tw4}} - \sigma_{\text{2v1,pt}}$ in
\eqref{coll-master}, because with the choice $\nu \sim Q$ the large
DPS logarithm $\log(Q/\Lambda)$ generated by 2v1 graphs is entirely
contained in $\sigma_{\text{DPS}}$, as discussed in the previous
subsection.


\subsection{Leading order analysis: TMD factorisation}
\label{sec:tmd-subtr}

We now show how our formalism works for TMD factorisation.  As explained
in section~\ref{sec:power}, the cross section for $q_1, q_2 \ll Q$
receives leading-power contributions from DPS, SPS and from their
interference.  The 2v1 contribution is power suppressed in this case
\rev{and will not be discussed further in this subsection.}  The
discussion of leading momentum regions is a bit more involved now, because
the two quark loops in the double box graph can be collinear or hard
independently of each other.

We start with the SPS/DPS interference, for which a graph is shown in
figure~\ref{fig:sps-dps-int}b.  Its contribution to the $V_1 V_2$
production cross section has the form
\begin{align}
  \label{sps-dps-int}
\frac{d\sigma_{\text{DPS/SPS}}}{dx_1\, dx_2\,
  d\bar{x}_1\, d\bar{x}_2\, d^2\tvec{q}_1\, d^2\tvec{q}_2} &\,\propto\,
  \sum_{\genfrac{}{}{0pt}{1}{\alpha_1,
    \alpha_2,\alpha_0}{\beta_1,\beta_2,\beta_0}}
  H_{\alpha_1 \beta_1}^{}\, H_{\alpha_2 \beta_2}^{}\, H^*_{\alpha_0 \beta_0}
  \int d^2\tvec{Z}\, d^2\tvec{y}_+\;
       e^{-i (\tvec{q}_1 + \tvec{q}_2) \tvec{Z} -i \tvec{q} \tvec{y}_+}
\nonumber \\
  &\qquad \times \Phi(y_+ \nu)\,
   D_{\beta_1 \beta_2| \beta_0}(\bar{x}_i, \tvec{y}_+, \tvec{Z}) \,
   D_{\alpha_1 \alpha_2| \alpha_0}(x_i, \tvec{y}_+, \tvec{Z}) \,,
\nonumber \\[0.6em]
\frac{d\sigma_{\text{SPS/DPS}}}{dx_1\, dx_2\,
  d\bar{x}_1\, d\bar{x}_2\, d^2\tvec{q}_1\, d^2\tvec{q}_2} &\,\propto\,
  \sum_{\genfrac{}{}{0pt}{1}{\alpha_0,
                 \alpha_1,\alpha_2}{\beta_0,\beta_1,\beta_2}}
            H^{}_{\alpha_0 \beta_0}\,
            H^*_{\alpha_1 \beta_1}\, H^*_{\alpha_2 \beta_2}
  \int d^2\tvec{Z}\, d^2\tvec{y}_-\;
       e^{-i (\tvec{q}_1 + \tvec{q}_2) \tvec{Z} +i \tvec{q} \tvec{y}_-}
\nonumber \\
  &\qquad \times \Phi(y_- \nu)\,
   D_{\beta_0| \beta_1 \beta_2}(\bar{x}_i, \tvec{y}_-, \tvec{Z}) \,
   D_{\alpha_0| \alpha_1 \alpha_2}(x_i, \tvec{y}_-, \tvec{Z})
\end{align}
\rev{for DPS in the amplitude or in its complex conjugate.}  Here $D$ are
the twist-three TMDs introduced in section~\ref{sec:tmd-split} and the
proportionality is up to kinematic and numerical factors.  We have
different hard scattering amplitudes $H$, with $\alpha_0$ and $\beta_0$
being gluons and the remaining partons quarks or antiquarks .  The regions
of small $y_+$ or $y_-$ are regulated by $\Phi(y_+ \nu)$ and $\Phi(y_-
\nu)$.

Notice that the graph in figure~\ref{fig:sps-dps-int}b has a leading
contribution when the $g\to q\bar{q}$ splittings in the conjugate
amplitude become collinear, in which case we are back in the DPS region.
From the first \rev{expression} in \eqref{sps-dps-int} one should
therefore subtract
\begin{align}
  \label{sps-dps-subt}
& \frac{d\sigma_{\text{DPS},\, y_-\to 0}}{dx_1\, dx_2\,
  d\bar{x}_1\, d\bar{x}_2\, d^2\tvec{q}_1\, d^2\tvec{q}_2} \,\propto\,
  \sum_{\genfrac{}{}{0pt}{1}{\alpha_1,\alpha_2}{\beta_1,\beta_2}}
  H^{}_{\alpha_1 \beta_1}\, H^{}_{\alpha_2 \beta_2}\,
  H^*_{\alpha_1 \beta_1}\, H^*_{\alpha_2 \beta_2}\,
\nonumber \\
  &\qquad\qquad \times \int d^2\tvec{Z}\, d^2\tvec{y}_+\, d^2\tvec{y}_-\;
       e^{-i (\tvec{q}_1 + \tvec{q}_2) \tvec{Z} 
          -i \tvec{q} (\tvec{y}_+ - \tvec{y}_-)}\,
       \Phi(y_+ \nu)\, \Phi(y_- \nu)\,
\nonumber \\[0.2em]
  &\qquad\qquad\quad \times
   F_{\beta_1 \beta_2,\ms
          y_- \to 0}(\bar{x}_i, \tvec{y}_\pm, \tvec{Z}) \,
   F_{\alpha_1 \alpha_2,\ms
          y_- \to 0}(x_i, \tvec{y}_\pm, \tvec{Z})
\end{align}
with $F_{y_- \to 0}$ given in \eqref{TMD-one-split} and represented in
figure~\ref{fig:more-split}b.  From the second \rev{expression} in
\eqref{sps-dps-int} one should subtract \rev{an analogous term} with $y_+
\to 0$.

According to \eqref{TMD-angular-int}, the computation of the subtraction
terms involves the integrals
\begin{align}
  \label{tmd-sub-ints}
\int_0^\infty \frac{dy}{y}\, J_n(y q)\, \Phi(y\nu)
  &= \int_0^\infty \frac{du}{u}\, J_n(u q/\nu)\, \Phi(u)
 & \text{with~~} n=0,2,
\end{align}
which are given in the appendix.  We see in table~\ref{tab:phi} that for
$n=0$ both our choices for $\Phi$ give a logarithmic behaviour
$\log(\nu/q)$ when $\nu \gg q$, as expected for a regulated
logarithmically divergent integral.

Let us now turn to the double box graph.  It has leading contributions
when the loops are both hard (SPS region) or both collinear (DPS region),
or when one loop is hard and the other one collinear (region of SPS/DPS
interference).  The terms to be added to the SPS cross section for
removing the contribution from the SPS/DPS interference regions are
\begin{align}
  \label{tmd-box-int-subt}
& - \biggl[ \frac{d\sigma_{\text{DPS/SPS},\, y_+ \to\ms 0}}{dx_1\ms dx_2\,
     d\bar{x}_1\ms d\bar{x}_2\, d^2\tvec{q}_1\, d^2\tvec{q}_2}
  - \frac{d\sigma_{\text{DPS},\, y_\pm \to\ms 0}}{dx_1\, dx_2\,
     d\bar{x}_1\, d\bar{x}_2\, d^2\tvec{q}_1\, d^2\tvec{q}_2} \biggr]
\nonumber \\[0.1em]
& - \biggl[ \frac{d\sigma_{\text{SPS/DPS},\, y_- \to\ms 0}}{dx_1\ms dx_2\,
     d\bar{x}_1\ms d\bar{x}_2\, d^2\tvec{q}_1\, d^2\tvec{q}_2}
  - \frac{d\sigma_{\text{DPS},\, y_\pm \to\ms 0}}{dx_1\, dx_2\,
     d\bar{x}_1\, d\bar{x}_2\, d^2\tvec{q}_1\, d^2\tvec{q}_2} \biggr] \,.
\end{align}
\rev{The first terms in the square brackets are} obtained from
\eqref{sps-dps-int} by taking the perturbative splitting form for all
distributions, the one for $D_{\alpha_1 \alpha_2| \alpha_0}(x_i,
\tvec{y}_+, \tvec{Z})$ being given in \eqref{tw3-split}.  The second terms
are obtained from \eqref{sps-dps-subt} and its analogue for $y_+ \to 0$ by
taking both splitting vertices perturbative, using the DPD given in
\eqref{TMD-two-splits}.  Notice that the subtraction terms
\eqref{tmd-box-int-subt} contain subtractions themselves, showing the
recursive nature of the construction in \eqref{subtr-def}.
Finally, one should remove the contribution from the DPS region by
adding
\begin{align}
  \label{tmd-box-DPS-subt}
{}- \frac{d\sigma_{\text{DPS},\, y_\pm \to\ms 0}}{dx_1\, dx_2\,
     d\bar{x}_1\, d\bar{x}_2\, d^2\tvec{q}_1\, d^2\tvec{q}_2} \,,
\end{align}
where again both splitting vertices in each DPD are taken as perturbative.
Note that in both \eqref{tmd-box-int-subt} and \eqref{tmd-box-DPS-subt}
all proton matrix elements are expressed in terms of ordinary twist-two
TMDs, as is the case for the SPS cross section.

Adding up contributions and double counting subtractions, we finally
obtain
\begin{align}
  \label{tmd-master}
\sigma = \sigma_{\text{DPS}\rule{0pt}{1.3ex}}
 &+ \bigl[\ms \sigma_{\text{DPS/SPS}}
       - \sigma_{\text{DPS},\, y_-\to 0 \rule{0pt}{1.3ex}}
       + \sigma_{\text{SPS/DPS}}
       - \sigma_{\text{DPS},\, y_+\to 0 \rule{0pt}{1.3ex}} \ms\bigr]
\nonumber \\
 &+ \bigl[\ms \sigma_{\text{SPS}\rule{0pt}{1.3ex}}
             - \sigma_{\text{DPS/SPS},\, y_+ \to 0}
             - \sigma_{\text{SPS/DPS},\, y_- \to 0}
             + \sigma_{\text{DPS},\, y_\pm \to 0} \rule{0pt}{1.3ex}
    \ms\bigr] \,,
\end{align}
where we have again omitted differentials $dx_1\, dx_2\, \ldots$ for
brevity.  \rev{The DPS cross section $\sigma_{\text{DPS}}$ contains
  contributions with and without splitting in each of the DPDs, which are
  defined by operator matrix elements as in the case of collinear
  factorisation.}

Let us now discuss the large logarithms appearing in the different terms.
The hard scattering kernel $H_{\alpha_0 \beta_0}$ in \eqref{sps-dps-int}
has a DPS logarithm $\log(Q/q)$, and correspondingly there is a
$\log^2(Q/q)$ in the unsubtracted SPS cross section.  We take $\nu \sim
Q$, so that these logarithms are fully removed by the double counting
subtractions.  They reappear in the logarithmic integrals over
$\tvec{y}_+$ and $\tvec{y}_-$ in the DPS cross section and in the SPS/DPS
interference terms \eqref{sps-dps-int}.  In the full cross section
\eqref{tmd-master} one then has a squared logarithm $\log^2(Q/q)$ in the
DPS term, and a single $\log(Q/q)$ in the subtracted SPS/DPS interference
in the first line, whereas the subtracted SPS contribution in the
last line has no leading logarithm.  If $q$ is in the nonperturbative
region, one has $\log(Q/\Lambda)$ instead of $\log(Q/q)$.

At a practical level, the computation of the SPS/DPS interference requires
twist-three TMDs as input, and the DPS term requires transverse-momentum
dependent DPDs.  In general, the modelling of these functions is very
difficult and largely unconstrained.  The situation is much better when
$q_1, q_2 \gg \Lambda$: in this case one can compute these functions in
terms of collinear ones \cite{Buffing:2016wip}.

On the perturbative side, the calculation of the hard scattering cross
sections in the DPS term is easiest, while the SPS and interference terms
require the full computation of box graphs.  If one is satisfied with
keeping only the leading double logarithm in $Q/q$, then it is sufficient
to retain only the DPS contribution with $\nu \sim Q$.


\subsection{Subtraction formalism in momentum space}
\label{sec:mom-subtr}

In \cite{Collins:2011zzd} it was proven that the subtraction formalism
leading to \eqref{subtr-def} correctly approximates a graph up to power
suppressed terms, provided that each approximant $T_R$ correctly
reproduces the graph in its design region $R$, up to power suppressed
terms.  This proof works with regions in momentum space.  However, our
UV regulator $\Phi$ for DPS singularities is multiplicative in $\tvec{y}$
space, which gives convolutions in transverse momenta that do not appear
in the original Feynman graphs and hence are not part of the proof in
\cite{Collins:2011zzd}.  We now show that this does not cause problems, so
that our setup gives adequate approximations when transformed to momentum
space, with corrections that are subleading in powers of $\Lambda/\nu$.
With $\nu \sim Q$ these corrections are of the same order as other
approximations in the factorisation formula.

We begin with the case of collinear factorisation, using dimensional
regularisation for UV and IR divergences as usual.  As mentioned
earlier, we choose $\Phi$ such that the Fourier transform
\begin{align}
  \label{Psi-def}
\Psi(\tvec{\Delta},\nu) &= 
  \int \frac{d^{D-2} \tvec{y}}{(2\pi)^{D-2}}\;
    e^{-i \tvec{y} \tvec{\Delta}}
    \bigl[ 1 - \Phi^2(y \nu) \bigr]
 = \frac{1}{\nu^{D-2}}
     \int \frac{d^{D-2}\tvec{u}}{(2\pi)^{D-2}}\;
     e^{-i \tvec{u} \tvec{\Delta}/\nu} \bigl[ 1 - \Phi^2(u) \bigr] \,.
\end{align}
is finite for all values of $\Delta/\nu$.  In the following we show that
terms that depend on $\Psi$ in the cross section are power suppressed and
can hence be discarded when establishing the validity of the subtraction
formalism.

Let us first discuss $\sigma_{\text{2v2}}$, which is proportional to
\begin{align}
  \label{2v2-mom}
 (2\pi)^{D-2}\! \int & d^{D-2} \tvec{y}\;
   \Phi^2(y \nu)\ms F_{\text{int}}(\tvec{y})\ms F_{\text{int}}(\tvec{y})
 = \int d^{D-2} \tvec{\Delta}\, 
    F_{\text{int}}(\tvec{\Delta}) F_{\text{int}}(-\tvec{\Delta})
\nonumber \\
& \quad - \int d^{D-2}\tvec{\Delta}\, d^{D-2}\tvec{\Delta}'\;
        \Psi(\tvec{\Delta}-\tvec{\Delta}')\,
        F_{\text{int}}(\tvec{\Delta})\ms
        F_{\text{int}}(-\tvec{\Delta}') \,.
\end{align}
The product $\hat{\sigma}_1\ms \hat{\sigma}_2$ of hard scattering cross
sections does not affect our argument and is hence omitted.  To keep the
notation simple, we only display transverse-momentum arguments in this
subsection.  For simplicity we perform the power counting for $D=4$
dimensions; changes in $D=4-2\epsilon$ are by fractional powers and do not
alter our conclusions.  We then have
\begin{align}
  \label{pow-mom}
\Psi(\tvec{\Delta}) &\sim 1/\nu^{2} \,,
&
F_{\text{int}}(\tvec{\Delta}) &\sim \Lambda^2/\Delta^{2}
\end{align}
for both $\Delta \sim \Lambda$ and $\Delta \sim \nu$, which are the
two scales present in \eqref{2v2-mom}.\footnote{In the ultraviolet
  region $\Delta \gg \nu$ there is a further suppression of
  $\Psi(\tvec{\Delta})$ compared with $1/\nu^2$, the exact form
  of which depends on the function $\Phi$.}
The large $\Delta$ behaviour of $F_{\text{int}}$ follows from dimensional
analysis ($\Lambda^2$ in the numerator comes from a collinear twist-four
matrix element and $1/\Delta^2$ from the hard splitting kernel in the
Fourier transformed version of \eqref{match-tw4}).  The $\Psi$ independent
term in \eqref{2v2-mom} receives its leading contribution of order
$\Lambda^2$ from the region $\Delta \sim \Lambda$, which is the design
region of the DPS approximation.  By contrast, the $\Psi$ dependent term
behaves like $\Lambda^4/\nu^2$, with contributions from both $\Delta \sim
\Lambda$ and $\Delta \sim \nu$.  We thus see that our UV regulator (which
is not necessary in $\sigma_{\text{2v2}}$ as noted earlier) only gives
changes suppressed by $\Lambda^2/\nu^2$.  For $\nu\sim Q$ this does not
degrade the overall accuracy of the calculation.

Next we consider the contribution $\sigma_{\text{2v1}} -
\sigma_{\text{2v1,pt}} + \sigma_{\text{tw4}}$ from 2v1 graphs.  Here we
have a subtraction term, which is obtained by replacing the DPDs in
$\tvec{y}$ space with their perturbative approximations.  In momentum
space this reads
\begin{align}
F_{\text{spl}}(\tvec{\Delta})
  &\to F_{\text{spl,pt}}(\tvec{\Delta})
\nonumber \\
F_{\text{int}}(\tvec{\Delta}) 
  &\to F_{\text{int,pt}}(\tvec{\Delta})
   + \delta^{(D-2)}(\tvec{\Delta})\,
     \int d^{D-2}\tvec{\Delta}'\,
         \bigl[ F_{\text{int}}(\tvec{\Delta}')
              - F_{\text{int,pt}}(\tvec{\Delta}') \bigr] \,.
\end{align}
The second line is obtained by replacing $F$ with $F_{\text{int}}$ in the
first term and with $F_{\text{int,pt}}$ in the second term on the
r.h.s.\ of \eqref{match-master} and then taking the Fourier transform with
respect to $\tvec{y}$.  We thus find that $\sigma_{\text{2v1}} -
\sigma_{\text{2v1,pt}}$ is proportional to
\begin{align}
  \label{2v1-mom}
& (2\pi)^{D-2}\! \int d^{D-2} \tvec{y}\;
   \Phi^2(y \nu)\ms \bigl[
      F_{\text{spl}}(\tvec{y})\ms F_{\text{int}}(\tvec{y})
    - F_{\text{spl,pt}}(\tvec{y})\ms F_{\text{int,pt}}(\tvec{y}) \bigr]
\nonumber \\
 &\quad = \int d^{D-2} \tvec{\Delta}\, \bigl[
    F_{\text{spl}}(\tvec{\Delta})\ms F_{\text{int}}(-\tvec{\Delta})
  - F_{\text{spl,pt}}(\tvec{\Delta}) F_{\text{int,pt}}(-\tvec{\Delta}) \bigr]
\nonumber \\
 &\qquad - F_{\text{spl,pt}}(\tvec{\Delta} \!=\! \tvec{0})
     \int d^{D-2} \tvec{\Delta}'\;
        \bigl[ F_{\text{int}}(\tvec{\Delta}')
             - F_{\text{int,pt}}(\tvec{\Delta}') \bigr]
\nonumber \\
 &\qquad
   - \int d^{D-2}\tvec{\Delta}\, d^{D-2}\tvec{\Delta}'\;
        \Psi(\tvec{\Delta}-\tvec{\Delta}')\, \bigl[
    F_{\text{spl}}(\tvec{\Delta})\ms F_{\text{int}}(-\tvec{\Delta}')
  - F_{\text{spl,pt}}(\tvec{\Delta})\ms F_{\text{int,pt}}(-\tvec{\Delta}')
  \bigr]
\nonumber \\
 &\qquad + \int d^{D-2}\tvec{\Delta}\,
    \Psi(\tvec{\Delta})\, F_{\text{spl,pt}}(\tvec{\Delta})    
    \int d^{D-2}\tvec{\Delta}'\;
       \bigl[ F_{\text{int}}(\tvec{\Delta}')
            - F_{\text{int,pt}}(\tvec{\Delta}') \bigr] \,.
\end{align}
For power counting we use \eqref{pow-mom} and additionally
\begin{align}
  \label{spl-pow-mom}
F_{\text{spl}}(\tvec{\Delta}) &\sim \Delta^{0} \,,
\end{align}
which follows from dimensional analysis of the splitting graph.  The
perturbative approximations $F_{\text{int,pt}}$ and $F_{\text{spl,pt}}$
scale like their un-approximated counterparts.  The $\Psi$ independent
terms in \eqref{2v1-mom} are then found to be of order $\Lambda^2$, which
after multiplication with $\hat{\sigma}_1\ms \hat{\sigma}_2$ gives the
same power behaviour in the cross section as $\sigma_{\text{tw4}}$.  To
analyse the $\Psi$ dependent terms, we rearrange them as follows:
\begin{align}
  \label{2v1-mom-nonloc}
& - \int d^{D-2}\tvec{\Delta}\, F_{\text{spl,pt}}(\tvec{\Delta})
    \int d^{D-2}\tvec{\Delta}'\;
    \bigl[ \Psi(\tvec{\Delta}-\tvec{\Delta}')
         - \Psi(\tvec{\Delta}) \bigr]
    \bigl[ F_{\text{int}}(-\tvec{\Delta}')
         - F_{\text{int,pt}}(-\tvec{\Delta}') \bigr]
\nonumber \\
& \quad - \int d^{D-2}\tvec{\Delta}\;
    \bigl[ F_{\text{spl}}(\tvec{\Delta})
         - F_{\text{spl,pt}}(\tvec{\Delta}) \bigr]
    \int d^{D-2}\tvec{\Delta}'\;
         \Psi(\tvec{\Delta}-\tvec{\Delta}')\,
         F_{\text{int}}(-\tvec{\Delta}') \,.
\end{align}
The first line is suppressed by the integration phase space for $\Delta
\sim \Lambda$, by $\Psi(\tvec{\Delta}-\tvec{\Delta}') -
\Psi(\tvec{\Delta})$ for $\Delta \sim \nu$ and $\Delta' \sim \Lambda$, and
by $F_{\text{int}}(-\tvec{\Delta}') - F_{\text{int,pt}}(-\tvec{\Delta}')$
for $\Delta, \Delta' \sim \nu$.  The second line is again suppressed by
the integration phase space for $\Delta \sim \Lambda$, and by
$F_{\text{spl}}(\tvec{\Delta}) - F_{\text{spl,pt}}(\tvec{\Delta})$ for
$\Delta \sim \nu$.  The sum of all $\Psi$ dependent terms is hence power
suppressed compared with the $\Psi$ independent terms.  One readily finds
that the relative suppression is by $\Lambda^2/\nu^2$, as it was in the
2v2 case.  Notice that this suppression is obtained for the combination
$\sigma_{\text{2v1}} - \sigma_{\text{2v1,pt}}$ but not for the two terms
separately.

Finally, we discuss the contribution $\sigma_{\text{1v1}} -
\sigma_{\text{1v1,pt}} + \sigma_{\text{SPS}}$ from 1v1 graphs.  The
combination $\sigma_{\text{1v1}} - \sigma_{\text{1v1,pt}}$ is proportional
to
\begin{align}
  \label{1v1-mom}
(2\pi)^{D-2}\! \int & d^{D-2} \tvec{y}\;
   \Phi^2(y \nu)\bigl[
      F_{\text{spl}}(\tvec{y})\ms F_{\text{spl}}(\tvec{y})
    - F_{\text{spl,pt}}(\tvec{y})\ms F_{\text{spl,pt}}(\tvec{y}) \bigr]
\nonumber \\
& \quad = \int d^{D-2} \tvec{\Delta}\, \bigl[
      F_{\text{spl}}(\tvec{\Delta})\ms F_{\text{spl}}(-\tvec{\Delta})
    - F_{\text{spl,pt}}(\tvec{\Delta})\ms
      F_{\text{spl,pt}}(-\tvec{\Delta}) \bigr]
\nonumber \\
& \qquad - \int d^{D-2}\tvec{\Delta}\, d^{D-2}\tvec{\Delta}'\;
    \Psi(\tvec{\Delta}-\tvec{\Delta}')\,
\nonumber \\
& \qquad\qquad \times
 \bigl[ F_{\text{spl}}(\tvec{\Delta})\ms
        F_{\text{spl}}(-\tvec{\Delta}')
      - F_{\text{spl,pt}}(\tvec{\Delta})\ms
        F_{\text{spl,pt}}(-\tvec{\Delta}') \bigr] \,.
\end{align}
One finds that both the $\Psi$ dependent and the $\Psi$ independent terms
on the r.h.s.\ scale like $\Lambda^2$.  Multiplied with $\hat{\sigma}_1\ms
\hat{\sigma}_2$ this is power suppressed by $\Lambda^2/Q^2$ compared with
$\sigma_{\text{SPS}}$.  Notice that the SPS approximation itself has
relative corrections of order $\Lambda^2/Q^2$.  This means that the
contribution in \eqref{1v1-mom} is beyond the accuracy of the calculation
of 1v1 graphs.  The fact that the $\Psi$ dependent part is of similar size
as the $\Psi$ independent one is hence of no concern.  We recall from
section~\ref{sec:power} that the rationale to evaluate the 2v2 and 2v1
terms, which are also suppressed by $\Lambda^2/Q^2$, is that they may be
enhanced by other factors.

The case of TMD factorisation can be discussed along the same lines.  In
analogy to \eqref{Psi-def} we define $\Psi$ as the Fourier transform of
$[\ms 1 - \Phi(y \nu)]$ instead of $[\ms 1 - \Phi^2(y \nu)]$.  Let us
discuss the SPS/DPS interference graph in figure~\ref{fig:sps-dps-int}b.
Its contribution to $\sigma_{\text{DPS/SPS}}$ in \eqref{sps-dps-int} is
proportional to
\begin{align}
  \label{TMD-mom}
\int & d^2\tvec{Z}\, d^2\tvec{y}_+\;
   e^{-i (\tvec{q}_1 + \tvec{q}_2) \tvec{Z} -i \tvec{q} \tvec{y}_+}\;
   \Phi(y_+ \nu)\,
     D(\tvec{y}_+, \tvec{Z})\ms D(\tvec{y}_+, \tvec{Z})
\nonumber \\
 & \quad = \int d^{D-2}\tvec{K}\, d^{D-2}\tvec{k}_-\;
     D(\tvec{k}_-, \tvec{K})\,
     D(\tvec{q} - \tvec{k}_-, \tvec{q}_1 + \tvec{q}_2 - \tvec{K})
\nonumber \\
 & \qquad
   - \int d^{D-2}\tvec{K}\, d^{D-2}\tvec{k}^{}_-\, d^{D-2}\tvec{k}'_-\; 
     \Psi(\tvec{k}^{}_- - \tvec{k}'_-)\,
\nonumber \\[0.1em]
 & \qquad\qquad \times
     D(\tvec{k}^{}_-, \tvec{K})\,
     D(\tvec{q} - \tvec{k}'_-, \tvec{q}_1 + \tvec{q}_2 - \tvec{K}) \,,
\end{align}
where on the r.h.s.\ we have twist-three TMDs in momentum representation.
An analogous relation holds for the contribution to
$\sigma_{\text{DPS/SPS},\, y_+ \to 0}$, which according to
\eqref{tmd-master} is to be subtracted from $\sigma_{\text{DPS/SPS}}$ in
the overall cross section.  In this case one should take the perturbative
splitting approximation $D_{\text{spl,pt}}$ of the distributions, which in
$\tvec{y}_+$ representation is given by \eqref{tw3-split}.  For power
counting we take $q_1, q_2 \sim \Lambda$ and use
\begin{align}
D(\tvec{k}_-, \tvec{K}) & \sim 1 \big/ \bigl( k_-\ms K^2 \bigr) \,,
\end{align}
where $k_-$ and $K$ can be of order $\Lambda$ or $\nu$.  This can be
derived by a general analysis of graphs as in section 5.2 of
\cite{Diehl:2011yj}; the $1/k_-$ dependence is also directly obtained by
Fourier transforming \eqref{tw3-split}.  $D_{\text{spl,pt}}$ behaves like
$D$, and of course we have $\Psi \sim 1/\nu^2$ as before.  

We see that for $k_- \sim K \sim \Lambda$ the $\Psi$ independent term in
\eqref{TMD-mom} is of order $1/\Lambda^2$, whilst the $\Psi$ dependent one
is of order $1/\nu^2$.  The only region in which the $\Psi$ dependent term
gives a leading contribution is when $k^{}_- \sim k'_- \sim \nu$ and $K
\sim \Lambda$.  However, in this region one has $D \approx
D_{\text{spl,pt}}$, so that its contribution is suppressed in the
combination $\sigma_{\smash{\text{DPS/SPS}}} -
\sigma_{\smash{\text{DPS/SPS},\, y_+ \to 0}}$.  We thus find that in the
overall cross section, $\Psi$ dependent terms are power suppressed, as in
the case of collinear factorisation.  \rev{The same holds of course for
  $\sigma_{\smash{\text{SPS/DPS}}} - \sigma_{\smash{\text{SPS/DPS},\, y_-
      \to 0}}$.}  Finally, a corresponding analysis (involving the momenta
$\tvec{k}_+$, $\tvec{k}_-$ and $\tvec{K}$) can be given for the
combination $\sigma_{\text{DPS}} - \sigma_{\text{DPS},\, y_-\to 0} -
\sigma_{\text{DPS},\, y_+\to 0} + \sigma_{\text{DPS},\, y_\pm \to 0}$,
which covers all terms in \eqref{tmd-master} that depend on our UV
regulator.

\section{Subtraction terms at higher orders}
\label{sec:higher-order}

Our construction of the double counting subtractions for DPS regions
is not limited to the leading-order graphs discussed so far.  In the
present section we illustrate how the formalism works at highers in
$\alpha_s$, taking as an example the 2v1 mechanism in collinear
factorisation.  In the master formula \eqref{subtr-def} we encounter
nested subtractions in this case.  We show for selected graphs how the
formalism gives the different contributions to the cross section,
namely $\sigma_{\text{tw4}}$, $\sigma_{\text{2v1}}$ and the
subtraction term $\sigma_{\text{2v1,pt}}$.  In particular we find that
the latter can be obtained from $\sigma_{\text{2v1}}$ by replacing the
DPDs with their short-distance expansions $F_{\text{int,pt}}$ and
$F_{\text{spl,pt}}$, as found at leading order in
section~\ref{sec:2v1}.

\begin{figure}
\begin{center}
\subfigure[]{\includegraphics[width=0.3\textwidth]{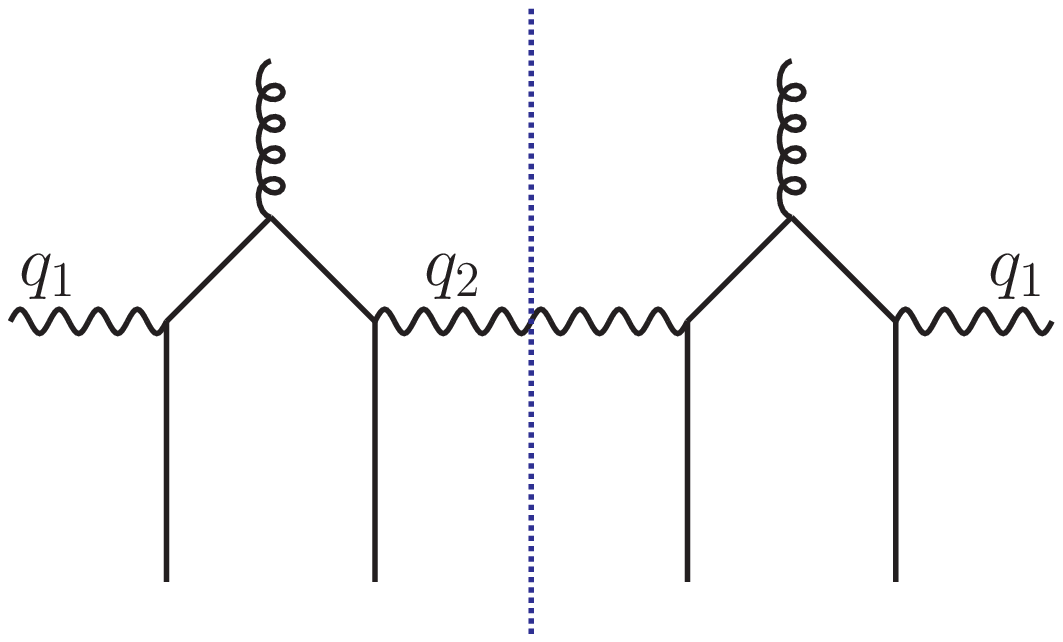}}
\hspace{1em}
\subfigure[]{\includegraphics[width=0.3\textwidth]{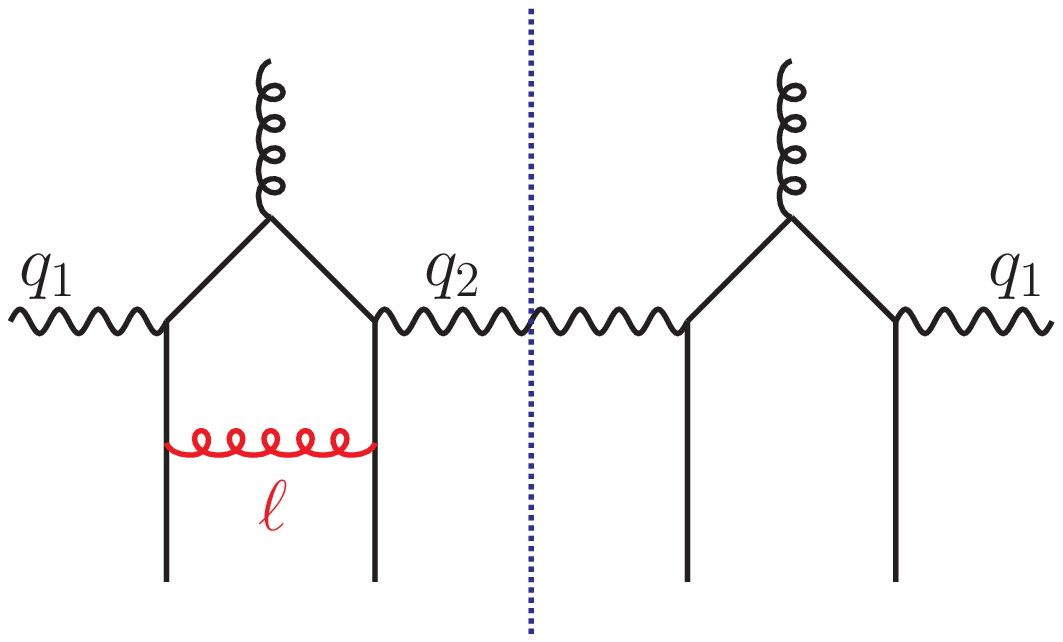}}
\hspace{1em}
\subfigure[]{\includegraphics[width=0.3\textwidth]{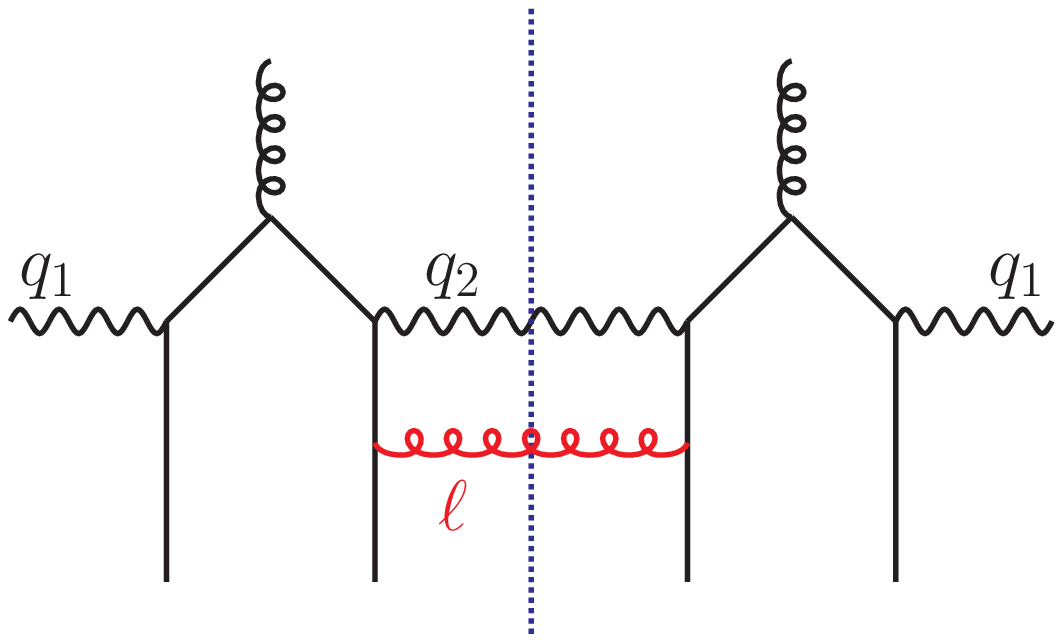}}
\\
\subfigure[]{\includegraphics[width=0.3\textwidth]{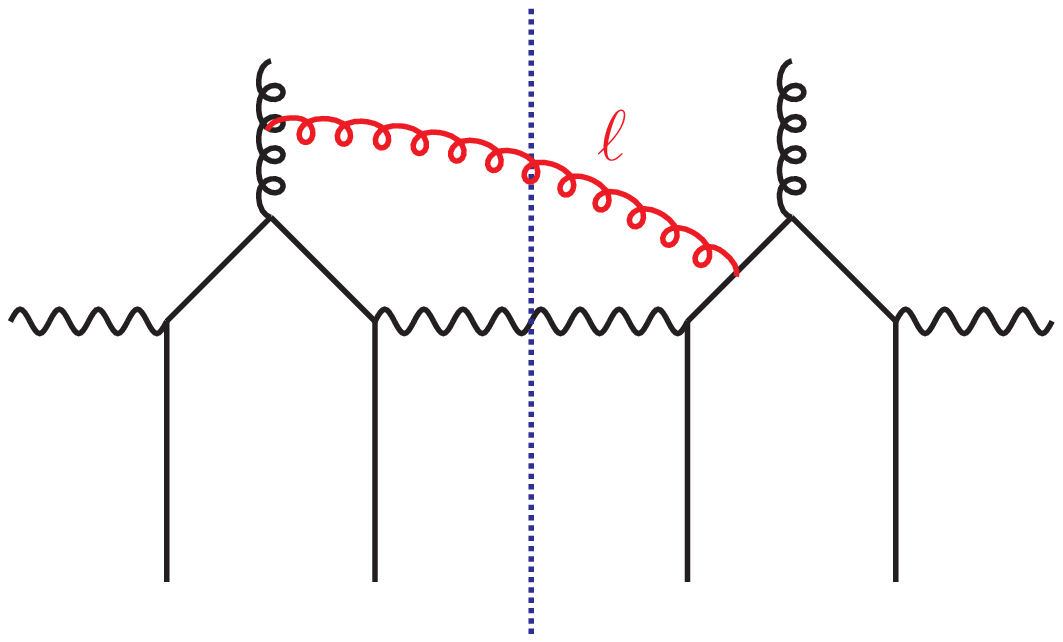}}
\hspace{1em}
\subfigure[]{\includegraphics[width=0.3\textwidth]{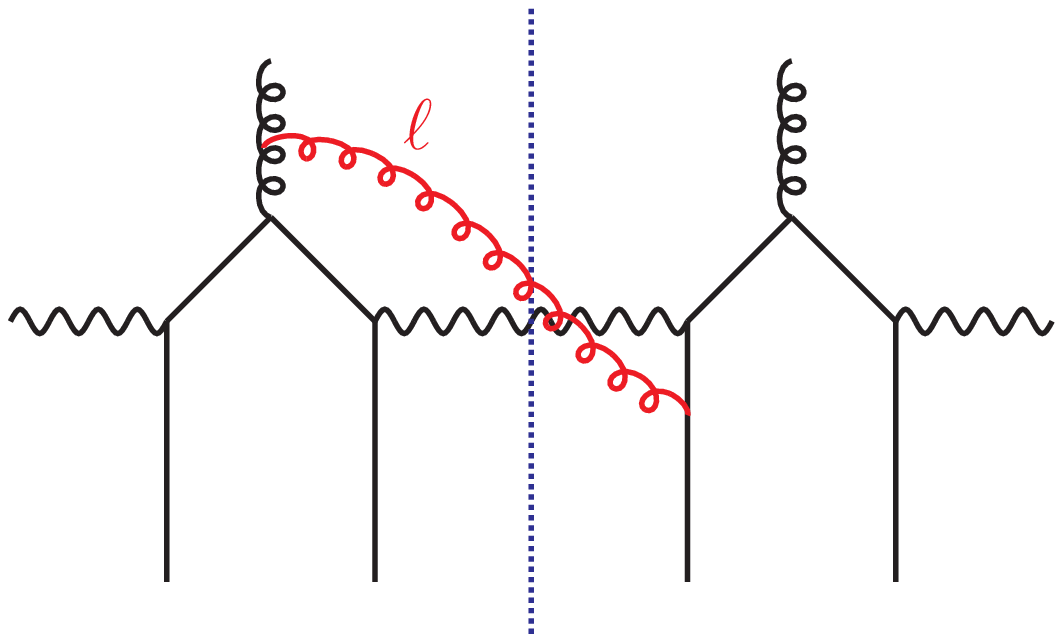}}
\hspace{1em}
\subfigure[]{\includegraphics[width=0.3\textwidth]{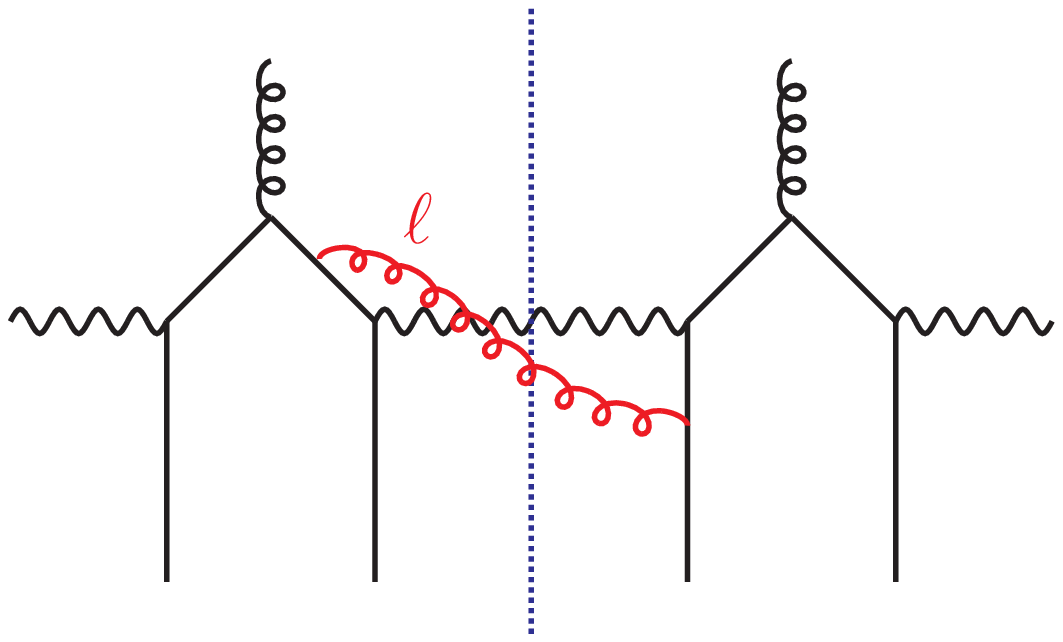}}
\caption{\label{fig:nlo-graphs} LO and NLO graphs for the 2v1
  mechanism, with the blobs for the hadronic matrix elements omitted
  as in figure~\protect\ref{fig:2v1-labels}.}
\end{center}
\end{figure}

We focus here on the general structure and use a schematic notation,
in particular suppressing indices and arguments that are not essential
for the discussion.

\paragraph{Graph \protect\ref{fig:nlo-graphs}a.}  To introduce this
notation, let us briefly review the leading-order graph in
figure~\ref{fig:nlo-graphs}a, which was discussed in detail in
section~\ref{sec:2v1}.  It has two leading momentum regions, which we
denote by $C$ and $H$.  In region $C$ the gauge bosons have small
transverse momenta and the $g\to q\bar{q}$ splitting is near
collinear.  This is the region of DPS, and its approximated
contribution to the cross section reads
\begin{align}
C_C\ms \Gamma^{a} &= \hat{\sigma}^{(0)}_1 \hat{\sigma}^{(0)}_2
  \int_y \Phi^2(y \nu)\, F^{(1)}_{\text{spl}}(y)\,
   F^{(0)}_{\text{int}}(y) \,.
\end{align}
Here $\hat{\sigma}^{(0)}_i$ is the tree-level cross section for
$q\bar{q}\to V_i$, and we write $\int_y$ as a shorthand for
$\int d^{D-2}\tvec{y}$.  The contribution of this graph to the
splitting DPD at the top is denoted by $F^{(0)}_{\text{spl}}$, and its
contribution to the intrinsic DPD at the bottom by
$F^{(0)}_{\text{int}}$.  Here and in the rest of the section,
superscripts $(0), (1), (2)$ indicate the power of $\alpha_s$ in the
considered quantity.  For simplicity we suppress longitudinal momentum
arguments throughout.

In region $H$, the transverse momenta of the gauge bosons are large
and thus the $q\bar{q}$ pair into which the gluon splits is far off
shell on both sides of the final state cut.  This is described by the
twist-four contribution and requires a subtraction for the smaller
region $C$.  The subtracted contribution is
\begin{align}
  \label{2v1-lo-H}
C_H\ms \Gamma^{a} &= \int_q f^{(0)}\, H^{(1)} \otimes G^{(0)} 
  - \hat{\sigma}^{(0)}_1 \hat{\sigma}^{(0)}_2 \int_y \Phi^2(y \nu)\,
    f^{(0)}\, V^{(1)}(y)\, G^{(0)}  \,.
\end{align}
Here $\int_q$ is shorthand for $\int d^{D-2}\tvec{q}$, the tree-level
hard scattering cross section for $q\bar{q} + g \to V_1 V_2$ is
denoted by $H^{(1)}$, whilst $f^{(0)}$ and $G^{(0)}$ respectively are
the zeroth order contributions to the gluon distribution at the top
and the twist-four $q\bar{q}$ distribution at the bottom of the graph.
As seen in \eqref{tw4-start}, there is a convolution integral over
longitudinal momentum fractions in $G$, denoted here by $\otimes$.
In the subtraction term, $V^{(1)}$ represents the lowest-order
perturbative $g\to q\bar{q}$ splitting kernel, including all terms in
\eqref{split-lo} except for the single-parton density $f_{a_0}$.  In
this term, the momentum fractions of $G^{(0)}$ are fixed by external
kinematics, so that the convolution $\otimes$ is absent.

The perturbative expressions of the DPDs for small $y$ are
\begin{align}
  \label{Fspl-pt-lo}
F_{\text{spl,pt}}^{(1)}(y) &= f^{(0)}\, V^{(1)}(y) \,,
\\
  \label{Fint-pt-lo}
F_{\text{int,pt}}^{(0)}(y) &= G^{(0)}
\end{align}
at leading order, so that the combined contribution of the $C$ and $H$
regions can be written as
\begin{align}
  \label{2v1-lo}
\Gamma^{a} &= \int_q f^{(0)}\, H^{(1)} \otimes G^{(0)}
  - \hat{\sigma}^{(0)}_1 \hat{\sigma}^{(0)}_2 \int_y \Phi^2(y \nu)\,
    F^{(1)}_{\text{spl,pt}}(y)\, F^{(0)}_{\text{int,pt}}(y)
\nonumber \\
 &\quad
  +  \hat{\sigma}^{(0)}_1 \hat{\sigma}^{(0)}_2
  \int_y \Phi^2(y \nu)\, F^{(1)}_{\text{spl}}(y)\,
   F^{(0)}_{\text{int}}(y) \,.
\end{align}
As already mentioned, the double counting subtraction in the second
term is simply obtained from the DPS term by replacing each DPD with
its small-$y$ expansion at the appropriate order in $\alpha_s$.

\paragraph{Graph \protect\ref{fig:nlo-graphs}b.}  We now analyse the
NLO graph in figure~\ref{fig:nlo-graphs}b.  The $q\bar{q}$ splitting
can be hard or collinear, and likewise the additional gluon with
momentum $\ell$.  The leading momentum regions thus are $CC$, $HC$ and
$HH$, where the first letter refers to the $q\bar{q}$ splitting and
the second to the additional gluon.  There is also a region where the
gluon is hard whereas the $q\bar{q}$ splitting to the right of the cut
is collinear.  The splitting at the left of the cut is then hard by
momentum conservation.  One finds that this region is not leading: the
loss of phase space for having the gauge bosons at low transverse
momenta is stronger than the gain from having one but not both
splitting vertices collinear.  Let us discuss the leading regions in
turn.

In the $CC$ region, the extra gluon is part of the DPD for the right
moving particles, and one simply has
\begin{align}
  \label{2v1-CC-b}
C_{CC}\ms \Gamma^{b} &= \hat{\sigma}^{(0)}_1 \hat{\sigma}^{(0)}_2
  \int_y \Phi^2(y \nu)\,
  F^{(1)}_{\text{spl}}(y)\, F^{(1)}_{\text{int}}(y) \,,
\end{align}
where the superscript $(1)$ indicates a contribution to
$F_{\text{int}}$ at order $\alpha_s$.  Likewise, in the $HC$ region,
the extra gluon becomes part of the twist-four distribution $G$, so
that one has
\begin{align}
  \label{2v1-HC-b}
C_{HC}\ms \Gamma^{b} &= \int_q f^{(0)}\, H^{(1)} \otimes
    G^{(1)}_{\text{ren}\rule{0pt}{1.1ex}}
  - \hat{\sigma}^{(0)}_1 \hat{\sigma}^{(0)}_2 \int_y \Phi^2(y \nu)\,
    f^{(0)}\, V^{(1)}(y)\, G^{(1)}_{\text{ren}\rule{0pt}{1.1ex}} \,.
\end{align}
Note that the integral over the gluon momentum in the graph giving
$G^{(1)}_{\text{ren}\rule{0pt}{1.1ex}}$ contains an ultraviolet
divergence.  It is understood that
$G^{(1)}_{\text{ren}\rule{0pt}{1.1ex}}$ is renormalised, i.e.\
includes the appropriate UV counterterm, which is indicated by the 
subscript ``ren''.

For the $HH$ region, the recursive character of the subtraction formalism
becomes evident.  We have a subtraction for the next smallest region $HC$,
which itself contains a subtraction for the region $CC$.  In addition, a
$CC$ subtraction has to be made for the overall graph.  This gives
\begin{align}
C_{HH}\ms \Gamma^{b} &= \int_q f^{(0)}\, H^{(2)} \otimes G^{(0)} 
\nonumber \\
 &\quad -
   \biggl[ \int_q f^{(0)}\, H^{(1)} \otimes
           \widetilde{K}^{(1)}_{\text{ren}}(0)
    - \hat{\sigma}^{(0)}_1 \hat{\sigma}^{(0)}_2
      \int_y \Phi^2(y \nu)\, f^{(0)}\, V^{(1)}(y)\,
      \widetilde{K}^{(1)}_{\text{ren}}(0) \biggr] \otimes G^{(0)}
\nonumber \\
 &\quad - \hat{\sigma}^{(0)}_1 \hat{\sigma}^{(0)}_2
    \int_y \Phi^2(y \nu)\, f^{(0)}\, V^{(1)}(y)\,
    \widetilde{K}^{(1)}(y) \otimes G^{(0)} \,.
\end{align}
The term in the first line is for the hard region, with $H^{(2)}$
being a one-loop contribution to the cross section for
$q\bar{q} + g \to V_1 V_2$.  In the second line, the term for the $HC$
region (including its subtraction for the $CC$ region) involves a
non-diagonal splitting kernel $\widetilde{K}^{(1)}_{\text{ren}}(0)$
for the exchange of a gluon between the two parton lines with momentum
fractions $x_1$ and $x_2$, which are taken at relative transverse
position $y=0$ because this is part of the approximation for the $HC$
region.  In the last line, which is the overall subtraction for the
$CC$ region, the same kernel is to be taken at relative distance $y$,
because in this case the $q\bar{q}$ splitting is not assumed to be
harder than the extra gluon emission.  One therefore has to take a
convolution in transverse momenta, which turns into a product of two
$y$ dependent factors $V^{(1)}(y)$ and $\widetilde{K}^{(1)}(y)$ after
Fourier transformation.

Notice that in dimensional regularisation the limit $y\to 0$ in
$\widetilde{K}^{(1)}(y)$ is not smooth.  The graph for
$\widetilde{K}^{(1)}_{\text{ren}}(0)$ contains a scaleless momentum
integral, which is zero.  After renormalisation, i.e.\ subtraction of
the ultraviolet counterterm, one is left with an infrared
divergence.\footnote{The UV divergent part of
  $\widetilde{K}^{(1)}_{\text{ren}}(0)$ gives DGLAP splitting kernels
  generalised to non-diagonal momentum fractions, known from the
  evolution of twist-four distributions
  \protect\cite{Bukhvostov:1985rn} and of GPDs
  \protect\cite{Belitsky:2005qn}.}
By contrast, $\widetilde{K}^{(1)}(y)$ at finite $y$ does not have an
ultraviolet divergence (and hence no counterterm), but it does have the
same infrared divergence as $\widetilde{K}^{(1)}_{\text{ren}}(0)$.

Adding all contributions and using \eqref{Fspl-pt-lo} we obtain
\begin{align}
  \label{2v1-b}
\Gamma^{b} &= \int_q f^{(0)}\, \Bigl( H^{(1)} \otimes
      G^{(1)}_{\text{ren}\rule{0pt}{1.1ex}}
   + \bigl[ H^{(2)} - H^{(1)} \otimes
            \widetilde{K}^{(1)}_{\text{ren}}(0) \bigr]
       \otimes G^{(0)} \Bigr)
\nonumber \\
 &\quad - \hat{\sigma}^{(0)}_1 \hat{\sigma}^{(0)}_2 \int_y \Phi^2(y \nu)\,
   F_{\text{spl,pt}}^{(1)}(y)\, F_{\text{int,pt}}^{(1)}(y)
\nonumber \\
 &\quad + \hat{\sigma}^{(0)}_1 \hat{\sigma}^{(0)}_2
  \int_y \Phi^2(y \nu)\,
  F^{(1)}_{\text{spl}}(y)\, F^{(1)}_{\text{int}}(y)
\end{align}
with
\begin{align}
  \label{Fint-pt-b}
F_{\text{int,pt}}^{(1)}(y) & \underset{\text{graph } b}{=}
  G^{(1)}_{\text{ren}\rule{0pt}{1.1ex}}
  + \bigl[ \widetilde{K}^{(1)}(y)
         - \widetilde{K}^{(1)}_{\text{ren}}(0) \bigr] \otimes G^{(0)} \,.
\end{align}
In the first line of \eqref{2v1-b} we have an NLO contribution to the
twist-four term, with the NLO kernel $H^{(2)}$ having a subtraction
for the region where the gluon becomes collinear, as is standard in
collinear factorisation.  In the DPS subtraction term we have a
contribution to the small-$y$ expansion of $F_{\text{int}}$ at NLO,
with the three terms in \eqref{Fint-pt-b} corresponding respectively
to the three terms in the general expression \eqref{match-master}.
The infrared divergence of the kernel $K$ drops out here, as it must.
We see that the DPS subtraction term in \eqref{2v1-b} has the same
form as in \eqref{2v1-lo}, with $F_{\text{int,pt}}^{(1)}$ instead of
$F_{\text{int,pt}}^{(0)}$.

In \eqref{2v1-b} all ultraviolet and infrared subtractions have been
carried out, so that one can set $D=4$ in dimensional regularisation.
The quantities with subscript ``ren'' then depend on the associated
scale $\mu$.  In the first line, this dependence cancels between
$G^{(1)}_{\text{ren}\rule{0pt}{1.1ex}}$ (the graph under discussion
contributes to the scale evolution of $G$) and the subtracted
hard scattering kernel in the second term.  In \eqref{Fint-pt-b}, the
$\mu$ dependence cancels between
$G^{(1)}_{\text{ren}\rule{0pt}{1.1ex}}$ and the subtracted kernel of
the small-$y$ expansion, i.e.\ graph~\ref{fig:nlo-graphs}b does not
contribute to the evolution of $F_{\text{int}}$.

\paragraph{Graph \protect\ref{fig:nlo-graphs}c.}  We next consider the
graph in figure~\ref{fig:nlo-graphs}c.  The contribution from the
region $CC$ is
\begin{align}
C_{CC}\ms \Gamma^{c} &= \hat{\sigma}^{(0)}_1 \hat{\sigma}^{(0)}_2
  \int_y \Phi^2(y \nu)\,
  F^{(1)}_{\text{spl}}(y)\, F^{(1)}_{\text{int,ren}}(y) \,,
\end{align}
where in contrast to graph~\ref{fig:nlo-graphs}b the one-loop
contribution to $F_{\text{int}}$ now has an ultraviolet divergence
(since the two partons connected by the gluon are at the same
transverse position) and needs renormalisation.  The contribution from
region $HC$ is analogous to \eqref{2v1-HC-b}.  Unlike for
graph~\ref{fig:nlo-graphs}b, a leading contribution is now also
obtained for the momentum region $CH$, where the $g\to q\bar{q}$
splittings are collinear, whilst the gluon carries a large transverse
momentum (balanced by the vector boson momentum $\tvec{q}_2$).
Including the subtraction for the region $CC$, we have
\begin{align}
C_{CH}\ms \Gamma^{c} &= \hat{\sigma}^{(0)}_1
   \int_y \Phi^2(y \nu)\, F_{\text{spl}}^{(1)}(y)\,
   \bigl[ \sigma^{(1)}_2 - \sigma^{(0)}_2\, K^{(1)}_{\text{ren}}
   \bigr] \otimes F_{\text{int}}^{(0)}(y) \,,
\end{align}
where $\sigma^{(1)}_2$ is the cross section for $q\bar{q}\to V_2 + g$
and $K^{(1)}_{\text{ren}}$ is a diagonal splitting kernel including
the necessary UV counterterm.  Its divergent part involves the
contribution of this graph to the usual DGLAP splitting kernel
$P_{qq}$.  Finally, the region $HH$ gives
\begin{align}
C_{HH}\ms \Gamma^{c} &= \int_q f^{(0)}\, H^{(2)} \otimes G^{(0)} 
\nonumber \\
 &\quad - \biggl[ \int_q f^{(0)}\, H^{(1)}
      \otimes K^{(1)}_{\text{ren}}
    - \hat{\sigma}^{(0)}_1 \hat{\sigma}^{(0)}_2
      \int_y \Phi^2(y \nu)\, f^{(0)}\, V^{(1)}(y)\,
      K^{(1)}_{\text{ren}} \biggr] \otimes G^{(0)}
\nonumber \\
 &\quad - \hat{\sigma}^{(0)}_1
   \int_y \Phi^2(y \nu)\, f^{(0)}\, V^{(1)}(y)\,
   \bigl[ \sigma^{(1)}_2 - \sigma^{(0)}_2\, K^{(1)}_{\text{ren}}
   \bigr] \otimes G^{(0)}
\nonumber \\
 &\quad - \hat{\sigma}^{(0)}_1 \hat{\sigma}^{(0)}_2
    \int_y \Phi^2(y \nu)\, f^{(0)}\, V^{(1)}(y)\,
    K^{(1)}_{\text{ren}} \otimes G^{(0)} \,,
\end{align}
where the last three lines are the subtraction terms for the regions
$HC$, $CH$, $CC$, respectively.  Adding all contributions, we obtain
\begin{align}
  \label{2v1-c}
\Gamma^{c} &= \int_q f^{(0)}\, \Bigl( H^{(1)} \otimes
      G^{(1)}_{\text{ren}\rule{0pt}{1.1ex}}
   + \bigl[ H^{(2)} - H^{(1)} \otimes K^{(1)}_{\text{ren}} \bigr]
       \otimes G^{(0)} \Bigr)
\nonumber \\
 &\quad - \hat{\sigma}^{(0)}_1 \int_y \Phi^2(y \nu)\,
   F^{(1)}_{\text{spl,pt}}(y)\, \Bigl(
      \hat{\sigma}^{(0)}_2\, F^{(1)}_{\text{int,pt}}(y)
    + \bigl[ \sigma^{(1)}_2 - \sigma^{(0)}_2\, K^{(1)}_{\text{ren}}
      \bigr] \otimes F^{(0)}_{\text{int}}(y) \Bigr)
\nonumber \\
 &\quad + \hat{\sigma}^{(0)}_1 \int_y \Phi^2(y \nu)\,
  F^{(1)}_{\text{spl}}(y)\, \Bigl(
     \hat{\sigma}^{(0)}_2\, F^{(1)}_{\text{int,ren}}(y)
   + \bigl[ \sigma^{(1)}_2 - \sigma^{(0)}_2\, K^{(1)}_{\text{ren}}
     \bigr] \otimes F_{\text{int}}^{(0)}(y) \Bigr) \,,
\end{align}
where in the second line we used \eqref{Fspl-pt-lo}, \eqref{Fint-pt-lo}
and
\begin{align}
  \label{Fint-pt-c}
F^{(1)}_{\text{int,pt}}(y) & \;\underset{\text{graph } c}{=}\;
   G^{(1)}_{\text{ren}\rule{0pt}{1.1ex}} \,.
\end{align}
The latter relation holds because the exchanged gluon in
graph~\ref{fig:nlo-graphs}c gives exactly the same contribution to
$F_{\text{int}}(y)$ and to $G$ and hence cancels between the second
and third term in the general small-$y$ expansion \eqref{match-master}
of $F_{\text{int}}(y)$.
In each line of \eqref{2v1-c} we have hard scattering cross sections
subtracted for the collinear region, as is characteristic for collinear
factorisation.  The $\mu$ dependence cancels between the subtracted cross
sections and the renormalised distributions.

\begin{figure}
\begin{center}
\includegraphics[height=0.19\textwidth]{figures/nlo-2v1-gql.eps}
\includegraphics[height=0.19\textwidth]{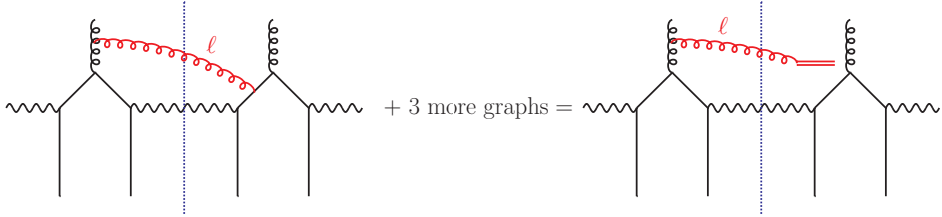}
\caption{\label{fig:eikonal} Graphical representation of applying the
  Grammer-Yennie approximation and a Ward identity to the sum of four
  graphs.  The three other graphs required are graph e in
  figure~\protect\ref{fig:nlo-graphs}, and the analogues to graphs d and
  e in which the gluon couples to the quark line at the very right.
  It is understood that the $g\to q\bar{q}$ splitting is hard, whereas
  the gluons are collinear left moving.}
\end{center}
\end{figure}

\paragraph{Graph \protect\ref{fig:nlo-graphs}d.}  We now discuss
figure~\ref{fig:nlo-graphs}d, where we have an additional gluon in the
upper part of the graph.  Its leading momentum regions are $CC$, $HC$
and $HH$.  As in the case of graph~\ref{fig:nlo-graphs}b, there is no
leading region $CH$.  For the $CC$ region we simply have
\begin{align}
C_{CC}\ms \Gamma^{d} &= \hat{\sigma}^{(0)}_1 \hat{\sigma}^{(0)}_2
  \int_y \Phi^2(y \nu)\,
  F^{(2)}_{\text{spl}}(y)\, F^{(0)}_{\text{int}}(y) \,.
\end{align}
In the region $HC$, the extra gluon in the graph is collinear and
couples to an off-shell quark, since the $g\to q\bar{q}$ splitting is
hard.  Among the approximations to be made in this case is the
Grammer-Yennie approximation (see
e.g.~\cite{Diehl:2011yj,Diehl:2015bca}).  After summing over an
appropriate set of graphs, one can apply a Ward identity, after which
the gluon couples to an eikonal line, as shown in
figure~\ref{fig:eikonal}.  We thus can write
\begin{align}
C_{HC}\ms \Gamma^{d} + C_{HC}\ms \Gamma^{e+}
  &= \int_q f^{(1)}_{\text{ren}}\,
  H^{(1)} \otimes G^{(0)} - \hat{\sigma}^{(0)}_1 \hat{\sigma}^{(0)}_2
  \int_y \Phi^2(y \nu)\, f^{(1)}_{\text{ren}}\, V^{(1)}(y)\, G^{(0)} \,,
\end{align}
where $\Gamma^{e+}$ denotes the sum of graph e and of two graphs
analogous to graphs d and e, with the gluon coupling to the rightmost
quark line.  $f^{(1)}_{\text{ren}}$ is an NLO contribution to the
gluon density, involving a gluon coupling to an eikonal line.  The
integral over the gluon transverse momentum diverges in the
ultraviolet and hence requires renormalisation.
For the $HH$ region we have
\begin{align}
  \label{2v1-HH-d}
C_{HH}\ms \Gamma^{d} & - T_{HH}\, C_{HC}\ms \Gamma^{e+}
  = \int_q f^{(0)}\, H^{(2)} \otimes G^{(0)} 
\nonumber \\
 &\quad - \biggl[ 
    \int_q f^{(0)} \otimes K^{(1)}_{\text{ren}}\; H^{(1)} \otimes G^{(0)}
  - \hat{\sigma}^{(0)}_1 \hat{\sigma}^{(0)}_2
      \int_y \Phi^2(y \nu)\, f^{(0)} \otimes K^{(1)}_{\text{ren}}\;
         V^{(1)}(y)\, G^{(0)} \biggr]
\nonumber \\
 &\quad - \hat{\sigma}^{(0)}_1 \hat{\sigma}^{(0)}_2
    \int_y \Phi^2(y \nu)\, f^{(0)}\, V^{(2)}(y)\, G^{(0)} \,.
\end{align}
In the subtraction term for the region $HC$, we have an NLO
contribution $K^{(1)}_{\text{ren}}$ to the $gg$ splitting kernel with
a gluon coupling to an eikonal line, in analogy to
$f^{(1)}_{\text{ren}}$ above.  Again, this requires summation over
several graphs as specified on the l.h.s.  In the last line
of \eqref{2v1-HH-d} we have the subtraction term for the region $CC$,
in which the gluon and the splitting are regarded as collinear, so
that no eikonal approximation is made and one obtains an NLO
contribution $V^{(2)}$ to the $g\to q\bar{q}$ splitting kernel.
Adding all contributions, we obtain
\begin{align}
\Gamma^{d} + (1 - T_{HH})\, C_{HC}\ms \Gamma^{e+}
 &= \int_q \Bigl( f^{(1)}_{\text{ren}}\, H^{(1)}
    + f^{(0)}\, \bigl[ H^{(2)} - K^{(1)}_{\text{ren}}\; H^{(1)} \bigr]
  \Bigr) \otimes G^{(0)}
\nonumber \\
 &\quad - \hat{\sigma}^{(0)}_1 \hat{\sigma}^{(0)}_2 \int_y
     \Phi^2(y \nu)\, F^{(2)}_{\text{spl,pt}}(y)\, F^{(0)}_{\text{int,pt}}(y)
\nonumber \\
 &\quad + \hat{\sigma}^{(0)}_1 \hat{\sigma}^{(0)}_2
  \int_y \Phi^2(y \nu)\,
  F^{(2)}_{\text{spl}}(y)\, F^{(0)}_{\text{int}}(y) \,,
\end{align}
where
\begin{align}
  \label{Fspl-pt-d}
F^{(2)}_{\text{spl,pt}}(y) & \;\underset{d \text{ and } e+}{=}\;
    f^{(1)}_{\text{ren}}\, V^{(1)}(y)
  + f^{(0)} \bigl[ V^{(2)}(y) - K^{(1)}_{\text{ren}}\; V^{(1)}(y) \bigr]
\end{align}
is a contribution to the perturbative splitting DPD at NLO.  The scale
dependence of $f^{(1)}_{\text{ren}}$ and $K^{(1)}_{\text{ren}}$
cancels in \eqref{Fspl-pt-d}, corresponding to the fact that
graph~\ref{fig:nlo-graphs}d and the other ones required for applying
the Ward identity do not contribute to the scale evolution of the DPD.

\paragraph{Other NLO graphs.}  The discussion of graphs where the
extra gluon is exchanged between a right-moving and a left-moving
line, such as in figure~\ref{fig:nlo-graphs}e and f, is more
complicated because it involves more momentum regions.  Depending on
whether the $g\to q\bar{q}$ splitting is collinear or hard, the extra
gluon can be hard ($H$), collinear left moving ($L$), collinear
right-moving ($R$), or soft ($S$).  In the three latter cases, one can
apply the Grammer-Yennie approximation and, after summation over an
appropriate set of graphs, apply Ward identities, similarly to the
example we have just discussed.  This leads to the gluon coupling to
eikonal lines, which in regions $L$ and $R$ are part of the Wilson
lines in the operator defining the DPDs, whereas in region $S$ they
are part of the Wilson lines in the soft factor.  For more details we
refer to \cite{Diehl:2015bca}.  Our formulation of the DPS cross
section in transverse position space is well suited for the treatment
of these Wilson lines, which are taken at the same transverse
positions as the quark or gluon fields in the DPD definitions.  In
transverse momentum space, one instead has a more complicated
structure with convolution integrals.

\section{Double DGLAP evolution}
\label{sec:dglap}

Collinear DPDs $F(x_i,\tvec{y}; \mu_i)$ evolve with $\mu_i$ according to
the double DGLAP equations.  In the present section, we discuss several
aspects of this scale dependence.

\subsection{Resummation of large logarithms}
\label{sec:dglap-logs}

As we have seen in section~\ref{sec:2v1}, the 2v1 graph in
figure~\ref{fig:nlo-graphs}a gives rise to a large logarithm
$\log(Q/\Lambda)$ when the cross section is integrated over the
transverse boson momenta.  This logarithm originates from the
$g\to q\bar{q}$ splitting vertices in the kinematic region of double
parton scattering and is correctly reproduced in the DPS cross
section, provided that we make a suitable choice of scales.

Radiative corrections to figure~\ref{fig:nlo-graphs}a give rise to
additional large logarithms of DGLAP type.  One can expect that these
logarithms are correctly resummed to all orders by the DGLAP evolution of
the DPDs, provided again that one makes suitable scale choices.  In this
section we show for explicit examples that this is indeed the case.
We will allow for different hard scales $Q_1, Q_2$ and set the scales
as
\begin{align}
  \label{scale-choice}
\mu_1 &= Q_1 \,, & 
\mu_2 &= Q_2 \,, &
\nu &= Q_{\text{min}} \,,
\end{align}
where we define
\begin{align}
Q_{\text{min}} &= \min{(Q_1, Q_2)} \,.
\end{align}
Of course, our conclusions regarding large logarithms will not change
if \eqref{scale-choice} is modified by factors of order unity.  For
definiteness, we will always take a rigid cutoff
$\Phi(u) = \Theta(u - b_0)$ in $y \nu$.  For later use we define the
scale
\begin{align}
\mu_y &= \frac{b_0}{y} \,.
\end{align}
DGLAP logarithms are generated by loop integrals in regions of strongly
ordered transverse momenta.  We will compare the logarithms obtained
directly from transverse momentum integrals with the ones obtained from
evolved DPDs in $y$ space.  

To set the scene, we start with the lowest-order 2v1 graph in
figure~\ref{fig:nlo-graphs}a.  In the DPS region, i.e.\ when the scale of
the $g\to q\bar{q}$ splittings is much smaller than $Q_1$ and $Q_2$, the
behaviour of a splitting vertex is easily recovered from the computation
of the splitting DPD.  As can be seen from \eqref{TMD-split}, a vertex
with quark momentum $k$ gives a factor $\tvec{k}^{j} /\tvec{k}^2$, where
$j$ is a transverse index.  The transverse momentum of the splitting
vertices must flow through the gauge boson lines.  Thus, the two splitting
vertices in figure~\ref{fig:nlo-graphs}a give a combined $1/q_1^2$
behaviour for $q_1 \ll Q_{\text{min}}$, as is evident in \eqref{tw4-final}
after angular integration.  The integrated cross section is then
proportional to
\begin{align}
  \label{dglap-2v1-lo}
\int_{\Lambda^2}^{Q_{\text{min}}^2} \frac{dq_1^2}{q_1^2}
 = \log\frac{Q_{\text{min}}^2}{\Lambda^2} \,.
\end{align}
Since the $q_1$ integral is logarithmic, we must include an infrared
cutoff $\Lambda$, which limits the integration to the perturbative
region where the preceding analysis is valid.  We take here the usual
procedure for extracting the leading behaviour of a logarithmic
integral $\int_a^b dk\; \Gamma(k)$, approximating the integrand
$\Gamma(k)$ for $a\ll k \ll b$ while integrating over the full range
$a\le k\le b$.

Let us now see how the logarithm \eqref{dglap-2v1-lo} arises in the DPS
cross section formula.  According to the perturbative expressions of the
DPDs, $F_{\text{spl,pt}} \sim 1/y^2$ and $F_{\text{int,pt}}$ is $y$
independent at leading order in $\alpha_s$, so that we have
\begin{align}
  \label{dglap-2v1-lo-y}
\int_{b_0^2/\nu^2}^{b_0^2/\Lambda^2} \frac{dy^2}{y^2}
 = \log\frac{Q_{\text{min}}^2}{\Lambda^2} \,.
\end{align}
With our choice for the infrared cutoff on $y$, the expressions
\eqref{dglap-2v1-lo} and \eqref{dglap-2v1-lo-y} agree including their
overall normalisation.

\begin{figure}
\begin{center}
\subfigure[]{\includegraphics[width=0.325\textwidth]{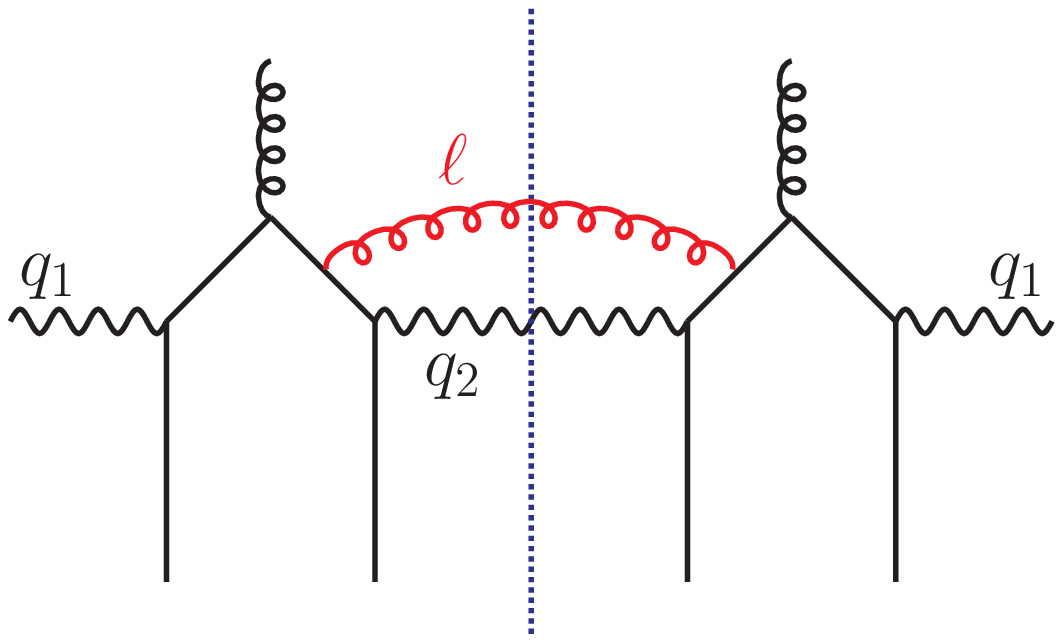}}
\subfigure[]{\includegraphics[width=0.325\textwidth]{figures/nlo-2v1-dia.eps}}
\subfigure[]{\includegraphics[width=0.325\textwidth]{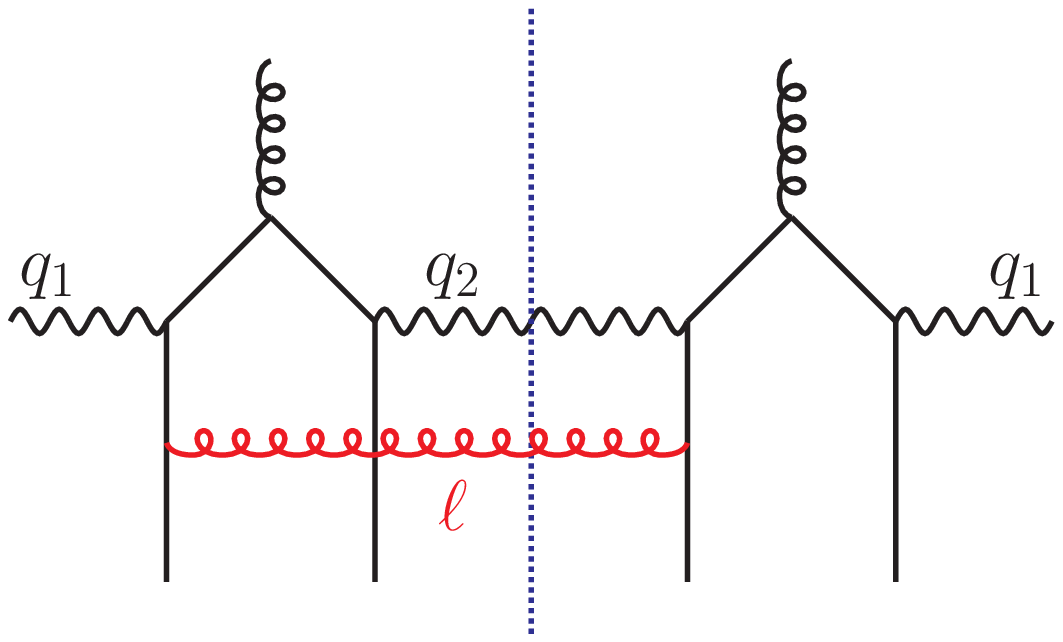}}
\caption{\label{fig:dglap-2v1} Radiative corrections to the
  hard scattering graph in figure~\protect\ref{fig:nlo-graphs}a that
  give rise to DGLAP logarithms.}
\end{center}
\end{figure}

Now consider the NLO graph in figure~\ref{fig:dglap-2v1}a.  We route
the gluon momentum $\ell$ back through the boson line with momentum
$q_2$, so that $\ell$ flows through exactly two quark propagators.  In
the strongly ordered region
\begin{align}
  \label{ordering-1v1}
\Lambda & \ll q_1 \ll Q_1 \,, &
q_1 & \ll \ell \ll Q_2
\end{align}
the graph has the same DPS logarithm as the leading-order result,
whilst developing an additional logarithm from the integral
$\int d\ell^2/\ell^2$, as is familiar from similar gluon emission
graphs in hard scattering processes.  In the case $Q_2 < Q_1$ this
gives
\begin{align}
  \label{dglap-2v1-nlo-Q2min}
\int_{\Lambda^2}^{Q_2^2} \frac{dq_1^2}{q_1^2}
  \int_{q_1^2}^{Q_2^2} \frac{d\ell^2}{\ell^2}
&= \frac{1}{2} \log^2\frac{Q_2^2}{\Lambda^2}
 = \frac{1}{2} \log^2\frac{Q_{\text{min}}^2}{\Lambda^2} \,,
\end{align}
whereas for $Q_1 < Q_2$ we have
\begin{align}
  \label{dglap-2v1-nlo-Q1min}
\int_{\Lambda^2}^{Q_1^2} \frac{dq_1^2}{q_1^2}
  \int_{q_1^2}^{Q_2^2} \frac{d\ell^2}{\ell^2}
&= \frac{1}{2}\, \biggl[ \log^2\frac{Q_2^2}{\Lambda^2} 
    - \log^2\frac{Q_2^2}{Q_1^2} \biggr]
 = \frac{1}{2} \log\frac{Q_{\text{min}}^2}{\Lambda^2}
     \biggl[ \log\frac{Q_{\text{min}}^2}{\Lambda^2}
             + 2 \log\frac{Q_2^2}{Q_1^2} \biggr] \,.
\end{align}
In both cases, we have an additional logarithm $\log(Q_{\text{min}}^2
/\Lambda^2)$ compared with the lowest-order result.  If $Q_1 \ll Q_2$,
then the logarithm of the ratio $Q_2/Q_1$ is also large.

Let us compare these results with the 2v1 part of the DPS cross
section formula when we take into account evolution of the splitting
DPD.  Focusing on the dependence on $y$ and $\mu$, we can write
\begin{align}
  \label{split-dpd-mu-y}
F_{\text{spl,pt}}(x_i, \tvec{y}; \mu_y,\mu_y) & \propto\, \frac{1}{y^2}\,
  \alpha_s(\mu_y)\, g(x_1+x_2; \mu_y)
\end{align}
at leading order, where the choice of scale $\mu_y$ ensures that there are
no large logarithms of $y \mu$ in the higher-order corrections.  This
should be evolved to $F_{\text{spl,pt}}(x_i, \tvec{y}; \mu_1,\mu_2)$ in
the cross section.  In \eqref{split-dpd-mu-y} there is a logarithmic
dependence on $y$ from the running of $\alpha_s$ and from the scale
dependence of the gluon density.  Graph \ref{fig:dglap-2v1}a does not
contribute to either of these, so that the corresponding dependence should
not be taken into account here.  Rather, the DGLAP logarithm of graph
\ref{fig:dglap-2v1}a corresponds to the evolution of
\eqref{split-dpd-mu-y} from $\mu_y$ to $\mu_2$, which gives a factor
proportional to $\alpha_s \log(\mu_2^2 /\mu_y^2)$.  The relevant $y$
integral in the DPD cross section thus is
\begin{align}
  \label{dglap-2v1-y}
\int_{b_0^2/\nu^2}^{b_0^2/\Lambda^2} \frac{dy^2}{y^2}\,
     \log\frac{y^2 \mu_2^2}{b_0^2}
 &= \frac{1}{2}\, \biggl[ \log^2\frac{Q_2^2}{\Lambda^2} 
    - \log^2\frac{Q_2^2}{Q_{\text{min}}^2} \biggr] \,,
\end{align}
where we have used the scale choice in \eqref{scale-choice}.
This agrees with both \eqref{dglap-2v1-nlo-Q2min} and
\eqref{dglap-2v1-nlo-Q1min}.

Note the ease with which \eqref{dglap-2v1-y} is obtained in $y$ space,
compared with the detailed considerations leading to
\eqref{dglap-2v1-nlo-Q2min} and \eqref{dglap-2v1-nlo-Q1min}.  We see that
with evolved DPDs in $\sigma_{\text{DPS}}$ one includes higher order
logarithms while using tree-level partonic cross sections.  Treating the
2v1 graphs fully within the collinear twist-four formalism would require
the explicit inclusion of the higher-order graphs.

Whereas in the previous example, the gluon is emitted close to the
splitting vertices, the opposite is true for the graphs in
figure~\ref{fig:dglap-2v1}b and c.  For figure~\ref{fig:dglap-2v1}b the
region generating two large logarithms is
\begin{align} 
\Lambda & \ll q_1 \ll Q_{\text{min}} \,, &
\Lambda & \ll \ell \ll Q_2 \,,
\end{align}
which gives
\begin{align}
  \label{dglap-2v1-nlo-b}
\int_{\Lambda^2}^{Q_{\text{min}}^2} \frac{dq_1^2}{q_1^2}
  \int_{\Lambda^2}^{Q_2^2} \frac{d\ell^2}{\ell^2}
 &= \log\frac{Q_{\text{min}}^2}{\Lambda^2}\,
    \log\frac{Q_2^2}{\Lambda^2} \,.
\end{align}
To see how these logarithms arise from the evolution of
$F_{\text{int,pt}}$, we use the short-distance expansion
\begin{align}
  \label{intr-dpd-mu-y}
F_{\text{int,pt}}(x_1,x_2,\tvec{y}; \mu_y,\mu_y)
 &= G(x_1,x_2,x_2,x_1; \mu_y)
\end{align}
at leading order, which we evaluate at scale $\mu_y$ to avoid large
logarithms of $y \mu$ in the higher-order corrections, as we did for
$F_{\text{spl,pt}}$ in \eqref{split-dpd-mu-y}.  Graph
\ref{fig:dglap-2v1}b contributes to the scale evolution of $G$, where
at leading order it gives $\alpha_s \log(\mu_y^2 /\Lambda^2)$ when
evolution is started from the generic soft scale $\Lambda^2$.  The
graph also contributes to the evolution of $F_{\text{int,pt}}$ from
$\mu_y$ to $\mu_2$, where it gives $\alpha_s \log(\mu_2^2 /\mu_y^2)$.
The two effects add up to a factor $\alpha_s \log(\mu_2 /\Lambda^2)$.
In the DPS cross section we thus have
\begin{align}
\int_{b_0^2/\nu^2}^{b_0^2/\Lambda^2} \frac{dy^2}{y^2}\,
     \log\frac{\mu_2^2}{\Lambda^2}
 &= \log\frac{Q_{\text{min}}^2}{\Lambda^2}\,
    \log\frac{Q_2^2}{\Lambda^2}
\end{align}
in agreement with \eqref{dglap-2v1-nlo-b}.

The graph in figure~\ref{fig:dglap-2v1}c also has a region where DGLAP
logarithms are generated, as was already discussed in \cite{Gaunt:2012dd}.
We route the loop momentum $\ell$ through two quark lines at one of the
two $g\to q\bar{q}$ splitting vertices.  In the strongly ordered region
\begin{align}
\Lambda & \ll \ell \ll q_1 \ll Q_{\text{min}}
\end{align}
the two lower fermion lines to which the gluon couples have virtualities
of order $\ell^2$, whereas both splitting vertices have virtualities of
order $q_1^2 \approx q_2^2$.  This gives\,\footnote{Note that in
  equation (3.13) of \protect\cite{Gaunt:2012dd} a factor $1/2$ is missing
  on the r.h.s.}
\begin{align}
  \label{dglap-2v1-nlo-c}
\int_{\Lambda^2}^{Q_{\text{min}}^2} \frac{dq_1^2}{q_1^2}
  \int_{\Lambda^2}^{q_1^2} \frac{d\ell^2}{\ell^2}
 &= \frac{1}{2} \log^2\frac{Q_{\text{min}}^2}{\Lambda^2} \,.
\end{align}
Just as graph \ref{fig:dglap-2v1}b, graph c contributes to the
evolution of $G$ in \eqref{intr-dpd-mu-y}, giving a factor
$\alpha_s \log(\mu_y^2 /\Lambda^2)$ at leading order.  There is
however no contribution to the evolution of $F_{\text{int,pt}}$ from
$\mu_y$ to $\mu_2$ from this graph, so that in the cross section we
have
\begin{align}
\int_{b_0^2/\nu^2}^{b_0^2/\Lambda^2} \frac{dy^2}{y^2}\,
     \log\frac{b_0^2}{\Lambda^2 y^2}
 &= \frac{1}{2} \log^2\frac{Q_{\text{min}}^2}{\Lambda^2} \,,
\end{align}
which again reproduces the result \eqref{dglap-2v1-nlo-c} obtained in
momentum space, including the overall factor.

A fully analogous discussion can be given for the graph in
figure~\ref{fig:nlo-graphs}b.  One finds the same type of DGLAP
logarithms as in \eqref{dglap-2v1-nlo-c}, which is again reproduced
correctly in the DPS cross section formula.  Note that in this case
the additional gluon is a virtual correction rather than emitted into
the final state -- a somewhat unusual situation in the context of
DGLAP logarithms.

\begin{figure}
\begin{center}
\subfigure[]{\includegraphics[width=0.325\textwidth]{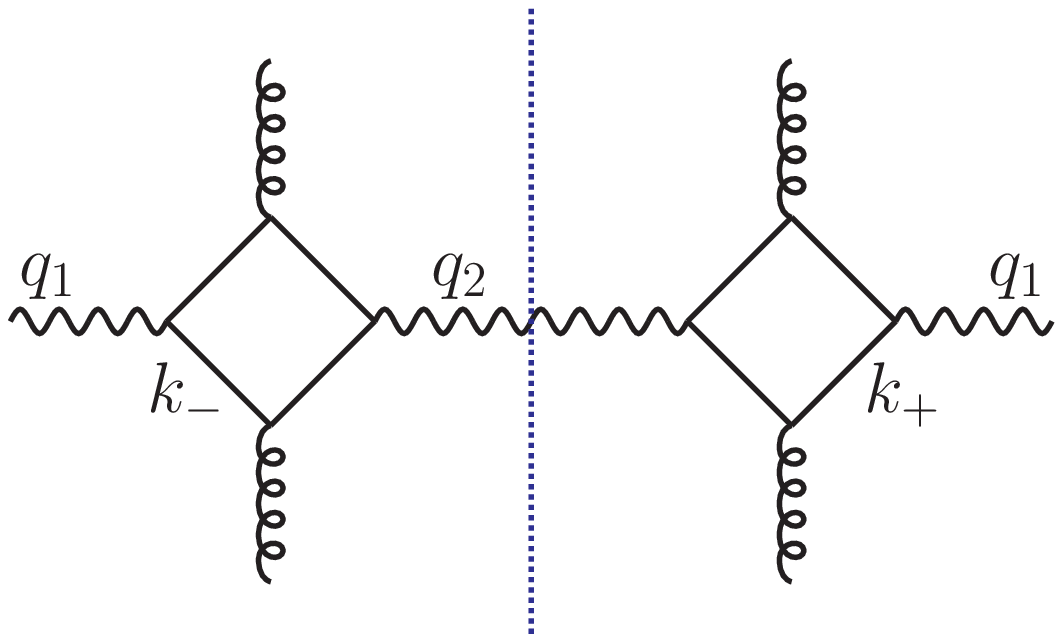}}
\subfigure[]{\includegraphics[width=0.325\textwidth]{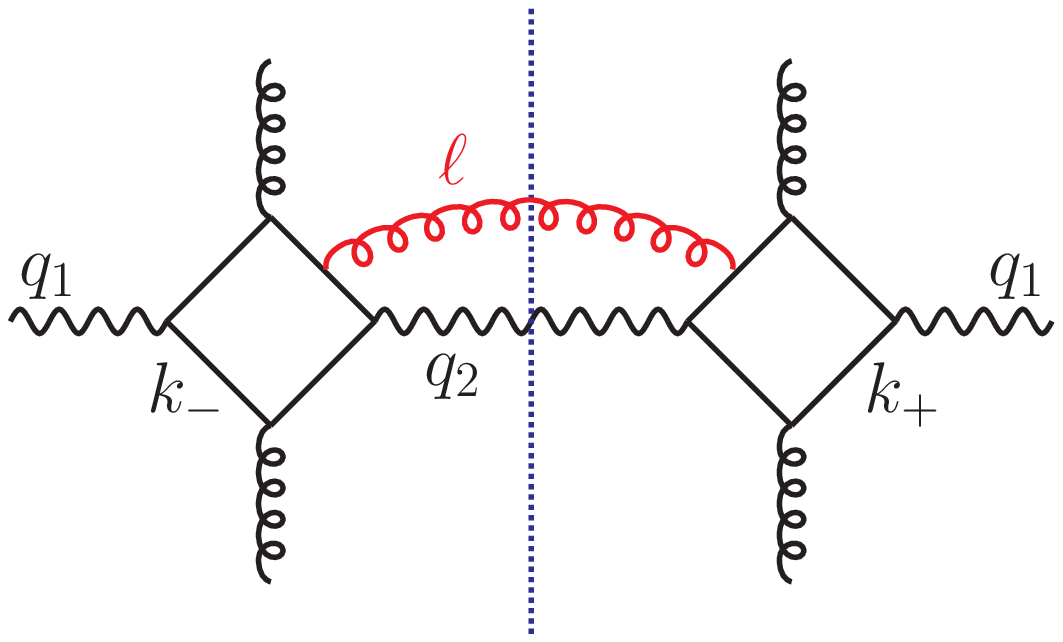}}
\caption{\label{fig:dglap-1v1} (a) Hard scattering part of the double
  box graph.  (b) Higher-order correction giving rise to a large DGLAP
  logarithm in the DPS region.}
\end{center}
\end{figure}

In conclusion, higher-order graphs with strong ordering of transverse
momenta produce DGLAP type logarithms in 2v1 graphs, which can depend
on $Q_{\text{min}}/\Lambda$ or $Q_2/Q_1$.  Our expression for the
regularised DPS cross section correctly reproduces these logarithms if
we make an adequate choice of DPDs and scales.  Specifically, we must
\begin{itemize}
\item use DPDs that have the correct short-distance behaviour, as
  discussed in section~\ref{sec:coll-all}.  Fixed-order perturbative
  results for this behaviour can be used reliably at scales of order
  $\mu_y$, as in \eqref{split-dpd-mu-y} and \eqref{intr-dpd-mu-y}.
\item take scales $\mu_1, \mu_2$ and $\nu$ in $\sigma_{\text{DPS}}$ as
  specified in \eqref{scale-choice}, with possible modifications by
  factors of order unity.
\end{itemize}
The situation is quite different for 1v1 graphs.  In the DPS region
\begin{align}
  \label{ordering-1v1-lo}
q_1 \ll k_\pm \ll Q_{\text{min}}
\end{align}
the double box in figure~\ref{fig:dglap-1v1}a develops a squared DPS
logarithm in $d\sigma /dq_1^2$,
\begin{align}
  \label{dglap-1v1-lo}
  \int_{q_1^2}^{Q_{\text{min}}^2} \frac{dk_+^2}{k_+^2} \,
  \int_{q_1^2}^{Q_{\text{min}}^2} \frac{dk_-^2}{k_-^2}
&= \log^2 \frac{Q_{\text{min}}^2}{q_1^2}
\end{align}
which disappears however when one integrates over $q_1$ up to
$Q_{\text{min}}$.  This is inevitable since the integrated 1v1 cross
section depends on a single dimensionful quantity $Q_{\text{min}}$ and
hence cannot have any large logarithm.  An infrared cutoff $\Lambda$ in
the $q_1$ integration does not play any role here since the cross section
is dominated by large $q_1 \sim Q_{\text{min}}$, where strong ordering as
in \eqref{ordering-1v1-lo} is not possible.

One finds that the higher-order graph in figure~\ref{fig:dglap-1v1}b
does have a DGLAP logarithm in the strong-ordering regime
\begin{align}
  \label{ordering-1v1-nlo}
q_1 \ll k_{\pm} & \ll Q_1 \,,
&
k_{\pm} & \ll \ell \ll Q_2 \,,
\end{align}
due to the additional loop integral $\int d\ell^2/\ell^2$.  But the region
\eqref{ordering-1v1-nlo} gives only a power suppressed contribution to the
cross section integrated over $q_1$, where DPS kinematics is not
dominant.\footnote{This point was overlooked in the discussion of DGLAP
  logarithms in \protect\cite{Diehl:2016khr,Diehl:2016rqr}.}

There is an exception to the statement just made.  If at least one of the
momentum fractions $x_1, x_2$ is very small and the corresponding scale
$\mu_1, \mu_2$ is sufficiently large, then the evolution of
$F_{\text{spl,pt}}$ from $\mu_y$ to $\mu_1$, $\mu_2$ can significantly
flatten the $1/y^2$ behaviour of the initial condition
\eqref{split-dpd-mu-y}.  The underlying physics is radiation of soft
gluons, which strongly enhances parton distributions at low $x$, as is
well known from the evolution of ordinary PDFs.  The enhancement by
evolution of $F_{\text{spl,pt}}$ increases with the distance between
$\mu_y$ and $\mu_1, \mu_2$, i.e.\ it increases with~$y$.  In the 1v1 (and
also the 2v1) contribution to the DPS cross section, this enhances the
importance of the region $y \gg 1/Q$, where the DPS approximation is
valid.  In this region, the evolution of the DPDs correctly describes the
DGLAP logarithms generated by graphs like the one in
figure~\ref{fig:dglap-1v1}b.  In the numerical studies presented in
section~\ref{sec:lumi}, we find that this happens for $F_{gg}$, $F_{qg}$
and $F_{\bar{q}g}$, as well as for $F_{q\bar{q}}$ when one momentum
fraction is much smaller than the other.


\subsection{Dependence on the cutoff scale}
\label{sec:dglap-subtr}

As already discussed in section~\ref{sec:coll-subtr}, our formalism
relies on the fact that in the region of small $y \sim 1/Q$ the
subtraction terms in the cross section \eqref{coll-master} cancel
against the DPS contribution, up to corrections that are beyond the
accuracy of the calculation.  Such corrections include higher-order
terms in $\alpha_s$, whose parametric size we now investigate.

Let $F_{\text{fo}}(x_i,\tvec{y}; \mu)$ denote the short distance
approximation of a DPD, computed in fixed-order perturbation theory.
It depends on $\mu$ via the scales of $\alpha_s$ and of PDFs or
twist-four distributions (and at higher orders also via powers of
$\log(y\mu)$ in the splitting kernels).  In the cross section,
this fixed-order approximation is used in two different ways:
\begin{enumerate}
\item For the small-$y$ limit of the DPDs used in the DPS cross
  section, $\sigma_{\text{DPS}}$, one takes
  $F_{\text{fo}}(x_i,\tvec{y}; \mu_y)$ as initial condition and
  evolves it to the scales $\mu_1 \sim Q_1$ and $\mu_2 \sim Q_2$ of
  the two hard scatters.
\item In the subtraction terms, $\sigma_{\text{1v1,pt}}$ and
  $\sigma_{\text{2v1,pt}}$, one takes
  $F_{\text{fo}}(x_i,\tvec{y}; \mu_h)$ evaluated at the same scale
  $\mu_h$ that is used in the hard scattering terms,
  $\sigma_{\text{SPS}}$ or $\sigma_{\text{tw4}}$, so as to cancel the
  contribution from the region $y \gg 1/Q$ order by order in
  perturbation theory.
\end{enumerate}
The scale $\mu_h$ may be a combination of $Q_1$ or $Q_2$ or an
independent third scale.  For now we assume that
$\alpha_s\ms L \ll 1$, where
\begin{align}
  \label{L-def}
  L &= \max\ms \bigl\{ |\log(Q_2/Q_1)|\ms, |\log(\mu_h/Q_1)|\ms,
                       |\log(\mu_h/Q_2)|  \ms\bigr\} \,.
\end{align}
The case where this is not satisfied because the hard scales are
widely different is discussed in the following subsection.

Consider now a calculation at N$^{k}$LO ($k=0$ for LO), using DPDs
evolved with N$^{k}$LO kernels $P$ and N$^k$LO expressions for the DPS
subprocess cross sections $\hat{\sigma}_{1,2}$ and the short-distance
approximations $\smash{F_{\text{fo}}}$.  We have
\begin{align}
  \label{sigma-mu-dep}
\frac{d}{d\log\mu_1^2}\, \hat{\sigma}_{1}
 + \hat{\sigma}_{1} \underset{x_1}{\otimes} P
 + \hat{\sigma}_{1} \underset{\bar{x}_1}{\otimes} P
 &= \mathcal{O}\bigl( \alpha_s^{k+1} \log^{k}(\mu_1/Q_1) \bigr)
      \times  \hat{\sigma}_{1}
\end{align}
and its analogue for $\hat{\sigma}_{2}$.  Here $\otimes$ denotes the
usual Mellin convolution for the indicated variable, and we omit
labels and sums for parton species.  To understand \eqref{sigma-mu-dep}
we observe that the l.h.s.\ would be exactly zero if $\hat{\sigma}_1$
and $P$ were calculated to all orders; the convolution of
$\hat{\sigma}_1$ with the relevant parton densities would then be
$\mu_1$ independent.  With N$^k$LO truncations on the l.h.s., the
first term beyond the accuracy of the calculation is of order
$\alpha_s^{k+1}\ms \hat{\sigma}_{1}^{}$.  Furthermore, the highest
power of $\log(\mu_1/Q_1)$ in $\hat{\sigma}_{1}$ is $k$ according to
the general pattern for logarithms related with the renormalisation or
factorisation scale.
In a similar fashion we can derive
\begin{align}
  \label{DGLAP-fo}
\frac{d}{d \log\mu^2}\,
  F_{\text{spl,fo}}(x_i,\tvec{y}; \mu)
  &= P \underset{x_1}{\otimes} F_{\text{spl,fo}}(\tvec{y}; \mu)
   + P \underset{x_2}{\otimes} F_{\text{spl,fo}}(\tvec{y}; \mu)
\nonumber \\
 &\quad + \mathcal{O}\bigl( \alpha_s^{k+1} \log^k(y \mu) \bigr)
        \times F_{\text{spl,fo}}(\tvec{y}; \mu) \,,
\end{align}
using that $k$ is the highest power of $\log(y\mu)$ in the N$^k$LO
approximation, according to~\eqref{V-series-zero}.
Finally, it is easy to see that for a quantity satisfying
\begin{align}
  \label{generic-evo}
\frac{d}{d\log\mu^2}\, S(\mu)
  &= \mathcal{O}\bigl( \alpha_s^{k+1} \log^k(\mu/\mu_0) \bigr)
       \times S(\mu)
\end{align}
one has
\begin{align}
S(\mu_b) - S(\mu_a) &= \mathcal{O}\bigl(
    \alpha_s^{k+1} \log^{k+1}(\mu_a/\mu_0),
    \alpha_s^{k+1} \log^{k+1}(\mu_b/\mu_0)
 \bigr) \times S(\mu_a) \,,
\end{align}
where the integration of the differential equation
\eqref{generic-evo} between $\mu_a$ and $\mu_b$ gives an extra
logarithm on the r.h.s.  We thus obtain for the small-$y$ contribution
to the 1v1 cross section
\begin{align}
  \label{1v1-logs-start}
& \bigl[ \sigma_{\text{1v1}} \bigr]_{y\sim 1/Q}
  = \hat{\sigma}_{1}(\mu_1) \otimes \hat{\sigma}_{2}(\mu_2)
     \otimes \!\! \int\limits_{y\sim 1/Q} \!\!\! d^2\tvec{y}\;
          F_{\text{spl,pt}}(\tvec{y}; \mu_1,\mu_2)\,
          F_{\text{spl,pt}}(\tvec{y}; \mu_1,\mu_2)
\nonumber \\
  &\quad = \int\limits_{y\sim 1/Q} \!\!\! d^2\tvec{y}\;
     \hat{\sigma}_{1}(\mu_y) \otimes \hat{\sigma}_{2}(\mu_y)
      \otimes
          F_{\text{spl,fo}}(\tvec{y}; \mu_y)\,
          F_{\text{spl,fo}}(\tvec{y}; \mu_y)
\nonumber \\
  &\qquad \times \Bigl[ 1 + \mathcal{O}\bigl(
       \alpha_s^{k+1} \log^{k+1}(Q_1/\mu_y),\,
       \alpha_s^{k+1} \log^{k+1}(Q_2/\mu_y)\bigr) \Bigr]
\nonumber \\[0.4em]
  &\quad = \hat{\sigma}_{1}(\mu_h) \otimes \hat{\sigma}_{2}(\mu_h)
       \otimes \!\! \int\limits_{y\sim 1/Q} \!\!\! d^2\tvec{y}\;
          F_{\text{spl,fo}}(\tvec{y}; \mu_h)\,
          F_{\text{spl,fo}}(\tvec{y}; \mu_h)
\nonumber \\
  &\qquad \times \Bigl[ 1 + \mathcal{O}\bigl(
       \alpha_s^{k+1} \log^{k+1}(y \mu_h),\,
       \alpha_s^{k+1} \log^{k+1}(y\ms Q_1),\,
       \alpha_s^{k+1} \log^{k+1}(y\ms Q_2)\bigr) \Bigr] \,,
\end{align}
using our initial condition $F_{\text{spl,pt}} = F_{\text{spl,fo}}$ at
scale $\mu_y$.  The longitudinal momentum arguments and convolutions
are as in \eqref{coll-Xsect-naive}.  The first line of the last
expression in \eqref{1v1-logs-start} is just the double counting
subtraction term.  We thus find
\begin{align}
  \label{1v1-logs}
\bigl[ \sigma_{\text{1v1}} - \sigma_{\text{1v1,pt}} \bigr]_{y\sim 1/Q}
  &= \bigl[ \sigma_{\text{1v1,pt}} \bigr]_{y\sim 1/Q} \,
     \times \mathcal{O}\bigl( \alpha_s^{k+1} L^{k+1}  \bigr)
\nonumber \\[0.1em]
  &= \bigl[ \sigma_{\text{SPS}} - \sigma_{\text{1v1,pt}} \bigr] \,
     \times \mathcal{O}\bigl( \alpha_s^{k+1} L^{k+1} \bigr) \,.
\end{align}
with $L$ defined in \eqref{L-def}.
In the second step we have used that the full SPS cross section is of the
same order of magnitude as the DPS subtraction term, regarding its power
behaviour in $Q^2$ as well as the relevant coupling constants and PDFs.
In full analogy one can derive
\begin{align}
  \label{2v1-logs}
\bigl[ \sigma_{\text{2v1}} - \sigma_{\text{2v1,pt}} \bigr]_{y\sim 1/Q}
  &= \bigl[ \sigma_{\text{2v1,pt}} \bigr]_{y\sim 1/Q} \,
     \times \mathcal{O}\bigl( \alpha_s^{k+1} L^{k+1} \bigr)
\nonumber \\[0.1em]
  &= \bigl[ \sigma_{\text{tw4}} - \sigma_{\text{2v1,pt}} \bigr] \,
     \times \mathcal{O}\bigl( \alpha_s^{k+1} L^{k+1} \bigr) \,.
\end{align}
Notice that the difference $\sigma_{\text{tw4}} - \sigma_{\text{2v1,pt}}$
is by construction dominated by short distances $y \sim 1/Q$, whilst the
individual terms receive contributions from a wide range of $y$.
We thus find that, for both 1v1 and 2v1 contributions, the DPS term
cancels against the double counting subtractions in the region
$y \sim 1/Q$, up to corrections of relative order $\alpha_s^{k+1}$.
This is beyond the accuracy of an N$^k$LO calculation, as it should
be.  The only logarithms multiplying $\alpha_s$ in this context are
logarithms of the ratios between different hard scales, which at this
point we must require to be of moderate size.

We can now also quantify the dependence of the overall cross section
$\sigma$ in \eqref{coll-master} on the scale $\nu$.  With two choices
$\nu_1 \sim \nu_2 \sim Q$ the difference $\sigma(\nu_1) - \sigma(\nu_2)$
is just of the size given by \eqref{1v1-logs} and \eqref{2v1-logs}.  In
analogy to the variation of renormalisation and factorisation scales, one
may thus use the variation of $\nu$ to estimate uncalculated higher-order
corrections.

An important consequence of \eqref{1v1-logs} is that
\begin{align}
\bigl[ \sigma_{\text{1v1}} \bigr]_{y\sim 1/Q}
 &\,\sim\,  \sigma_{\text{1v1,pt}}
    \sim    \sigma_{\text{SPS}} \,,
\end{align}
where we used that the subtraction term $\sigma_{\text{1v1,pt}}$ is
dominated by $y\sim 1/Q$, given the $1/y^2$ behaviour of
$F_{\text{spl,fo}}$.  This means that we can use the $\nu$ variation
of the 1v1 contribution to $\sigma_{\text{DPS}}$ to estimate the size
of the subtraction term $\sigma_{\text{1v1,pt}}$ and thus of the SPS
contribution to the cross section at the corresponding order in
$\alpha_s$.  This is of great practical value, given the considerable
difficulty to compute SPS at this order.  We will come back to this
point in section~\ref{sec:lumi}.


\subsection{Multiscale problems}
\label{sec:sub-multi}

The arguments leading to \eqref{1v1-logs} and \eqref{2v1-logs} are based
on an expansion in $\alpha_s$ and break down when $\alpha_s\ms L$ is not
small, for instance because there is a large hierarchy $Q_1 \ll Q_2$
between the two hard scales.  We now show that a modified choice of DPDs
in the subtraction terms $\sigma_{\text{1v1,pt}}$ and
$\sigma_{\text{2v1,pt}}$ can cope with such a situation.  

Let us for definiteness describe the 1v1 subtraction; the 2v1 case can be
treated in full analogy.  We use the notation
$F_{\text{spl,pt}}(x_i,\tvec{y}; \mu_b, \mu_c \ms|\ms \mu_a)$ for the
distribution that is initialised to
\begin{align}
F_{\text{spl,pt}}(x_i,\tvec{y}; \mu_a, \mu_a \ms|\ms \mu_a)
 &= F_{\text{spl,fo}}(x_i,\tvec{y}; \mu_a)
\end{align}
and evolved according to the double DGLAP equations to the scale $\mu_b$
($\mu_c$) for the parton with momentum fraction $x_1$ ($x_2$).  The
small-$y$ limit of the distributions used in $\sigma_{\text{DPS}}$ is
$F_{\text{spl,pt}}(x_i,\tvec{y}; \mu_1, \mu_2 \ms|\ms \mu_y)$ and its
analogue for $F_{\text{int,pt}}$.  Let us now introduce a
profile function $p(u; \mu_a,\mu_b)$ that interpolates monotonically
between the limiting values $\mu_a$ at $u=0$ and
$\mu_b$ at $u\to \infty$, satisfying
\begin{align}
  \label{profile-scales-gen}
  p(u;\mu_a,\mu_b) &\approx \mu_a & & \text{for $u \sim 1$,}
\nonumber \\
  p(u;\mu_a,\mu_b) &\approx \mu_b & & \text{for $u \gg 1$.}
\end{align}
We can then take distributions
\begin{align}
  \label{profile-DPD}
F_{\text{spl,pt}}\bigl( x_i,\tvec{y};
  p(y \nu; \mu_1,\mu_h), p(y \nu; \mu_2,\mu_h) \,|\,
  p(y \nu; \nu,\mu_h) \bigr)
\end{align}
in $\sigma_{\text{1v1,pt}}$.  These distributions tend to
$F_{\text{spl,fo}}(x_i,\tvec{y}; \mu_h)$ for $y\gg 1/\nu$, so that the
subtraction term cancels the contribution of $\sigma_{\text{SPS}}$ in
this region.  For $y \sim 1/\nu$ the distributions tend to
$F_{\text{spl,pt}}(x_i,\tvec{y}; \mu_1,\mu_2 \ms|\ms \nu) \approx
F_{\text{spl,pt}}(x_i,\tvec{y}; \mu_1,\mu_2 \ms|\ms \mu_y)$, so that
in that region the subtraction term cancels against
$\sigma_{\text{DPS}}$.  Distances $y \ll 1/\nu$ are of course removed
in the cross section by the cutoff function $\Phi(y\nu)$.  Using the
same technique as in section~\ref{sec:dglap-subtr}, one can show that
the cancellations just described are up to corrections of order
$\alpha_s^{k+1}$ at N$^k$LO, without any large logarithms.  In
particular, this ensures that the $\nu$ dependence of the overall
cross section is beyond the accuracy of the computation.

As an alternative, one may replace $\nu$ in \eqref{profile-DPD} with
$\min(\mu_1, \mu_2)$, which requires less computation when $\nu$ is
varied in the cross section (while $\mu_1, \mu_2, \mu_h$ are kept
fixed).  With this choice, the DPDs in the subtraction term reduce to
$F_{\text{spl,fo}}(x_i,\tvec{y}; \mu_h)$ in the case of equal hard
scales, $\mu_1 = \mu_2 = \mu_h$.

Profile functions for the renormalisation scale have been used in
other contexts, see e.g.~\cite{Ligeti:2008ac,Abbate:2010xh}.  An
suitable function for our purpose is
\begin{align}
  \label{profile-ex}
p(u; \mu_a,\mu_b) &=
\begin{cases}
   \mu_a & \text{for $u \le u_1$} \\[0.1em]  
   \dfrac{\mu_b + \mu_a}{2} + \dfrac{\mu_b - \mu_a}{2} \biggl[
           \dfrac{4\ms (u-u_2)^3}{(u_1-u_3)^3}
         - \dfrac{3\ms (u-u_2)}{u_1-u_3} \biggr] &
             \text{for $u_1 < u < u_3$} \\
   \mu_b & \text{for $u \ge u_3$}
\end{cases}
\end{align}
where $u_1 \sim 1$, $u_3 \gg 1$ and $u_2 = \half (u_1 + u_3)$.
Functions that are piecewise polynomials have been used for instance
in \cite{Abbate:2010xh,Stewart:2013faa}.  To estimate theoretical
uncertainties, one can vary the profile function $p$, for instance by
varying $u_1$ and $u_3$.  For a detailed discussion we refer to
\cite{Stewart:2013faa}.


\subsection{Approximation of the intrinsic distribution}
\label{sec:simpler}

As discussed earlier, the correct small-$y$ limit $F_{\text{int,pt}}$
at leading order in $\alpha_s$ is obtained by making the ansatz
\eqref{intr-dpd-mu-y} at scale $\mu_y$ and then evolving
$F_{\text{int,pt}}$ to the scales $\mu_1, \mu_2$ relevant for the DPS
cross section.  In practical terms, this requires an ansatz or model
for the twist-four distribution $G$ at some reference scale $\mu_0$
and subsequent evolution of $G$ to the scale $\mu_y$.  The technical
implementation of twist-four evolution is rather involved, and one may
wish to avoid it in situations where the corresponding loss of
precision is tolerable.

In the model used later in this paper we indeed follow this path, taking
as limiting form for small $y$
\begin{align}
  \label{intr-dpd-fixed-mu}
F_{\text{int,pt}}(x_1,x_2,\tvec{y}; \mu_0,\mu_0)
 &= G(x_1,x_2,x_2,x_1; \mu_0)
\end{align}
at some fixed reference scale $\mu_0$ and then evolving
$F_{\text{int,pt}}$ to the final scales $\mu_1, \mu_2$.  For the
r.h.s.\ we will take a form proportional to the product $f(x_1,\mu_0)\,
f(x_2,\mu_0)$ of ordinary PDFs (this model is restricted to DPDs in the
colour singlet channel).  Multiplying with a suitable function of $y$, one
readily obtains a model for the intrinsic DPD at any value of $y$.  This
is indeed standard in the existing literature.

As follows from the discussions below \eqref{match-tw4} and below
\eqref{intr-dpd-mu-y}, evolution of the ansatz \eqref{intr-dpd-fixed-mu}
to scales $\mu_1, \mu_2$ correctly takes into account diagonal
interactions as in figure~\ref{fig:intr-nlo}c whilst neglecting
non-diagonal interactions as in figure~\ref{fig:intr-nlo}a and b, which in
the correct treatment contribute to the evolution of $G$ from $\mu_0$ to
$\mu_y$.  For $\sigma_{\text{2v2}}$, these interactions are irrelevant
since the small-$y$ region is power suppressed.  Furthermore, we can
expect them to be of limited importance for $\sigma_{\text{2v1}}$ in the
situation described at the end of section~\ref{sec:dglap-logs}.  Evolution
then leads to a $y$ dependence of $F_{\text{spl,pt}}$ that is
significantly flatter than the $1/y^{2}$ behaviour obtained at fixed
order.  This shifts the dominant region of the $y$ integration in
$\sigma_{\text{2v1}}$ to larger values, so that on the $F_{\text{int,pt}}$
side there remains little room for the evolution of $G$ between $\mu_0$ to
$\mu_y$.  The limited numerical importance of non-diagonal interactions
for 2v1 diagrams in such situations was also observed in
\cite{Gaunt:2012dd}, where calculations were done in transverse momentum
space.

\section{Collinear DPDs in momentum space}
\label{sec:mom-space-dpds}

As we argued in section~\ref{sec:scheme}, there are several advantages
to work with DPDs in $\tvec{y}$ space when setting up a factorisation
formula for the overall cross section.  We now consider collinear DPDs
depending on the Fourier conjugate transverse momentum
$\tvec{\Delta}$, showing that they can be connected with our formalism
in a useful way.  The motivation for this is twofold.
\begin{itemize}
\item Sum rules for momentum space DPDs at $\Delta=0$ have been proposed
  in \cite{Gaunt:2009re}, where it was stated that they are preserved by
  the leading-order inhomogeneous evolution equations in
  \cite{Kirschner:1979im,Shelest:1982dg,Snigirev:2003cq,%
    Gaunt:2009re,Ceccopieri:2010kg}.  If it can be shown that they are
  valid at some starting scale, this will provide valuable (albeit
  nontrivial) constraints on DPDs.  Of course, no such sum rules can hold
  for $\int d^2\tvec{y}\, F(x_i,\tvec{y})$ since this integral is
  divergent at small $y$.
\item Previous work on DPS, in particular on the perturbative splitting
  contributions
  \cite{Blok:2011bu,Blok:2013bpa,Ryskin:2011kk,Ryskin:2012qx}, is
  formulated in transverse momentum space.  The results derived below will
  facilitate the comparison of that work with ours, which is the subject
  of section~\ref{sec:compare}.
\end{itemize}
Both aspects only concern colour singlet DPDs, to which the rest of
this section is restricted.  Colour non-singlet DPDs have an
additional rapidity dependence described by Collins-Soper equations,
which are multiplicative in $\tvec{y}$ space (see
\cite{Buffing:2016wip}) and hence more complicated in momentum space.

The function $\Phi$ that regulates the DPS cross section in our
formalism can also be used to introduce momentum space DPDs.  We
define
\begin{align}
  \label{F-phi-def}
F_{\Phi}(x_i,\tvec{\Delta};\mu,\nu) &= 
  \int d^{2}\tvec{y}\; e^{i\tvec{y}\tvec{\Delta}}\,
    \Phi(y \nu) F(x_i,\tvec{y},\mu) \,,
\end{align}
omitting the superscript indicating the colour singlet channel for
brevity.  With the conditions on $\Phi$ stated in
\eqref{Phi-conditions}, the integral converges at $y=0$.  Throughout
this section, we restrict ourselves to the two choices of $\Phi$ in
table~\ref{tab:phi} and assume that $\nu \gg \Lambda$.  The dependence
of $F_\Phi$ on $\nu$ is then given by
\begin{align}
  \label{nu-dep}
\frac{d F_{\Phi}(x_i,\tvec{\Delta};\mu,\nu)}{d\log\nu^2}
 &= \int d^2\tvec{y}\; e^{i\tvec{y}\tvec{\Delta}}\,
    \frac{d\Phi(y \nu)}{d\log\nu^2}\;
    F_{\text{spl,pt}}(x_i,\tvec{y};\mu) \,.
\end{align}
The derivative of $\Phi(y \nu)$ strongly
suppresses the nonperturbative region $y\sim 1/\Lambda$ in the integral
(or removes it altogether if $\Phi$ is a step function).  This allows us
to replace $F(x_i,\tvec{y},\mu)$ with its perturbative splitting part on
the r.h.s.  The contribution from $F_{\text{int,pt}}(x_i,\tvec{y},\mu)$ is
suppressed by $\Lambda^2/\nu^2$.
Inserting the expressions \eqref{split-lo} or \eqref{split-ho} for
$F_{\text{spl,pt}}(x_i,\tvec{y},\mu)$ at $\epsilon=0$ one obtains
\begin{align}
  \label{nu-dep-lo}
\frac{d F_{\Phi}(x_i,\tvec{\Delta};\mu,\nu)}{d\log\nu^2}
 &= \frac{f(x_1+x_2;\mu)}{x_1+x_2}\, 
      \frac{\alpha_s(\mu)}{2\pi}\,
      P\biggl( \frac{x_1}{x_1+x_2}, 0 \biggr)\, I_0(\Delta,\mu,\nu)
\intertext{at leading order and}
   \label{nu-dep-ho}
\frac{d F_{\Phi}(x_i,\tvec{\Delta};\mu,\nu)}{d\log\nu^2}
 &= \int\limits_{x_1+x_2}^1 \!\! \frac{dv}{v}\, \frac{f(v;\mu)}{v}\,
    \sum_{n=1}^\infty \alpha_s^n(\mu) \sum_{m=0}^{n-1}
    V^{[n,m]}\biggl( \frac{x_1}{v}, \frac{x_2}{v} \biggr)\,
      I_m(\Delta,\mu,\nu)
\end{align}
in general, where
\begin{align}
I_m(\Delta,\mu,\nu)
 &= \int\limits_0^\infty du\, \frac{d\Phi(u)}{du}\,
      J_0(u\ms \Delta/\nu)\, \log^m(u \mu/\nu) \,.
\end{align}
For brevity, we have omitted labels and sums for parton species and
polarisations.

For $\Delta=0$ we have $I_0 = 1$ and recognise in \eqref{nu-dep-lo} the
inhomogeneous term of the evolution equation for a conventional momentum
space DPD $F(x_i,\tvec{\Delta},\mu)$ defined by $\overline{\text{MS}}$
subtraction.  At leading order, our momentum space DPD
$F_{\Phi}(x_i,\tvec{0},\mu,\mu)$ thus obeys the same evolution equation as
$F(x_i,\tvec{\Delta},\mu)$.
At small $\Delta$ the inhomogeneous term is multiplied by
$I_0(\Delta,\mu,\mu) = 1 + \mathcal{O}(\Delta^2/\mu^2)$.  The higher-order
version \eqref{nu-dep-ho} at $\Delta=0$ has the same structure as the
inhomogeneous term in the evolution equation proposed in
\cite{Ceccopieri:2010kg}, although the evolution kernel may not be the
same.

Let us now compute the momentum space analogue of
$F_{\text{spl,pt}}(x_i,\tvec{y},\mu)$, both with the cutoff regularisation
\eqref{F-phi-def} and with $\overline{\text{MS}}$ renormalisation of the
splitting singularity.  This gives the correct result for the full
momentum-space DPD in the limit $\Delta \gg \Lambda$, when the exponent
$e^{i\tvec{y}\tvec{\Delta}}$ suppresses large $y \sim 1/\Lambda$ in the
Fourier integral, so that the perturbative splitting approximation can be
used.

Restricting ourselves to leading order, we start with \eqref{split-lo} and
use the Fourier integrals given in \eqref{bessel-step} and
\eqref{bessel-exp}.  For $\Delta \ll \nu$ we find
\begin{align}
  \label{dpd-mom-lo}
\frac{F_{\Phi, \text{spl,pt}}(x_i,\tvec{\Delta};\mu,\nu)}{R(x_i,\mu)}
 &\approx \begin{cases}
            \log\dfrac{\nu^2}{\Delta^2 \rule[-1ex]{0pt}{0pt}}
              & \text{for $\Phi(u) = \Theta(u-b_0)$} \\
            \log\dfrac{\nu^2}{\Delta^2} - \gamma
              & \text{for $\Phi(u) = 1-e^{-u^2/4}$}
          \end{cases}
\intertext{with corrections of $\mathcal{O}(\Delta^2 /\nu^2)$, whereas
  for $\Delta \gg \nu$ we obtain}
  \label{dpd-mom-hi}
\frac{F_{\Phi, \text{spl,pt}}(x_i,\tvec{\Delta};\mu,\nu)}{R(x_i,\mu)}
 &\approx \begin{cases}
    \dfrac{e^{3\gamma/2}}{\sqrt{\pi}}\,
    \biggl( \dfrac{\nu}{\Delta} \biggr)^{3/2}
    \cos\biggl( b_0\dfrac{\Delta}{\nu} + \dfrac{\pi}{4
      \rule[-1.5ex]{0pt}{0pt}} \biggr)
      & \text{for $\Phi(u) = \Theta(u-b_0)$} \\
    \dfrac{\nu^2}{\Delta^2}\,
        \exp\biggl[ - \dfrac{\Delta^2}{\nu^2} \biggr]
      & \text{for $\Phi(u) = 1-e^{-u^2/4}$}
          \end{cases}
\end{align}
with
\begin{align}
R(x_i,\mu) &= \frac{f(x_1+x_2;\mu)}{x_1+x_2}\,
      \frac{\alpha_s(\mu)}{2\pi}\,
      P\biggl( \frac{x_1}{x_1+x_2}, 0 \biggr) \,.
\end{align}

To compute the usual momentum space DPD, we take the Fourier transform
\eqref{Delta-FT} of $F_{\text{spl,pt}}(x_i,\tvec{y})$ in
$D=4-2\epsilon$ dimensions, subtract the UV divergence and then set
$\epsilon=0$.  This gives
\begin{align}
F_{\text{spl,pt}}(x_i,\tvec{\Delta};\mu) &= \lim\limits_{\epsilon\to 0}\,
  \mu^{2\epsilon} \! \int d^{2-2\epsilon}\tvec{y}\;
    e^{i \tvec{y}\tvec{\Delta}}\, \frac{1}{y^{2-4\epsilon}}\,
      \frac{\Gamma^2(1-\epsilon)}{\pi^{1-2\epsilon}}\,
      \frac{\alpha_s}{2\pi}\, P(v, \epsilon) 
      \frac{f(x)}{x}
  - C_{\epsilon} \,,
\end{align}
where we have abbreviated $x=x_1+x_2$ and $v=x_1/(x_1+x_2)$.  Here
$C_{\epsilon}$ denotes the $\overline{\text{MS}}$ subtraction term for the
splitting singularity.  Using \eqref{power-FT} we get
\begin{align}
& F_{\text{spl,pt}}(x_i,\tvec{\Delta};\mu) =
  \lim\limits_{\epsilon\to 0}\,
  \biggl(\frac{\mu}{\Delta}\biggr)^{2\epsilon}\,
  \frac{\Gamma(\epsilon)\, \Gamma^2(1-\epsilon)}{\Gamma(1-2\epsilon)}\,
  (4\pi)^{\epsilon}\,
  \frac{\alpha_s}{2\pi}\, P(v, \epsilon) \frac{f(x)}{x}
  - C_{\epsilon}
\nonumber \\
 &\qquad =  \lim\limits_{\epsilon\to 0}\,
    \biggl[ \frac{1}{\epsilon} + \log(4\pi) - \gamma
             + \log\frac{\mu^2}{\Delta^2} \biggr]\,
    \frac{\alpha_s}{2\pi}\, P(v, 0)\, \frac{f(x)}{x}
      + \frac{\alpha_s}{2\pi}\, P'(v, 0)\, \frac{f(x)}{x}
  - C_{\epsilon} \,,
\end{align}
where $P'(v,\epsilon) = \partial P(v,\epsilon) /\partial\epsilon$.  The
$\overline{\text{MS}}$ counterterm is equal to
\begin{align}
C_\epsilon &= \biggl[ \frac{1}{\epsilon} + \log(4\pi) - \gamma \ms\biggr]\,
                \frac{\alpha_s}{2\pi}\, P(v, 0)\, \frac{f(x)}{x} \,,
\end{align}
so that we obtain
\begin{align}
  \label{dpd-msbar}
F_{\text{spl,pt}}(x_i,\tvec{\Delta};\mu) &=
  \biggl[ \log\frac{\mu^2}{\Delta^2}
          + \frac{P'(v, 0)}{P(v, 0)} \ms\biggr]\, R(x_i,\mu) \,.
\end{align}

We can use the preceding results to compute the matching between the
DPD defined with a $\Phi$ cutoff and the one obtained by
$\overline{\text{MS}}$ renormalisation.  Since the integral in
\eqref{F-phi-def} is finite, it can be smoothly continued to
$D=4-2\epsilon$ dimensions.  We then have
\begin{align}
  \label{matching-general}
F(x_i,\tvec{\Delta};\mu) - F_{\Phi}(x_i,\tvec{\Delta};\mu,\nu)
 = \lim\limits_{\epsilon\to 0}\,
   \int d^{2-2\epsilon}\tvec{y}\; e^{i \tvec{y}\tvec{\Delta}}\,
   F(x_i,\tvec{y};\mu)\, \bigl[ 1 - \Phi(y \nu) \bigr]
   - C_{\epsilon} \,.
\end{align}
The factor $[1 - \Phi(y \nu)]$ strongly suppresses the region $y \gg
1/\nu$, so that on the r.h.s.\ we can use the fixed-order perturbative
splitting contribution, evaluated in $4-2\epsilon$ dimensions.  This gives
\begin{align}
  \label{matching-master}
F(x_i,\tvec{\Delta};\mu) - F_{\Phi}(x_i,\tvec{\Delta};\mu,\nu)
 = F_{\text{spl,pt}}(x_i,\tvec{\Delta};\mu)
   - F_{\Phi, \text{spl,pt}}(x_i,\tvec{\Delta};\mu,\nu)
\end{align}
for all $\Delta$, and we have a smooth limit for $\Delta\to 0$.  For
$\Delta \ll \nu$ we can use \eqref{dpd-mom-lo} for $\Phi(u) =
\Theta(u-b_0)$ with \eqref{dpd-msbar} and obtain
\begin{align}
  \label{matching-small-delta}
F(x_i,\tvec{\Delta};\mu) - F_{\Phi}(x_i,\tvec{\Delta};\mu,\nu)
 &= \biggl[ \log\frac{\mu^2}{\nu^2}
            + \frac{P'(v, 0)}{P(v, 0)} \ms\biggr]\, R(x_i,\mu)
  + \mathcal{O}\biggl( \frac{\Delta^2}{\nu^2} \biggr)
  + \mathcal{O}(\alpha_s^2) \,.
\end{align}
At $\Delta=0$ we can in particular achieve equality of the two DPDs by
choosing the scale
\begin{align}
\nu = \mu \exp \biggl[ \frac{P'(v,0)}{2 P(v,0)} \biggr] \,.
\end{align}
With the splitting kernel $P_{g\to q\bar{q}}$ from \eqref{split-gqq} this
gives for instance a scale varying between $\nu = \mu\ms e^{-1/2} \approx
0.6\ms \mu$ and $\nu = \mu$.  These values respectively pertain to $v=1/2$
(i.e.\ $x_1 = x_2$) and to $v=0$ or $1$ (i.e.\ $x_1=0$ or $x_2=0$).


\section{Comparison with other work}
\label{sec:compare}

Three groups have previously studied the consequences of the perturbative
splitting mechanism on DPS cross sections, namely Blok et
al.~\cite{Blok:2011bu,Blok:2013bpa}, Ryskin and Snigirev
\cite{Ryskin:2011kk,Ryskin:2012qx}, and Manohar and Waalewijn
\cite{Manohar:2012pe}.  We refer to them as BDFS, RS and MW, respectively.
In this section we compare the results obtained in our formalism with
theirs.  We restrict ourselves to collinear factorisation.
To begin with, we rearrange our master formula \eqref{coll-master} as
\begin{align}
  \label{new-coll-master}
\sigma &= \sigma_{\text{SPS}}
  + (\ms \sigma_{\text{tw4}} - \sigma_{\text{2v1,pt}} \ms)
  + (\ms \sigma_{\text{DPS}} - \sigma_{\text{1v1,pt}} \ms)
\end{align}
and recall that one should take $\nu \sim \min(Q_1, Q_2)$ for the cutoff
parameter.  Let us briefly characterise the different terms in
\eqref{new-coll-master}.

In all approaches just mentioned, the SPS cross section
$\sigma_{\text{SPS}}$ is defined in the standard way.  It is obtained by
multiplying two PDFs with a hard scattering cross section and summing over
all parton channels.

The term $\sigma_{\text{tw4}} - \sigma_{\text{2v1,pt}}$ does not have any
equivalent in the work of BDFS and of RS.  As discussed in
section~\ref{sec:2v1}, this term is lacking the large $\log(Q/\Lambda)$
that is contained in $\sigma_{\text{DPS}}$ and built up by parton
splitting taking place over a wide region $1/Q \ll y \ll 1/\Lambda$ in the
2v1 graph~\ref{fig:boxed-graphs}b.  The approaches of BDFS and RS can thus
only be valid at leading logarithmic accuracy.  This limitation is
explicitly stated in the work of BDFS \cite{Blok:2011bu}.  The same
limitation holds of course in our approach if $\sigma_{\text{tw4}} -
\sigma_{\text{2v1,pt}}$ is not computed.

The term $\sigma_{\text{DPS}} - \sigma_{\text{1v1,pt}}$ includes the 2v2
and 2v1 parts of DPS, whereas the 1v1 part has its small $y$ approximation
subtracted.  As discussed at the end of section~\ref{sec:dglap-logs} and
illustrated in section~\ref{sec:lumi}, there are situations where strongly
ordered parton emissions in 1v1 graphs enhance the region $y \gg 1/Q$ for
the initial parton splitting.  In our approach, this effect
is included via the evolution of the DPDs in $\sigma_{\text{DPS}}$ (which
thus provides a small $x$ enhancement for $\sigma_{\text{2v2}}$,
$\sigma_{\text{2v1}}$ and $\sigma_{\text{1v1}}$).

If such an enhancement does not take place, then the 1v1 part of DPS has an
integrand going like $1/y^4$ at small $y$, so that the integral is dominated
by the region $y \sim 1/Q$.  In $\sigma_{\text{DPS}} -
\sigma_{\text{1v1,pt}}$ the contribution from this region then cancels up to
higher orders in $\alpha_s$, as shown in section~\ref{sec:dglap-subtr}.  In
turn, the contribution from $y \sim 1/\Lambda$ is suppressed by
$\Lambda^2/Q^2$ compared with $\sigma_{\text{SPS}}$ and thus of the same
order as other power corrections beyond the accuracy of the calculation.  In
this situation, the dominant terms in $\sigma_{\text{DPS}} -
\sigma_{\text{1v1,pt}}$ are the 2v2 and 2v1 parts of DPS, provided that they have a small $x$
enhancement compared with generic power corrections to SPS.


\subsection{The approach of Manohar and Waalewijn}

The master formula for the overall cross section in the MW approach is given
in eq.~(14) of \cite{Manohar:2012pe}.  Adapted to our notation, it reads
\begin{align}
  \label{master-mw}
\sigma &= \sigma_{\text{SPS}}
    + \hat{c}_3\, \bigl[\ms f\ms F(\tvec{y}=\tvec{0}) \ms\bigr]
    + \hat{c}_3\, \bigl[ F(\tvec{y}=\tvec{0})\ms f \ms\bigr]
    + \hat{c}_4\, \Bigl[\ms {\textstyle\int}
           d^2\tvec{y}\, F(\tvec{y}) F(\tvec{y}) \ms\Bigr] \,,
\end{align}
where $\hat{c}_3$ and $\hat{c}_4$ are hard scattering coefficients and
$[\ms \cdots \ms]$ indicates UV renormalisation of the enclosed quantity.
This means that the product $F(\tvec{y}) F(\tvec{y})$ of DPDs is
integrated over $\tvec{y}$ in $2-2\epsilon$ transverse dimensions, and the
resulting poles in $\epsilon$ are subtracted as usual in dimensional
regularisation (we assume $\overline{\text{MS}}$ renormalisation, although
this is not explicitly stated in \cite{Manohar:2012pe}).  The terms going
with $\hat{c}_3$ involve the renormalised product of a twist-two
distribution in one proton and a twist-four distribution in the other.  In
the following, the terms with $\hat{c}_4$ and $\hat{c}_3$ in
\eqref{master-mw} will be called DPS and twist four terms, respectively.
Via the perturbative splitting mechanism in one of the two protons, the
operator in the DPS term mixes under renormalisation with the operators in
the twist-four terms.  This is discussed at LO in \cite{Manohar:2012pe},
but we see no obstacle to extending the analysis to higher orders.

In section~\ref{sec:2v1} we have seen that the computation of the hard
scattering coefficient for graph~\ref{fig:2v1-labels} has a divergence
from the region where the $g\to q\bar{q}$ splittings become collinear.  In
our formalism the term $\sigma_{\text{2v1,pt}}$ in \eqref{new-coll-master}
subtracts the contribution from this region.  In the MW approach, the
general subtraction formalism recalled in section~\ref{sec:scheme} will
yield an $\overline{\text{MS}}$ subtraction term for the collinear
divergence, resulting from the renormalisation procedure described in the
previous paragraph.  For the choice $\Phi(u) = \Theta(u-b_0)$ we see in
\eqref{2v1-subtr} that $\sigma_{\text{2v1,pt}}$ is proportional to the
$\overline{\text{MS}}$ expression $1/\epsilon + \log(4\pi) - \gamma$ plus
an extra term $\log(\mu^2/\nu^2) + P'(0)/P(0)$, where $P(\epsilon)$ is the
relevant splitting function in $4-2\epsilon$ dimensions.  This extra term
corresponds to the difference between the momentum space DPD
$F_{\Phi,\text{spl,pt}}$ in \eqref{dpd-mom-lo} and its
$\overline{\text{MS}}$ counterpart in \eqref{dpd-msbar}.  Up to this
difference, the term $\sigma_{\text{tw4}} - \sigma_{\text{2v1,pt}}$ in our
master formula \eqref{new-coll-master} is thus equivalent to the twist
four term in \eqref{master-mw}.  The mismatch between the two versions may
be understood as a scheme difference.

It remains to compare the DPS term in the MW approach with our term
$\sigma_{\text{DPS}} - \sigma_{\text{1v1,pt}}$.  As the scale evolution of
$\bigl[\ms \int d^2\tvec{y}\, F(\tvec{y}) F(\tvec{y}) \ms\bigr]$ derived by MW
reflects perturbative splitting in one of the two protons, their DPS term
should contain the contributions from 2v2 and 2v1 graphs.  Our approach also
contains the contribution of 1v1 graphs, with a subtraction of their small $y$
approximation.  As discussed above, this part is negligible if DPS evolution
effects are moderate.  Both approaches should then give a valid representation
of the overall cross section, with the possibilities to go beyond the leading
logarithmic approximation and to systematically include higher orders in
$\alpha_s$.  On the other hand, strongly ordered emissions can significantly
enhance the the region $y \gg 1/Q$ in 1v1 graphs.  In our formalism, such
emissions are resummed to all orders by the evolution of the DPDs, but we see
no counterpart of this in the MW formulation.

MW argue that under renormalisation $\bigl[\ms \int d^2\tvec{y}\,
  F(\tvec{y}) F(\tvec{y}) \ms\bigr]$ does not mix with the product of two
PDFs by pointing out that the 1v1 graph in figure~\ref{fig:boxed-graphs}a
gives a quadratically divergent scaleless integral, which is zero in
dimensional regularisation.  This argument hinges on using the
perturbative approximation for the $g\to q\bar{q}$ splitting at all values
of~$\tvec{y}$ (or equivalently of the Fourier conjugate momentum
$\tvec{\Delta}$).  This includes the infrared region, where the
perturbative approximation does not work.  It would be interesting to
investigate this point further, but this is beyond the scope of the
present paper.


\subsection{The approaches of Blok et al.\ and of Ryskin and Snigirev}

\rev{Since BDFS and RS use the same $\sigma_{\text{SPS}}$ as we do, and
  since they have no counterpart to our $\sigma_{\text{tw4}} -
  \sigma_{\text{2v1,pt}}$ (which can be neglected at leading logarithmic
  accuracy), it remains to compare our term $\sigma_{\text{DPS}} -
  \sigma_{\text{1v1,pt}}$ with their expressions for the DPS cross
  section.  In this comparison, we can take the leading logarithmic
  approximation of our result.}

\rev{The approaches of BDFS and of RS are both} based on separating the
intrinsic and splitting contributions to a DPD at all values of
\rev{$\tvec{\Delta}$.}  As discussed in section~\ref{sec:scheme}, we avoid
such a separation when formulating the factorisation formula for the cross
section.  When modelling the DPDs -- which is of course unavoidable for
phenomenology -- it is however convenient to use this separation.  In
$\tvec{y}$ space we thus make the ansatz
\begin{align}
  \label{split-ansatz}
F(x_i,\tvec{y}, \mu_i) &= F_{\text{int}}(x_i,\tvec{y}, \mu_i)
                        + F_{\text{spl}}(x_i,\tvec{y}, \mu_i)
\end{align}
for all $\tvec{y}$.  We demand that $F_{\text{int}}$ and $F_{\text{spl}}$
separately follow the homogeneous evolution equation \eqref{DGLAP-mu1} and
that for $y \ll 1/\Lambda$ they tend to the correct short-distance limits
$F_{\text{int,pt}}$ and $F_{\text{spl,pt}}$.  \rev{At small $y$ the form
  \eqref{split-ansatz} is hence unambiguous, whereas for large $y$ it
  represents a model.}  For $F_{\text{int,pt}}$ we make the approximation
discussed in section~\ref{sec:simpler}.  Then $F_{\text{int}}$ has a
finite limit for $y=0$ rather than a weak singularity induced by
non-diagonal twist-four evolution.  \rev{Notice that $F_{\text{int}}$ and
  $F_{\text{spl}}$ now denote two terms in a DPD model, whereas in
  previous sections they were associated with different Feynman graphs in
  the discussion of factorisation.}

\rev{BDFS and RS only consider DPDs for unpolarised partons in the colour
  singlet sector, neglecting spin and colour correlations in both the
  intrinsic and the splitting parts.  We will therefore do the same when
  comparing to our formalism.  We note that spin correlations tend to be
  washed out by evolution to high scales \cite{Diehl:2014vaa} and that colour
  correlations are suppressed by Sudakov logarithms
  \cite{Mekhfi:1988kj,Manohar:2012jr}.  For sufficiently large hard scales,
  the importance of such correlations will therefore be reduced.}

In terms of the momentum space DPDs introduced in
section~\ref{sec:mom-space-dpds}, the DPS cross section at leading order in
$\alpha_s$ is given by
\begin{align}
  \label{Xsect-Delta}
\int \frac{d^2\tvec{\Delta}}{(2\pi)^2}\,
   F_\Phi(x_i, \tvec{\Delta}; \mu_i, \nu)\,
   F_\Phi(\bar{x}_i, -\tvec{\Delta}; \mu_i, \nu)
\end{align}
times the appropriate subprocess cross sections and a combinatorial
factor.  As was done in the papers cited above, we allow for different
factorisation scales $\mu_1, \mu_2$.  We note that with
the behaviour \eqref{dpd-mom-hi} of the splitting DPD at large $\Delta$,
the integral \eqref{Xsect-Delta} converges in the ultraviolet for both
choices of $\Phi$.  For definiteness, we take $\Phi(u) = \Theta(u-b_0)$ in
the following.  \rev{We now analyse the 2v2, 2v1 and 1v1 contributions in
  turn.}

The ultraviolet cutoff can be neglected in the intrinsic contribution:
\begin{align}
  \label{mom-dpd-int}
F_{\Phi, \text{int}}(x_i, \tvec{\Delta}; \mu_i, \nu) &=
    \int d^{2}\tvec{y}\; e^{i\tvec{y}\tvec{\Delta}}\,
    \Phi(y \nu) F_{\text{int}}(x_i,\tvec{y},\mu_i)  
 \approx \int d^{2}\tvec{y}\; e^{i\tvec{y}\tvec{\Delta}}\,
    F_{\text{int}}(x_i,\tvec{y},\mu_i) \,.
\end{align}
This is because $F_{\text{int}}(x_i,\tvec{y})$ is of generic size
$1/\Lambda^2$ for $y \sim 1/\Lambda$ and has a smooth behaviour for $y \ll
1/\Lambda$.  A cutoff $y > b_0/\nu$ in the integral therefore only removes
a contribution of order $\Lambda^2/\nu^2$.  If we make the same ansatz for
the intrinsic DPD as \rev{BDFS or RS, we will thus obtain the same
result for $\sigma_{\text{2v2}}$ as they do, up to power corrections that are
beyond the accuracy of all all approaches discussed here.}

For a typical ansatz, $F_{\Phi,\text{int}}(x_i, \tvec{\Delta})$ strongly
decreases when $\Delta\gg \Lambda$.  The 2v1 contribution to the integral
\eqref{Xsect-Delta} is therefore dominated by the region $\Delta \sim
\Lambda$.  In this region, the integral
\begin{align}
  \label{mom-dpd-spl-def}
F_{\Phi, \text{spl}}(x_i, \tvec{\Delta}; \mu_i, \nu) &=
   \int d^{2}\tvec{y}\; e^{i\tvec{y}\tvec{\Delta}}\,
    \Phi(y \nu)\, F_{\text{spl}}(x_i, \tvec{y}; \mu_i)
\end{align}
receives leading contributions from both small and large $y$.  However, a
large logarithm $\log(\nu/\Delta)$ is only built up in the region $1/\nu
\ll y \ll 1/\Delta$, where we can use the perturbative form
$F_{\text{spl,pt}}(x_i, \tvec{y}) \sim 1/y^2$.  To leading logarithmic
accuracy, we can therefore compute the splitting contribution without a
nonperturbative model, even for $\Delta \sim 1/\Lambda$, which we now do.
We want to improve on our result \eqref{dpd-mom-lo} by including evolution
effects.  To do so, we start from the first order perturbative expression
\eqref{split-lo} at scale $\mu_y = b_0/y$ and then evolve using the Green
function $D(x,\mu;\mu_y)$, which solves the single DGLAP equation in the
scale $\mu$ with the initial condition $D(x,\mu_y;\mu_y) = \delta(1-x)$.
For simplicity we omit labels for and sums over parton species.  We then
have
\begin{align}
  \label{mom-dpd-spl}
F_{\Phi, \text{spl}}(x_i, \tvec{\Delta}; \mu_i, \nu)
 &\approx
    \int_{b_0^2/\nu^2}^{\infty} \frac{dy^2}{y^2}\, J_0(y \Delta)
       E(x_i; \mu_i, \mu_y)
\end{align}
with
\begin{align}
   \label{mom-dpd-spl-start}
E(x_i; \mu_i, \mu_y) &=
   \int_{x_1}^{1-x_2} \frac{dz_1}{z_1}
   \int_{x_2}^{1-z_1} \frac{dz_2}{z_2}\;
     D\biggl(\frac{x_1}{z_1}, \mu_1; \mu_y \biggr)\,
     D\biggl(\frac{x_2}{z_2}, \mu_2; \mu_y \biggr)
\nonumber \\[0.1em]
 & \quad \times
   \frac{\alpha_s(\mu_y)}{2\pi}\, P\biggl( \frac{z_1}{z_1+z_2}, 0 \biggr)\,
     \frac{f(z_1+z_2, \mu_y)}{z_1+z_2} \,.
\end{align}
The function $E(x_i; \mu_i, \mu_y)$ depends weakly on $y$ due to
logarithmic corrections from evolution and the running of $\alpha_s$.  To
make this explicit, we expand the quantities depending on $\mu_y$ (the
Green functions, $\alpha_s$ and the parton density) around $\nu$.  Keeping
only leading logarithms, we obtain
\begin{align}
  \label{I-series}
E(x_i; \mu_i, \mu_y) &= \sum_{n=0}^{\infty} c^{(n)}(x_i,\mu_i,\nu)\,
    \alpha_s^{n+1}(\nu)\, \log^{n}\biggl( \frac{y \nu}{b_0} \biggr) \,.
\end{align}
We show in appendix \ref{app:fourier} that with such a $y$ behaviour, the
Bessel function $J_0(y \Delta)$ in \eqref{mom-dpd-spl} can be replaced by
an upper cutoff $\Theta(b_0 - y \Delta)$, up to relative corrections in
$\Delta^2/\nu^2$ and up to subleading logarithms $\alpha_s^{n+1}(\nu)
\log^{n-2}(\nu/\Delta)$.  We then have
\begin{align}
  \label{dpd-mom-spl-app}
F_{\Phi, \text{spl}}(x_i, \tvec{\Delta}; \mu_i, \nu) &\approx
    \int_{b_0^2/\nu^2}^{b_0^2/\Delta^2} \frac{dy^2}{y^2}\,
       E(x_i; \mu_i, \mu_y)
  = \int_{\Delta^2}^{\nu^2} \frac{d\mu^2}{\mu^2}\, E(x_i; \mu_i, \mu)
\end{align}
for $\Delta \ll \nu$.
With the scale choice \eqref{scale-choice} our result
\eqref{dpd-mom-spl-app} corresponds to the expressions of the perturbative
splitting DPD in equation (18) of \cite{Blok:2011bu} and in equation (8)
of \cite{Ryskin:2012qx}.  Note that at leading logarithmic accuracy, the
lower integration limit on $k$ in \cite{Blok:2011bu} is equivalent to
$\Delta$ in the region $\Delta \sim \Lambda$ that dominates the 2v1 cross
section.

To leading logarithmic accuracy, the results for $\sigma_{\text{2v1}}$
obtained by \rev{both BDFS \cite{Blok:2011bu,Blok:2013bpa} and RS
  \cite{Ryskin:2011kk,Ryskin:2012qx}} are therefore consistent with our
approach (provided of course we make a suitable ansatz for $F_{\text{int}}$).
In equation (6) of \cite{Ryskin:2012qx} an upper cutoff $\min(Q_1, Q_2)$ is
put on the $\Delta$ integration, but this is of no concern since as discussed
above, the 2v1 term is dominated by $\Delta \sim \Lambda$.

The situation is different for $\sigma_{\text{1v1}}$, \rev{which RS compute
  using the form \eqref{dpd-mom-spl-app} we just derived for $\Delta \ll \nu$.
  The integral over $\Delta$ in the cross section then requires an upper
  cutoff, which RS choose as $\min(Q_1, Q_2)$, as we do for our cutoff $\nu$
  in $y$ space.  The dominant integration region for the 1v1 cross section is
  $\Delta \sim \nu$ in their approach and in ours, given that} up to
logarithms the splitting DPD is of order unity for both $\Delta \sim \Lambda$
and $\Delta \sim \nu$.  For $\Delta \sim \nu$ our distribution
$F_{\Phi,\text{spl}}$ differs from \eqref{dpd-mom-spl-app} even if we neglect
evolution effects, which readily follows from the result \eqref{bessel-step}
for the $y$ integral in \eqref{mom-dpd-spl}.  Furthermore,
$F_{\Phi,\text{spl}}$ strongly depends on the choice of $\Phi$ in that region.
This just highlights that $\sigma_{\text{1v1}}$ has no physical meaning by
itself.  In our formalism, the contribution from $\Delta \sim \nu$ in
$\sigma_{\text{1v1}}$ is cancelled by the subtraction term
$\sigma_{\text{1v1,pt}}$ , as it should be since the DPS approximation is not
valid in that region at all.

\rev{The DPS cross section in the approach of BDFS is given entirely by
  2v2 and 2v1 terms.  To leading logarithmic accuracy, these terms agree
  with the ones in our approach.  The term $\sigma_{\text{DPS}} -
  \sigma_{\text{1v1,pt}}$ in our approach has an additional contribution
  $\sigma_{\text{1v1}} - \sigma_{\text{1v1,pt}}$, which as already
  mentioned may be neglected \emph{unless} DPD evolution enhances the
  region $y \gg 1/Q$ (or equivalently $\Delta \ll Q$)
  in~$\sigma_{\text{1v1}}$.  This possible enhancement is not captured by
  the BDFS approach.}

\rev{In the RS approach, the DPS cross section includes 2v2, 2v1 and 1v1
  terms, the first two of which agree with ours in the leading logarithmic
  approximation.  If DPD evolution effects are moderate, then the 1v1 term
  of RS is dominated by large $\Delta \sim Q$, strongly depends on the
  upper cutoff imposed on the $\Delta$ integral, and leads to a double
  counting problem with $\sigma_{\text{SPS}}$ that is not solved in the RS
  approach.  In the sense we have discussed in
  section~\ref{sec:dglap-subtr}, one may at best take the cutoff
  dependence of the 1v1 cross section as an estimate for the size of
  $\sigma_{\text{SPS}}$ and thus as an indicator for how serious the
  double counting problem is.  Only if DPD evolution enhances the region
  $\Delta \ll Q$ so strongly that it \emph{dominates} the 1v1 term does
  the RS approach give a valid representation of the overall cross
  section.  The cutoff dependence of the 1v1 cross section is then
  reduced, as was emphasised in \cite{Ryskin:2011kk}.}

\section{Quantitative illustrations}
\label{sec:numerics}

We have set up a general formalism for calculating and combining the
contributions from SPS, DPS and other mechanisms to the physical cross
section.  Whilst a detailed phenomenological application is beyond the
scope of this work, we wish to give some quantitative illustrations of the
formalism.  This will allow us to estimate the relative size of
contributions in specific channels and kinematics, and to explore the
interplay between SPS or DPS terms and the double counting subtractions.
We limit ourselves to collinear factorisation, where the construction of a
model for DPDs is much simpler than in the TMD case.

Convenient quantities for our purpose are double parton luminosities
\begin{align}
  \label{lumi-def}
  \mathcal{L}_{a_1 a_2 b_1 b_2}(x_i, \bar{x}_i, \mu_i, \nu) &=
  \int d^2\tvec{y}\; \Phi^2(y \nu)\,
  F_{a_1 a_2}(x_i,\tvec{y};\mu_i) \,
  F_{b_1 b_2}(\bar{x}_i,\tvec{y};\mu_i)\,,
\end{align}
which directly appear in the DPS cross section.  For definiteness, we use
a sharp cutoff $\Phi(u) = \Theta(u - b_0)$ to remove the UV region.  As a
model for colour singlet DPDs we take the sum of an intrinsic and a
splitting part, as specified in \eqref{split-ansatz} and in the paragraph
following that equation.  The luminosity \eqref{lumi-def} then naturally
splits into separate terms for 2v2, 2v1 and 1v1.  Colour non-singlet
channels will not be considered, except in subsection~\ref{sec:ggtoscal},
where they can be computed without recourse to a model.


\subsection{Simplified analytic estimates}
\label{sec:simple-est}

We shall see that DPD evolution has important quantitative effects on
double parton luminosities.  Nevertheless, let us start by deriving
simple analytic expressions where evolution is neglected, but where
the power behaviour, logarithms and other important factors can be
easily identified.

For the intrinsic contribution we make the well-known product ansatz
\begin{align}
  \label{simple-int}
  F_{\text{int}}(x_i, \tvec{y})
&= \frac{\Lambda^2}{\pi}\, e^{- y^2 \Lambda^2}\,
  f(x_1)\, f(x_2)
\end{align}
with a Gaussian form for the $y$ dependence.  For the splitting
contribution we take
\begin{align}
  \label{simple-split}
  F_{\text{spl}}(x_i, \tvec{y})
&= \frac{1}{\pi y^2}\, e^{- y^2 \Lambda^2}\;
  \frac{f(x_1+x_2)}{x_1+x_2}\,
  \frac{\alpha_s}{2\pi}\, P\biggl( \frac{x_1}{x_1+x_2} \biggr) \,,
\end{align}
which is the perturbative expression at $O(\alpha_s)$ times an exponential
to suppress the large $y$ region.  Sums and labels for the parton type are
omitted here.  For simplicity we use the same exponential in
$F_{\text{int}}$ and $F_{\text{spl}}$.  \rev{Since we ignore evolution for
  the time being, no scale dependence is taken into account in
  \eqref{simple-int} and \eqref{simple-split}.}  To simplify even further,
let us take equal momentum fractions $x_i = \bar{x}_i = x$ in both
protons.

The parton luminosity for the 2v2 contribution then is
\begin{align}
  \label{simple-2v2}
  \mathcal{L}_{\text{2v2}}
  &= \frac{\Lambda^4}{\pi}\, \bigl[ f(x) \bigr]^4
  \int_{b_0^2 /\nu^2}^\infty d y^2\, e^{- 2y^2 \Lambda^2}
  = \frac{\Lambda^2}{2\pi}\, e^{- 2 b_0^2 \Lambda^2/\nu^2}\, 
  \bigl[ f(x) \bigr]^4
  \nonumber \\
  &= \frac{\Lambda^2}{2\pi}\, \bigl[ f(x) \bigr]^4\,
  + \mathcal{O}(\Lambda^4/\nu^2) \,.
\end{align}
As mentioned earlier, the UV regulator $\Phi(y \nu)$ only gives rise to a
power suppressed effect.  The 2v1 luminosity reads
\begin{align}
  \label{simple-2v1}
  \mathcal{L}_{\text{2v1}}
  &= \frac{2 \Lambda^2}{\pi}\, \bigl[ f(x) \bigr]^2\, \frac{f(2x)}{2x}\,
  \frac{\alpha_s}{2\pi}\, P(1/2)\,
  \int_{b_0^2 /\nu^2}^\infty\!\!
  \frac{d y^2}{y^2}\, e^{- 2y^2 \Lambda^2}
  \nonumber \\
  &= \frac{2 \Lambda^2}{\pi}
  \log\biggl( \frac{\nu^2}{4b_0 \Lambda^2} \biggr)\,
  \frac{\alpha_s}{2\pi}\, \bigl[ f(x) \bigr]^2\; \frac{f(2x)}{2x}\,
  P(1/2) + \mathcal{O}(\Lambda^4/\nu^2) \,.
\end{align}
This is enhanced over $\mathcal{L}_{\text{2v2}}$ by
$\log(\nu^2/\Lambda^2)$ but suppressed by $\alpha_s /(2\pi)$.
Furthermore, there are different factors regarding the parton densities.
We note that the size of the splitting function strongly depends on the
parton channel, as is exemplified by the cases $P_{g\to q\bar{q}}(1/2) =
1/4$, $P_{q\to qg}(1/2) = 10/3$ and $P_{g\to gg}(1/2) = 27/2$.
For the 1v1 luminosity we finally obtain
\begin{align}
  \label{simple-1v1}
  \mathcal{L}_{\text{1v1}}
  &= \frac{1}{\pi}\, \biggl[ \frac{f(2x)}{2x}\,
    \frac{\alpha_s}{2\pi}\, P(1/2) \biggr]^2\,
  \int_{b_0^2 /\nu^2}^\infty \frac{d y^2}{y^4}\, e^{- 2y^2 \Lambda^2}
  \nonumber \\
  &= \frac{\nu^2}{\pi b_0^2}\, \biggl( \frac{\alpha_s}{2\pi} \biggr)^2\,
  \biggl[ \frac{f(2x)}{2x}\, P(1/2) \biggr]^2
  + \mathcal{O}\bigl( \Lambda^2 \log(\nu^2/\Lambda^2) \bigr) \,.
\end{align}
We see the power enhancement over the 2v2 and 2v1 terms, as well as the
quadratic dependence on the cutoff $\nu$.

It is easy to repeat the above exercise for our alternative regulator
function $\Phi(u) = 1 - e^{-u^2/4}$.  The integrals can be performed
writing $\Phi(u) = u^2 \int_0^{1/4} d\tau\, e^{-\tau u^2}$.  For the
leading term of the expansion in $\Lambda^2/\nu^2$, one obtains the same
result for $\mathcal{L}_{\text{2v2}}$ as in \eqref{simple-2v2}, whereas in
\eqref{simple-2v1} and \eqref{simple-1v1} one should replace
\begin{align}
  \log\biggl( \frac{\nu^2}{4b_0 \Lambda^2} \biggr) &\to 
  \log\biggl( \frac{\nu^2}{16 \Lambda^2} \biggr) \,,
  &
  \frac{\nu^2}{b_0^2} &\to \frac{\nu^2 \log 2}{2} \,,
\end{align}
respectively.  The leading logarithm $\log(\nu^2/\Lambda^2)$ in
$\mathcal{L}_{\text{2v1}}$ is independent of the choice for $\Phi(u)$, but not
the constant term.  The corresponding ambiguity is compensated by the DPS
subtraction term, as is the change of the numerical prefactor in
$\mathcal{L}_{\text{1v1}}$.


\subsection{Collinear Parton Luminosities}
\label{sec:lumi}

We now present a numerical study of the parton luminosities, using a
slightly refined model for the input DPDs, and including evolution
effects.  We start by considering the unpolarised luminosities only,
turning to the issue of polarised DPDs and luminosities at the very
end of the section.

\subsubsection{Setup}
\label{sec:set-up}

We use scale evolution at leading order in $\alpha_s$ and take equal scales
$\mu_1 = \mu_2 = \mu$ for the two hard processes throughout.  In this
illustrative study, we fix the number of active quark flavours to $n_f=3$.
This is sufficient for exploring the broad qualitative features of the DPDs
and luminosities.  Implementation of heavy quark flavours, as required for a
realistic phenomenological investigation, would require some dedicated work on
flavour thresholds and on memory management in our numerical code, which we
postpone to future studies.

As in section \ref{sec:simple-est}, we write each DPD as a sum of intrinsic
and splitting pieces:
\begin{align}
  \label{eq:Fmodelunpol}
F_{a_1 a_2}(x_1, x_2, \tvec{y}; \mu,\mu)
  &=  F_{a_1 a_2,\ms \text{int}}(x_1, x_2, \tvec{y}; \mu,\mu)
    + F_{a_1 a_2,\ms \text{spl}}(x_1, x_2, \tvec{y}; \mu,\mu) \,.
\end{align}
We initialise $F_{\text{int}}$ at a low scale $\mu_0 = 1 \gev$ using a
product ansatz similar to \eqref{simple-int}:
\begin{align}
 \label{full_intrinsic_input}
F_{a_1 a_2,\ms \text{int}}(x_1,x_2, \tvec{y}; \mu_0,\mu_0)
 = \frac{1}{4\pi h_{a_1 a_2}}\, &
   \exp\biggl[ - \frac{y^2}{4h_{a_1 a_2}} \biggr]\,
  f_{a_1}(x_1;\mu_0)\, f_{a_2}(x_2;\mu_0) 
\nonumber \\
  & \times (1-x_1-x_2)^2\, (1-x_1)^{-2}\, (1-x_2)^{-2} \,.
\end{align}
For the single PDFs $f_{a}(x,\mu_0)$, we take the MSTW2008LO
distributions \cite{Martin:2009iq}.  In \eqref{full_intrinsic_input} we
have multiplied the
PDFs by a function of the $x_i$ that does not affect the DPD at small
$x_i$, but smoothly cuts it off near the kinematic boundary $x_1+x_2=1$.
The function we use is that given in equation~(3.12) of
\cite{Gaunt:2009re}, with $n=2$.  We also take a different $y$ dependence
for different parton species.  For this we use a simplified version of the
model in section 4.1 of \cite{Diehl:2014vaa}, taking the width $h$ to be
$x$ independent (corresponding to $h(x_1,x_2)$ of \cite{Diehl:2014vaa}
evaluated at $x_1=x_2=10^{-3}$) and setting each $h$ with $q^-$ indices to
be the same as the one with $q^+$.  Thus we have
\begin{align}
  h_{a_1 a_2} = h_{a_1} + h_{a_2}
\end{align}
with
\begin{align}
  h_{g} = 2.33 \gev^{-2} \,,  & &
  h_{q} = h_{\bar{q}} = 3.53 \gev^{-2} \,.
\end{align}
For the splitting piece of the DPD we generalise our ansatz in
\eqref{simple-split}, choosing an initialisation scale that goes to $b_0/y$ at
small $y$ but freezes to a constant value $b_0/y_{\text{max}}$ when $y$
exceeds a value $y_{\text{max}}$ that marks the transition between
perturbative and nonperturbative behaviour.  This ensures that the single PDF
and $\alpha_s$ in the splitting expression are never evaluated at too low
scales.  A suitable choice of scale is
\begin{align}
\mu_y &= \frac{b_0}{y^*} \,, &
y^{*} &= \frac{y}{\sqrt{ 1 + y^2 /y_{\text{max}}^2 }} \,.
\end{align}
Such a prescription is very similar to the $b^*$ prescription used in TMD
phenomenology \cite{Collins:1981va,Collins:1984kg}.  Here we take
$y_{\text{max}} = 0.5 \gev^{-1}$, which is one of the values considered in
the recent TMD study \cite{Collins:2014jpa}.  Using the same parton
dependent Gaussian damping as in \eqref{full_splitting_input}, we have
\begin{align} 
  \label{full_splitting_input}
& F_{a_1 a_2,\ms \text{spl}}(x_1,x_2, \tvec{y}; \mu_y,\mu_y)
\nonumber \\
  &\qquad =
  \frac{1}{\pi y^2}\, \exp\biggl[ - \frac{y^2}{4h_{a_1 a_2}} \biggr] \;
  \frac{f_{a_0}(x_1+x_2;\mu_y)}{x_1+x_2}\,
  \frac{\alpha_s(\mu_y)}{2\pi}\, 
  P_{a_0\to a_1 a_2}\biggl( \frac{x_1}{x_1+x_2} \biggr) \,.
\end{align}
The coupling $\alpha_s(\mu_y)$ is determined by $3$ flavour running,
starting with the MSTW2008LO value $\alpha_s(\mu_0) = 0.68183$.  The
single PDFs are obtained by taking the MSTW2008LO distributions at $\mu_0$
and evolving them according to the DGLAP equations for $n_f=3$.

To obtain the splitting and intrinsic DPDs at scale $\mu$, as in
\eqref{eq:Fmodelunpol}, the input forms just discussed must be evolved,
starting from $\mu_0$ for $F_{\text{int}}$ and from $\mu_y$ for
$F_{\text{spl}}$, according to the homogeneous double DGLAP equations.
For this we use a modified version of the code developed in
\cite{Gaunt:2009re}.  The modified code works on a grid in the $x_i$,
$\mu$ and $\mu_y$ directions (the grid of the original code is in $x_i$
and $\mu$ only).  The grid points in the $x_i$ directions are evenly
spaced in $\log(x_i/(1-x_i))$, whilst those in the $\mu$ and $\mu_y$
directions are evenly spaced in $\log\mu$ or $\log\mu_y$.  The integrals
appearing in the double DGLAP equations are computed from points in the
$x_i$ grids using Newton-Cotes rules (for details see Appendix A of
\cite{Gaunt:2009re}), and evolution from one point in the $\mu$ grid to
the next is carried out using the Runge-Kutta method.

The grid is set up with $88$ points in each $x_i$ direction, spanning
$5\times 10^{-5} < x_i < 1$, with $60$ points in the $\mu$ direction,
spanning $\mu_0 < \mu < 170\gev$, and with $60$ points in the $\mu_y$
direction, spanning $b_0/y_{\text{max}} < \mu_y < 340\gev$.  According to
the studies made in \cite{Gaunt:2009re}, this suggests an error on the
level of a few per cent in the DPD values obtained after evolution, which
is tolerable in this first study.  The DPDs computed on this grid are used
together with an interpolation code to produce numerical values for the
investigations below.


\subsubsection{Numerical results}

\begin{figure}
  \begin{center}
    \subfigure[]{\includegraphics[trim = 1.5cm 0 0 0,
        height=20em]{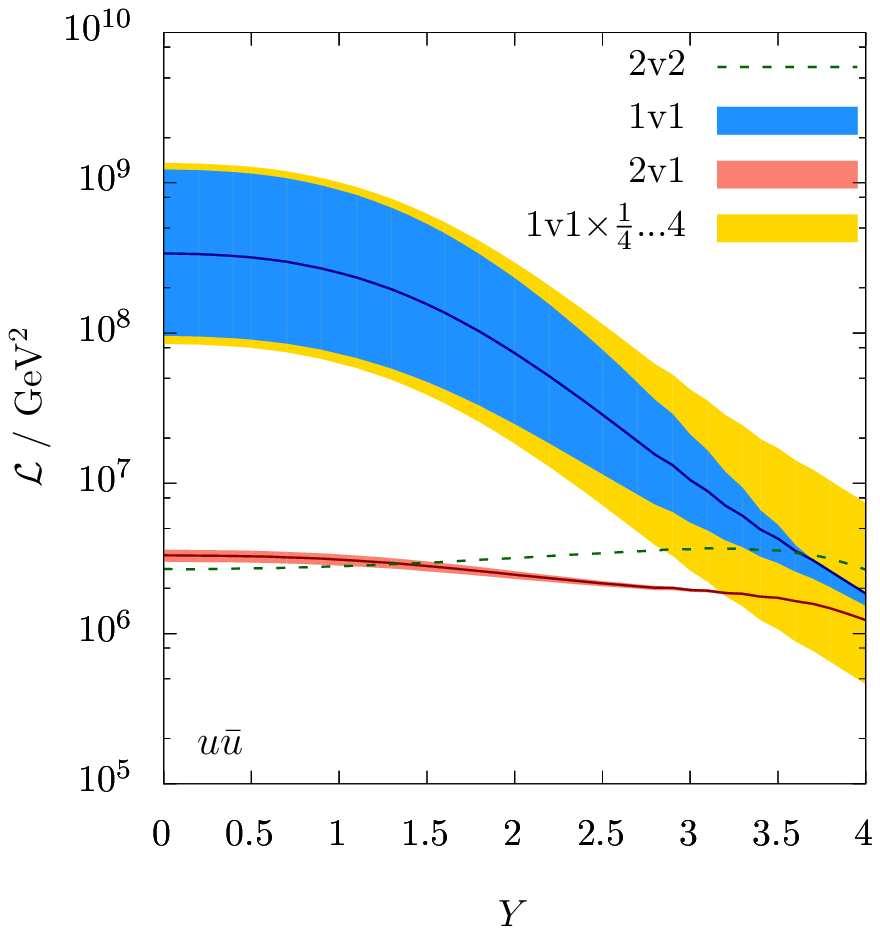}}
    \hspace{1em}
    \subfigure[]{\includegraphics[trim = 1.5cm 0 0 0,
        height=20em]{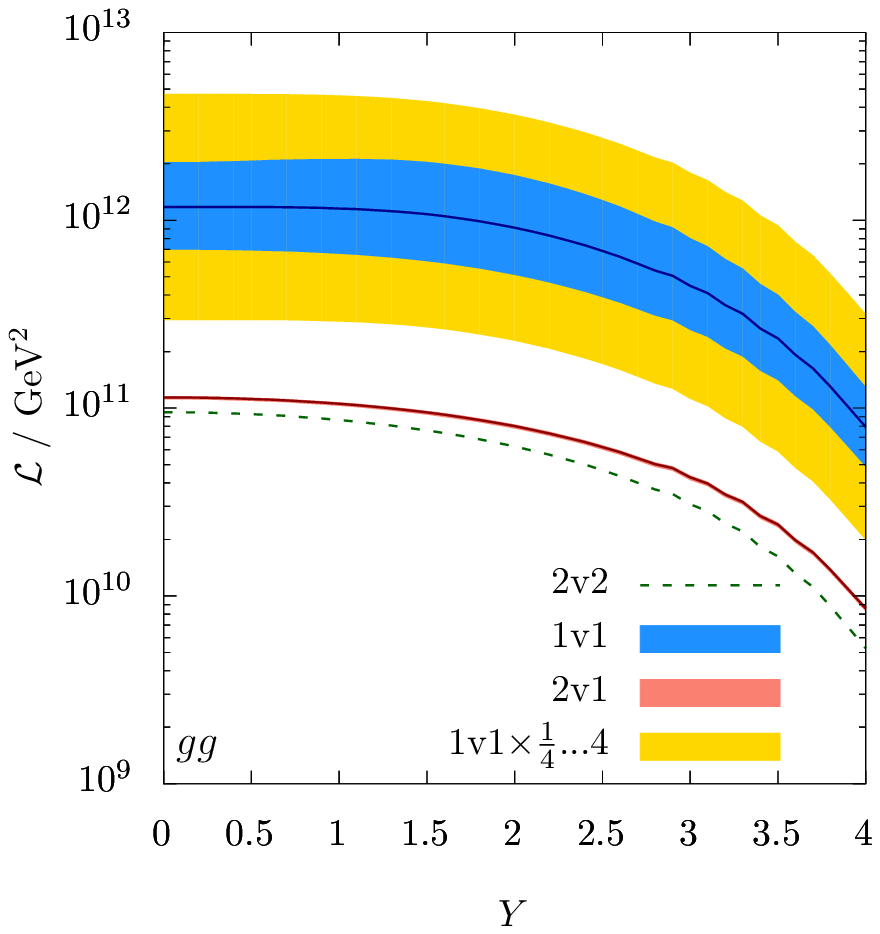}} \\[1em]
    \subfigure[]{\includegraphics[trim = 1.2cm 0 0 0,
        height=20em]{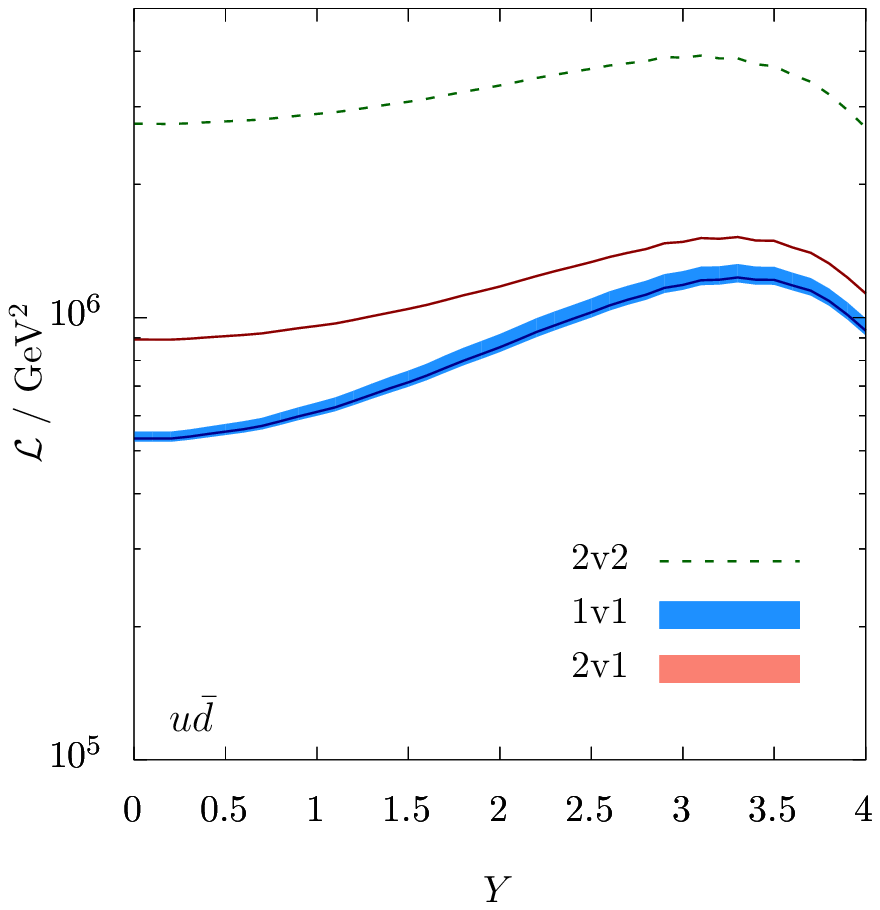}}
      \caption{\label{fig:DPSlumi} Double parton scattering luminosities
        $\mathcal{L}_{a_1 a_2 b_1 b_2}(Y)$ for the production of two systems
        with $Q_1 = Q_2 = 80\gev$ at $\sqrt{s} = 14\tev$, one with central
        rapidity and the other with rapidity $Y$.  The parton combinations
        $a_1 a_2\ms b_1 b_2$ are $u\bar{u}\bar{u}u + \bar{u}uu\bar{u}$ (a),
        $gggg$ (b) and $u\bar{d}\bar{d}u + \bar{d}uu\bar{d}$ (c).}
  \end{center}
\end{figure}

We begin with a study where the scale $\mu$ is equal to
$Q_1 = Q_2 = 80\gev$ (as in the production of a $W$ boson pair).
Taking the collider energy to be $\sqrt{s} = 14 \tev$, we set $x_1$
and $\bar{x}_1$ in \eqref{lumi-def} to correspond to central
production of the first system and $x_2$ and $\bar{x}_2$ to correspond
to the production of the second system with rapidity $Y$ (all
rapidities refer to the $pp$ centre of mass).  This gives
\begin{align}
  x_1 &= \bar{x}_1 = 5.7\times 10^{-3} \,, &
  x_2 &= 5.7\times 10^{-3}\ms \exp(Y) \,, &
  \bar{x}_2 &= 5.7\times 10^{-3}\ms \exp(-Y) \,.
\end{align}
In figure \ref{fig:DPSlumi} we plot $\mathcal{L}_{a_1 a_2 b_1 b_2}(Y)$ in the
range $0 \le Y \le 4$ for the parton combinations
$a_1 a_2\ms b_1 b_2 = u\bar{u}\bar{u}u + \bar{u}uu\bar{u}$,
$a_1 a_2\ms b_1 b_2 = gggg$ and
$a_1 a_2\ms b_1 b_2 = u\bar{d}\bar{d}u + \bar{d}uu\bar{d}$.  The first parton
combination appears e.g.\ in $ZZ$ production, the second is important in
four-jet production, and the last appears in $W^+W^+$.  For ease of language,
we will refer to these parton combinations as the $u\bar{u}$, $gg$ and
$u\bar{d}$ channels, respectively.  We split the overall luminosity into
contributions from 1v1 ($F_{\text{spl}} \times F_{\text{spl}}$), from 2v1
($F_{\text{spl}} \times F_{\text{int}} + F_{\text{int}} \times
F_{\text{spl}}$) and from 2v2 ($F_{\text{int}} \times F_{\text{int}}$).  We
vary $\nu$ by a factor of $2$ up and down around a central value of $80\gev$,
in order to see how DPS alone is affected by variation of this cutoff.  For
each contribution, the line in the plots denotes the luminosity with
$\nu = 80\gev$, whilst the band is generated from the envelope of the
functions with $\nu = 40\gev$, $\nu = 80\gev$ and $\nu = 160\gev$.  In the
case of the 2v2 contribution, the $\nu$ scale variation is negligible (as
expected from basic considerations, see \eqref{simple-2v2}), so this appears
as a dashed line in each plot.

For the 1v1 contribution of the $gg$ and $u\bar{u}$ channels, we also plot
a band generated by varying the prediction with $\nu = 80\gev$ by a factor
$4$ up and down.  This corresponds to a strictly quadratic cutoff
dependence of $\mathcal{L}$ , i.e.\ to the variation of $\nu$ by a factor
$2$ in the naive formula \eqref{simple-1v1}, where DPD evolution is
neglected.  Any discrepancy between this band and the actual 1v1 band is
therefore due to evolution effects.  We do not plot this band for the
$u\bar{d}$ channel: there is no LO splitting process giving $u\bar{d}$, so
that the scale variation (and central value) from \eqref{simple-1v1}
is zero in this case.

We immediately notice in figures \ref{fig:DPSlumi}a and b that the 1v1
contribution is generally much larger than the 2v2 and 2v1 contributions, and
that it has an enormous $\nu$ variation.  To obtain a sensible prediction in
these cases, one must include the SPS corrections up to the order that
includes the double box graph in figure~\ref{fig:boxed-graphs}a, so that the
associated subtraction term can approximately cancel the $\nu$ dependence of
DPS.  We also notice that at central rapidities in the $u\bar{u}$ channel, the
1v1 $\nu$ variation band essentially fills the band corresponding to a
quadratic $\nu$ dependence, indicating small evolution effects in this case.
By contrast, the 1v1 $\nu$ variation band for $gg$ is clearly smaller than the
naive expectation, which indicates significant evolution effects.  We shall
explore the reasons for these differences below.

In figure \ref{fig:DPSlumi}c, the 1v1 contribution is small compared to the
2v2 piece and has a small $\nu$ dependence.  This is because, as already
mentioned, there is no LO splitting directly giving $u\bar{d}$ (generation of
a $u\bar{d}$ pair requires at least two steps, such as
$u \to u + g \to u + d + \bar{d} \,$).  In this case, there is less of a need
to compute $\sigma_{\text{SPS}}$ and the subtraction term
$\sigma_{\text{1v1,pt}}$ up to the order that contains the lowest-order
DPS-type loop (in both amplitude and conjugate).  This is fortunate, since
this order is two powers of $\alpha_s$ higher than for
graph~\ref{fig:boxed-graphs}a (two-step rather than one-step splittings are
required in both protons), and the corresponding SPS calculation will not be
available for some time.

In the $gg$ channel, the 2v1 and 2v2 contributions are of similar magnitude
and shape.  This is consistent with what has been found in previous
phenomenological studies \cite{Blok:2013bpa, Gaunt:2014rua} at similar scales
and $x$ values.  In the $u\bar{u}$ channel, the magnitude of 2v1 and 2v2 is
broadly the same but their shape is somewhat different.  In the $u\bar{d}$
channel, the shape of the two contributions is similar, but 2v1 lies well
below 2v2 (and is close to 1v1).

We can examine the degree to which the $\nu$ variation in the $gg$ and
$u\bar{u}$ channels is reduced by adding the remaining fixed-order terms in
the cross section, even without considering a specific final state or having
to compute $\sigma_{\text{SPS}}$.  This is because only $\sigma_{\text{DPS}}$
and the subtraction terms depend on $\nu$.  Just as we did for
$\sigma_{\text{DPS}}$, we can factor the two parton-level cross sections out
of $\sigma_{\text{1v1,pt}}$ and introduce a ``subtraction luminosity''
$\mathcal{L}_{\text{1v1,pt}}$.  This is defined as in \eqref{lumi-def} except
that the DPDs are replaced by the fixed-order splitting expression evaluated
at scale $\mu$,
\begin{align} 
  \label{full_splitting_input_sub}
F_{a_1 a_2,\ms \text{spl,fo}}(x_1,x_2, \tvec{y}; \mu)
 &= \frac{1}{\pi y^2}\, 
  \frac{f_{a_0}(x_1+x_2; \mu)}{x_1+x_2}\,
  \frac{\alpha_s(\mu)}{2\pi}\, 
  P_{a_0\to a_1 a_2}\biggl( \frac{x_1}{x_1+x_2} \biggr) \,.
\end{align}
For a given parton combination $a_1 a_2\ms b_1 b_2$, one can directly subtract
$\mathcal{L}_{\text{1v1,pt}}$ from the DPS luminosity
$\mathcal{L}_{\text{DPS}}$ and compare the $\nu$ dependence of the result to
that of $\mathcal{L}_{\text{DPS}}$ alone.  Up to multiplication by the
appropriate parton-level cross sections, the $\nu$ dependence of
$\mathcal{L}_{\text{DPS}} - \mathcal{L}_{\text{1v1,pt}}$ is equal to that of
the full cross section.  On the other hand, the overall magnitude of
$\mathcal{L}_{\text{DPS}} - \mathcal{L}_{\text{1v1,pt}}$ (which can be
positive or negative) has no direct significance, since one needs to add the
SPS contribution to obtain the physical cross section.

\begin{figure}
  \begin{center}
    \subfigure[]{\includegraphics[trim = 0cm 0 -4cm 0,
        height=16.5em]{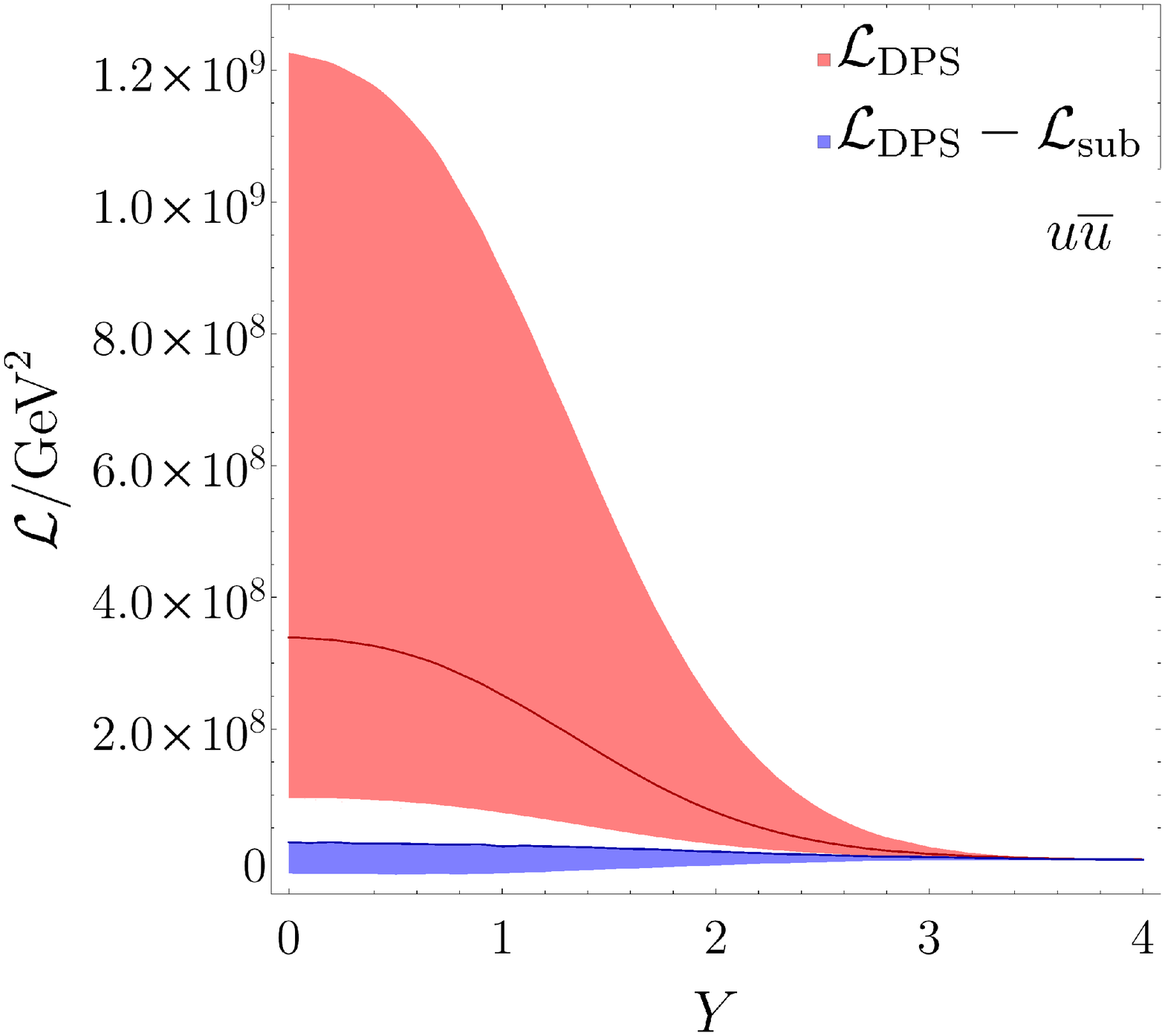}}
    \hspace{1em}
    \subfigure[]{\includegraphics[trim = 4cm 0 0 0,
        height=16.5em]{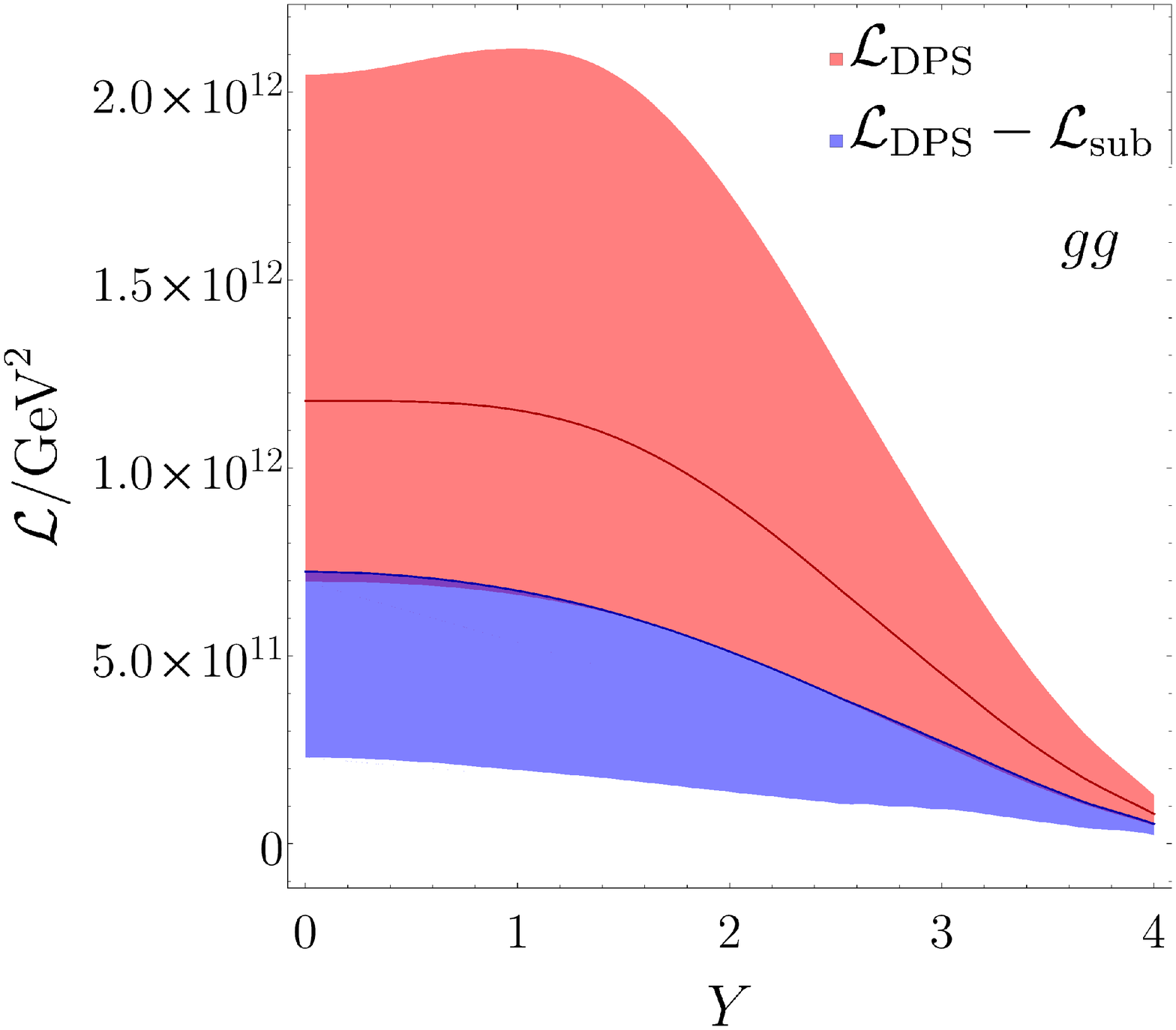}}
      \caption{\label{fig:DPSLmL} DPS luminosity $\mathcal{L}_{\text{DPS}}$
        and the difference
        $\mathcal{L}_{\text{DPS}} - \mathcal{L}_{\text{1v1,pt}}$ between the
        DPS and subtraction luminosities, shown for the $u\bar{u}$ (a) and
        $gg$ (b) channels.  Kinematic conditions are as in figure
        \protect\ref{fig:DPSlumi}.  In the case of
        $\mathcal{L}_{\text{DPS}} - \mathcal{L}_{\text{1v1,pt}}$, the value
        for $\nu= 80\gev$ is always at the top of the scale variation band.}
  \end{center}
\end{figure}

We plot $\mathcal{L}_{\text{DPS}}$ and
$\mathcal{L}_{\text{DPS}} - \mathcal{L}_{\text{1v1,pt}}$ for the $u\bar{u}$
and $gg$ channels in figure \ref{fig:DPSLmL}.  The lines are the values for
$\nu = 80\gev$, whilst the bands show the variation when $\nu$ is varied by a
factor of $2$ up and down.  Note that for
$\mathcal{L}_{\text{DPS}} - \mathcal{L}_{\text{1v1,pt}}$, the curve with
$\nu = 80\gev$ is always at the very top of the scale variation band.  We see
that the $\nu$ scale variation of
$\mathcal{L}_{\text{DPS}} - \mathcal{L}_{\text{1v1,pt}}$ is indeed reduced
compared to $\mathcal{L}_{\text{DPS}}$, with the reduction being much stronger
for $u\bar{u}$ than for $gg$.  The latter is consistent with our previous
observation that evolution effects are weaker in the $u\bar{u}$ channel at
central rapidities: if evolution effects are weak, $F_{\text{spl}}$ and
$F_{\text{spl,fo}}$ have a similar $y$ dependence, so that
$(F_{\text{spl}} \times F_{\text{spl}} - F_{\text{spl,fo}} \times
F_{\text{spl,fo}})$ is flat in $y$ space and
$\mathcal{L}_{\text{DPS}} - \mathcal{L}_{\text{1v1,pt}}$ varies weakly with
$\nu$.

In section~\ref{sec:dglap-subtr} we argued on generic grounds that the
$\nu$ variation of the 1v1 contribution to $\sigma_{\text{DPS}}$ should be
of the same order as the subtraction term $\sigma_{\text{1v1,pt}}$.  This
is confirmed in figure~\ref{fig:DPSLmL}, where the size of
$\mathcal{L}_{\text{1v1,pt}}$ (evaluated at central $\nu$) can be read off
from the distance between the central curve for $\mathcal{L}_{\text{DPS}}$
and the top of the band for $\mathcal{L}_{\text{DPS}} -
\mathcal{L}_{\text{1v1,pt}}$.  This finding allows us to sharpen the
argument we already made in the discussion of figure~\ref{fig:DPSlumi}:
not only does a large $\nu$ variation of the 1v1 contribution to
$\sigma_{\text{DPS}}$ indicate the need to include SPS and the associated
subtraction term in the cross section, but the size of the $\nu$ variation
may even serve as a rough estimate for the size of these terms.

\begin{figure}
  \begin{center}
    \subfigure[]{\includegraphics[trim = 0cm 0 -4cm 0,
        height=16.5em]{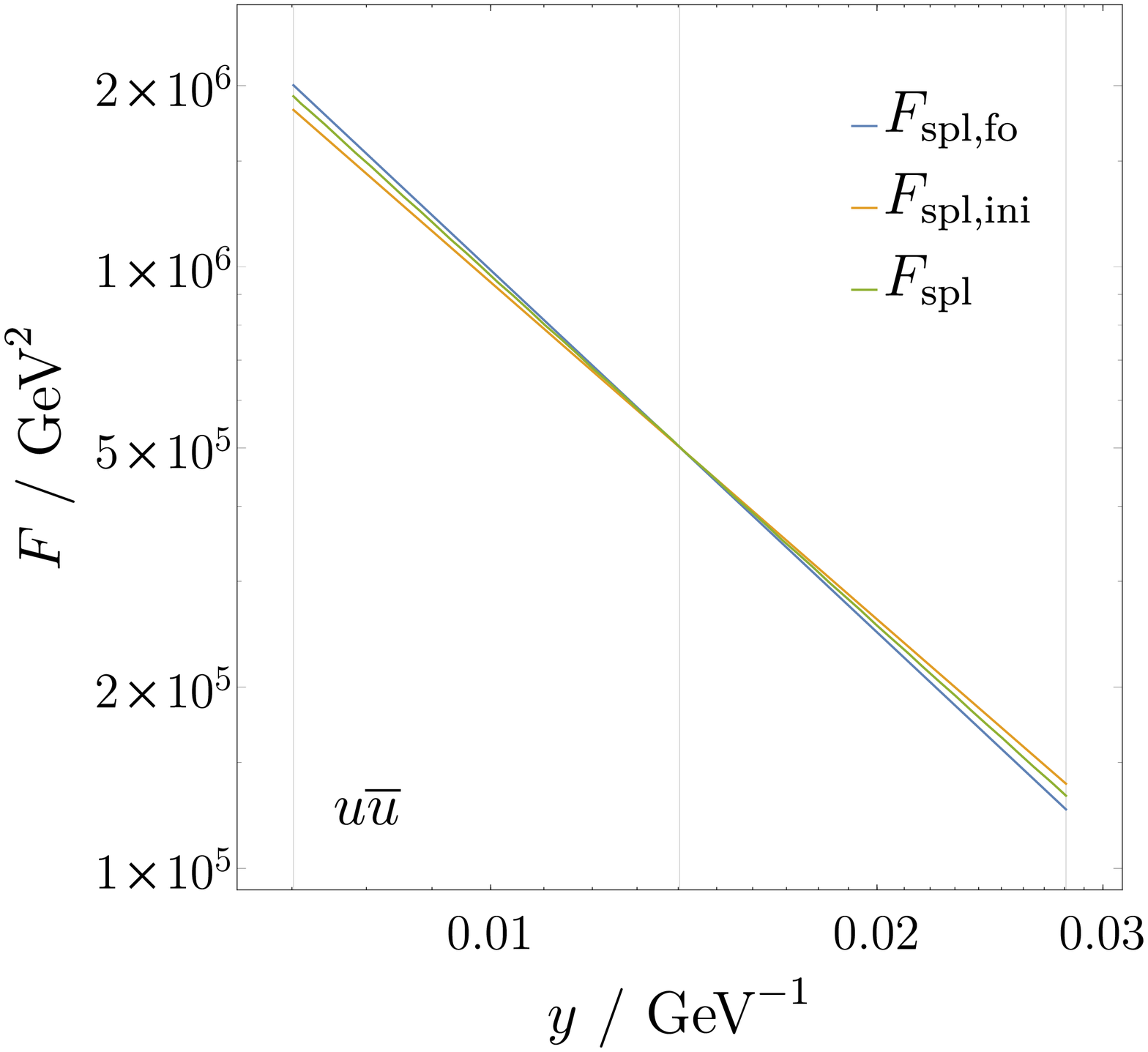}}
    \hspace{1em}
    \subfigure[]{\includegraphics[trim = 4cm 0 0 0,
      height=16.5em]{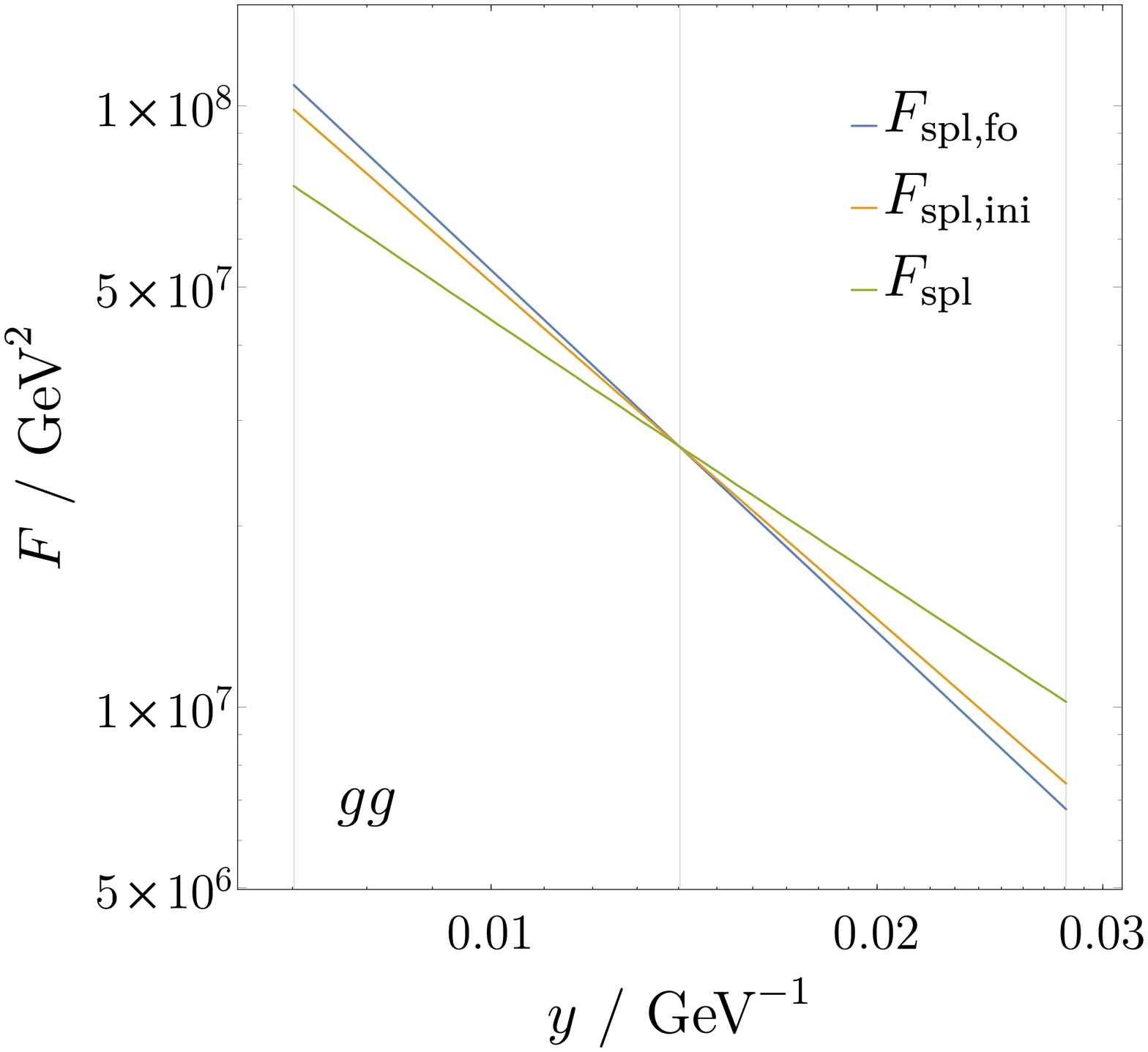}} \\[1em]
    \subfigure[]{\includegraphics[trim = 4cm 0 0 0,
        height=16.5em]{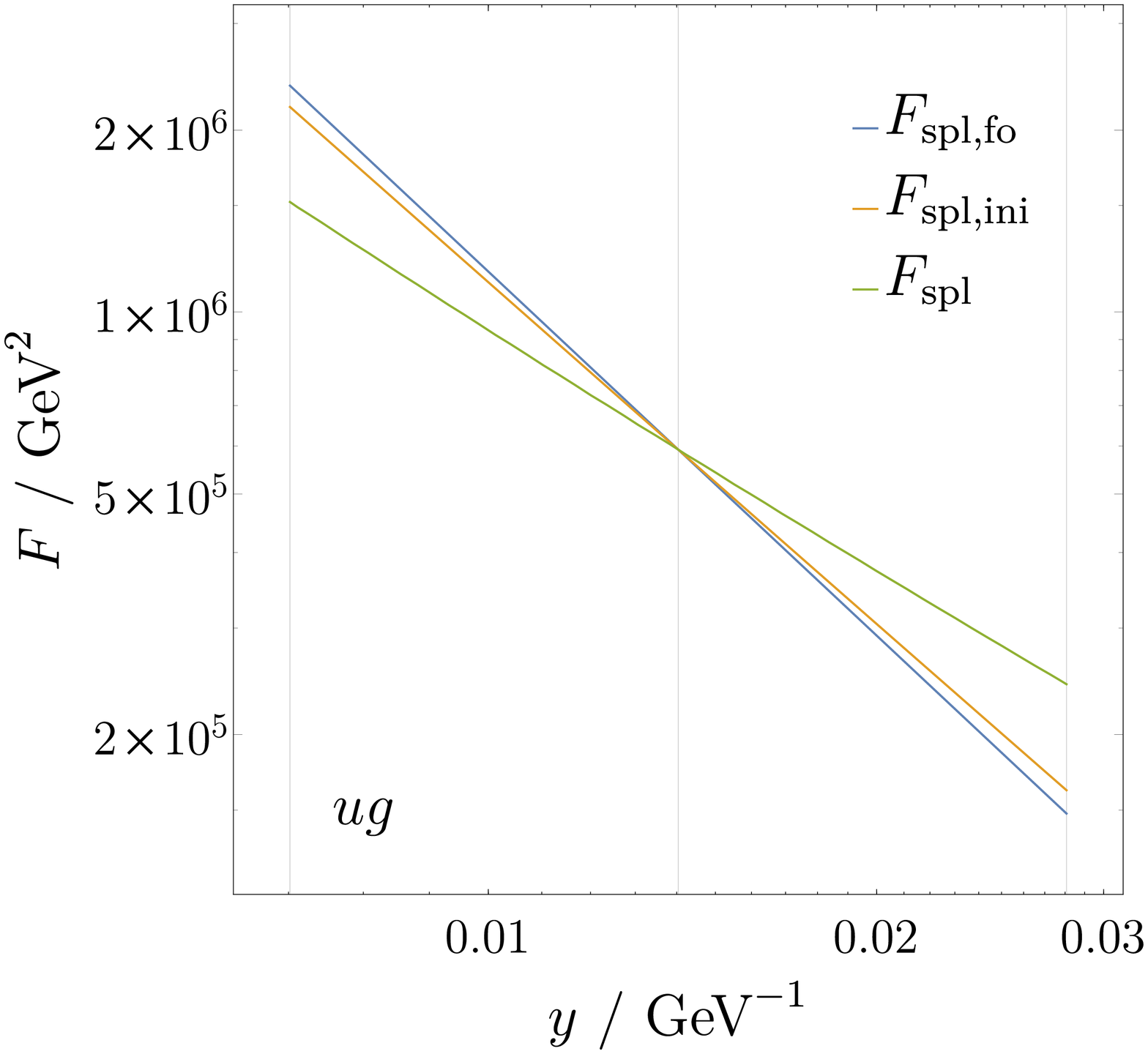}}
      \caption{\label{fig:DPDsvsy} Splitting DPDs $F_{\text{spl}}$ as a
        function of $y$, with $x_1 = x_2 = 5.7\times 10^{-3}$ and
        $\mu = 80\gev$.  The parton combinations shown are $u\bar{u}$ (a),
        $gg$ (b) and $ug$ (c).  The vertical grey lines correspond to
        $y=b_0/(160 \gev)$, $y=b_0/(80 \gev)$ and $y=b_0/(40 \gev)$.}
  \end{center}
\end{figure}

To better understand the behaviour we have seen in the 1v1 luminosities for
the $u\bar{u}$ and $gg$ channels, we now take a closer look at how evolution
affects the $y$ dependence of splitting DPDs.  In figure \ref{fig:DPDsvsy} we
plot $F_{\text{spl}}(\tvec{y};\mu)$ against $y$, setting
$x_1 = x_2 = 5.7\times 10^{-3}$ so that the $u\bar{u}$ and $gg$ DPDs are the
ones used for the point $Y=0$ in figures \ref{fig:DPSlumi} and
\ref{fig:DPSLmL}.  We also plot the $ug$ DPD, which mixes with $u\bar{u}$ and
$gg$ under evolution.  For comparison we show
$F_{\text{spl,fo}}(\tvec{y};\mu)$, as well as the initial condition
$F_{\text{spl,ini}}(\tvec{y}) = F_{\text{spl}}(\tvec{y};\mu_y)$ given in
\eqref{full_splitting_input}, from which $F_{\text{spl}}(\tvec{y};\mu)$ is
obtained by evolution.  The distributions are plotted in the range
$b_0/(160 \gev) < y < b_0/(40 \gev)$, i.e.\ in the range over which the lower
integration limit in $y$ is varied when we vary $\nu$ between $40\gev$ to
$160 \gev$. In this region, the exponential damping factor in our DPD model is
irrelevant, so that $F_{\text{spl,ini}}$ and $F_{\text{spl,fo}}$ only differ
by the scales taken in $\alpha_s$ and $f_{a_0}(x_1+x_2)$.

For the initial conditions $F_{\text{spl,ini}}$ we note that the $u\bar{u}$
and $ug$ distributions are of similar size; the former is initialised by a
larger PDF ($f_g$ instead of $f_u$) but has a smaller splitting coefficient
$P(1/2)$ as we noted before \eqref{simple-1v1}. By contrast, the $gg$
distribution is much bigger; here both the initialising PDF and the splitting
coefficient are large.

By construction, all three curves in each plot are equal at
$y = b_0/(80 \gev)$, when $\mu_y = \mu$.  In all plots, $F_{\text{spl,ini}}$
has a more shallow $y$ behaviour than $F_{\text{spl,fo}}$. This difference is
mainly driven by $\alpha_s$, as the PDFs do not vary so much between $\mu_y$
and $\mu$ at momentum fraction $x_1+x_2 = 1.14 \times 10^{-2}$.  One observes
that the difference between $F_{\text{spl}}$ and $F_{\text{spl,ini}}$ is more
significant for $gg$ and $ug$ than for $u\bar{u}$, i.e.\ that DPD evolution
has a much stronger effect on the former two channels.  This is to be expected
at small $x$, since the $1/v$ behaviour of the splitting kernels $P_{gg}(v)$
and $P_{gq}(v)$ at small $v$ favours the radiation of a gluon with much
smaller momentum than its parent parton, whereas the kernels $P_{qq}(v)$ and
$P_{qg}(v)$ giving a quark stay finite for $v\to 0$.

An interesting point to note is that, whilst for $gg$ and $ug$ the curves for
$F_{\text{spl}}$ are more shallow than for $F_{\text{spl,ini}}$, in the case
of $u\bar{u}$ the $F_{\text{spl}}$ curve is actually steeper.  The latter is
surprising since --- based on the experience with single PDFs at small $x$ ---
one may expect that forward evolution for $y > b_0/(80 \gev)$ would always
increase a DPD, and backward evolution for $y < b_0/(80 \gev)$ would always
decrease it.  This indeed happens in the $gg$ and $ug$ channels, whilst
forward evolution results in a decrease of the $u\bar{u}$ DPD.  The reason for
this is that $F_{ug, \ms \text{spl}}$ and $F_{g \bar{u},\ms \text{spl}}$,
which directly feed into $F_{u\bar{u},\ms \text{spl}}$ during evolution, are
comparatively small.  In the case of PDFs, $f_g$ is much larger than $f_u$ and
hence can drive its small-$x$ evolution although the splitting function
$P_{qg}$ does not favour the radiation of low-momentum quarks.  The evolution
of the $gg$ and $ug$ DPDs is driven by the large distribution
$F_{gg,\ms \text{spl}}$ and enhanced by the $P_{gg}$ splitting function.

Note that in all three channels, $F_{\text{spl}}$ has a smaller slope in $y$
than $F_{\text{spl,fo}}$.  This implies that the $y$ integrand for the
computation of $\mathcal{L}_{\text{DPS}} - \mathcal{L}_{\text{1v1,pt}}$ is
positive for $y > b_0/(80 \gev)$ and negative for $y < b_0/(80 \gev)$.
Therefore, the $\mathcal{L}_{\text{DPS}} - \mathcal{L}_{\text{1v1,pt}}$ curves
for $\nu = 40\gev$ and $\nu = 160\gev$ lie below that for $\nu = 80\gev$.  The
curve for the central $\nu$ value thus lies at the top of the $\nu$ variation
band, as we already observed in figure \ref{fig:DPSLmL}.

For $u\bar{u}$, the evolution from $\mu_y$ to $\mu$ turns out to have much the
same quantitative effect as adjusting the scale in the fixed order expression
from $\mu_y$ to $\mu$, such that $F_{\text{spl}}$ and $F_{\text{spl,fo}}$ end
up extremely close together.  This explains why at central rapidities the
$\nu$ variation of the 1v1 $u\bar{u}$ contribution in figure
\ref{fig:DPSlumi}a coincides almost exactly with the naive prediction from
\eqref{simple-1v1}, which assumes that $F_{\text{spl}}$ depends on $y$ like
$y^{-2}$ (as does $F_{\text{spl,fo}}$).  Furthermore, if $F_{\text{spl}}$ and
$F_{\text{spl,fo}}$ are very close to each other, then
$\mathcal{L}_{\text{DPS}} - \mathcal{L}_{\text{1v1,pt}}$ is much smaller than
$\mathcal{L}_{\text{DPS}}$, which we indeed see in figure \ref{fig:DPSLmL}a.

In the $gg$ channel, the modification of the $y$ slope by evolution is
significant.  In the range of figure \ref{fig:DPDsvsy}, the $y$ dependence is
changed from $y^{-2}$ at the starting scale to approximately $y^{-1.4}$.  One
may expect this flattening effect to become even stronger as the $x$ values in
the DPDs decrease.  The dependence of $\mathcal{L}_{\text{DPS}}$ on the scale
$\nu$ should then decrease.  This was already anticipated in
\cite{Ryskin:2011kk, Ryskin:2012qx}, where studies in the double leading
logarithm approximation were performed.

\begin{figure}
  \begin{center}
    \subfigure[]{\includegraphics[trim = 1.6cm 0 0 0,
      height=18em]{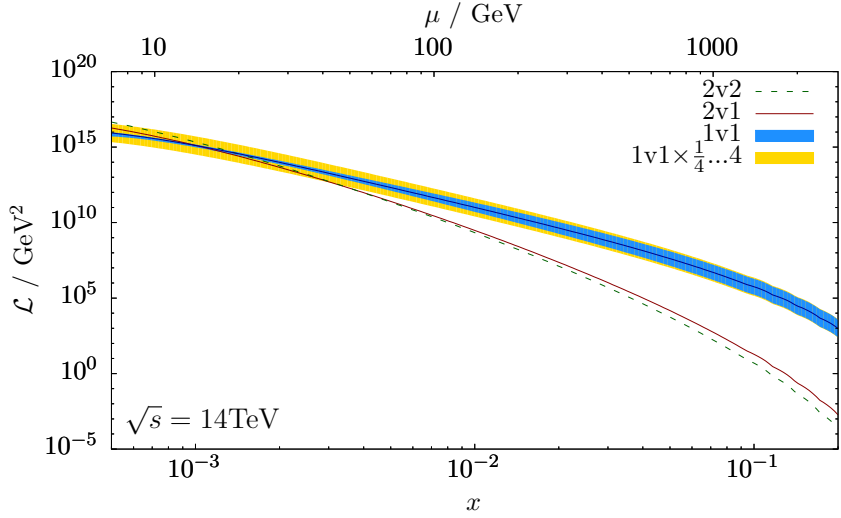}} \\[1em]
    \subfigure[]{\includegraphics[trim = 1.6cm 0 0 0,
      height=18em]{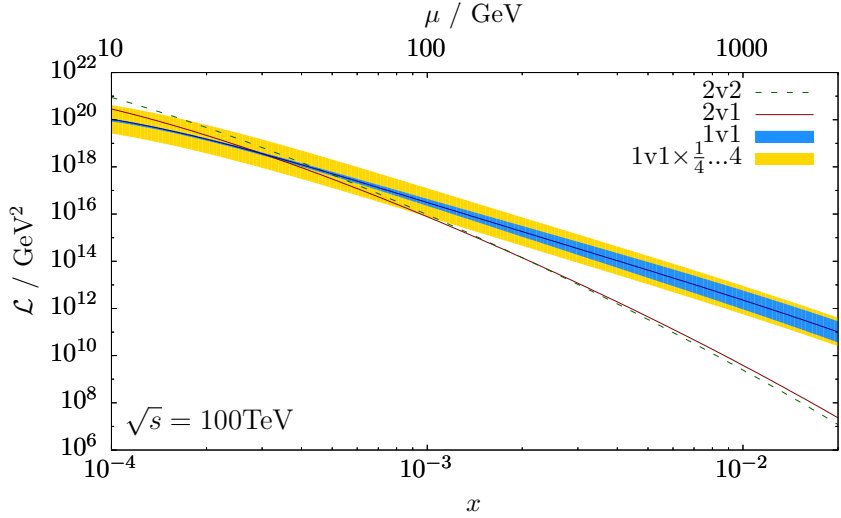}}
    \caption{\label{fig:DPSggvx} DPS luminosities in the $gg$ channel with all
      $x$ values in the DPDs set equal.  The collider energy $\sqrt{s}$ is
      fixed to $14\tev$ (a) or to $100\tev$ (b).  The scales of the DPDs, set
      to $\mu = x \sqrt{s}$, are given at the top of the plot.}
  \end{center}
\end{figure}

Here we investigate this effect using our full LO DGLAP set-up.  We consider
the $gg$ channel with all $x$ values set equal and study the DPS luminosity as
a function of the common $x$ value.  We fix the collider energy $\sqrt{s}$, so
that the scale $\mu = Q_1 = Q_2 = x \sqrt{s}$ also varies with $x$.  We make
one plot with $\sqrt{s} = 14\tev$ and $5\times10^{-4} < x < 0.2$
(corresponding to $7 \gev < \mu <2800 \gev$) and another one with higher
collider energy $\sqrt{s} = 100\tev$ and $10^{-4} < x < 0.02$ (corresponding
to $10 \gev < \mu <2000 \gev$).  In our numerical code this requires a new DPD
grid with larger $\mu$ and $\mu_y$ ranges, which was generated using 60 points
in the $\mu$ and $\mu_y$ directions (as in the original grid).  The resulting
1v1 luminosity is shown in figure \ref{fig:DPSggvx}, with bands for the $\nu$
variation and its naive version as in figure~\ref{fig:DPSlumi}.  For
comparison we also show the 2v1 and 2v2 luminosities.

We see that the $\nu$ variation does indeed become progressively smaller
compared to the central value as $x$ (and $\mu$) decreases.  At the lowest $x$
values it is much smaller than the naive expectation.  At larger $x$, where
evolution has a smaller effect on the DPDs, the $\nu$ variation becomes
larger: for $\sqrt{s} = 14 \tev$ the actual and naive $\nu$ bands essentially
coincide towards the right of the plot, where values of the order of
$x \sim 0.1$ are reached.  Towards the left of each plot, where $x$ and $\mu$
become small, the 2v2 and 2v1 contributions begin to dominate over 1v1.  For
given $\mu$, this effect is more pronounced for $\sqrt{s} = 100 \tev$ than for
$\sqrt{s} = 14 \tev$, since the $x$ values in the former case are smaller.  We
note that in the $u\bar{u}$ channel (not shown here) the actual and naive
$\nu$ bands remain very similar throughout the kinematics of
figure~\ref{fig:DPSggvx}.

\begin{figure}
  \begin{center}
    \includegraphics[trim = 4cm 0 0 0,
        height=16.5em]{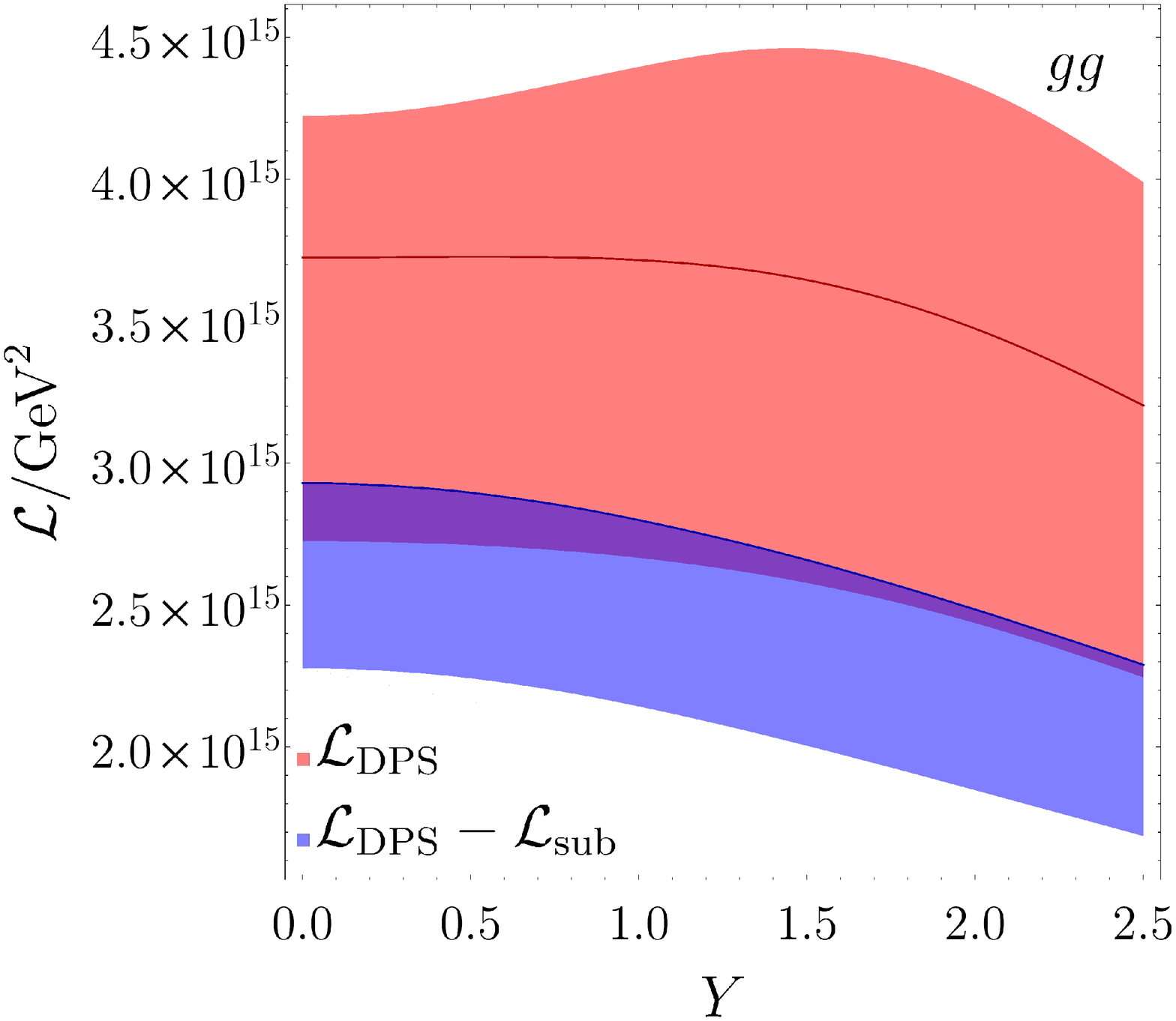}
    \caption{\label{fig:DPSLmL-10} As in figure~\protect\ref{fig:DPSLmL}b, but
      for different kinematics, with $\mu = 10\gev$ and all $x$ set equal to
      $7.1 \times 10^{-4}$, corresponding to $Q_1 = Q_2 = 10 \gev$ at
      $\sqrt{s} = 14\tev$.  Notice the suppressed zero on the $\mathcal{L}$
      axis.}
  \end{center}
\end{figure}

For the small $x$ values where the $\nu$ variation is small in figure
\ref{fig:DPSggvx}, the $gg$ splitting DPD is significantly shallower in $y$
than the fixed-order $y^{-2}$ form.  For $\sqrt{s} = 14 \tev$ we find that
around $x=2\times 10^{-3}$ it reaches $y^{-1}$, implying a logarithmic
dependence on $\nu$ in $\mathcal{L}_{DPS}$, whilst a behaviour like $y^{-0.5}$
is reached around $5\times 10^{-4}$. For $\sqrt{s} = 100 \tev$ the $y$
behaviour is like $y^{-1}$ around $x=4\times 10^{-4}$ and like $y^{-0.5}$
around $x=10^{-4}$.  In such a regime, the bulk of the contribution to the 1v1
part of $\mathcal{L}_{\text{DPS}}$ comes from large distances $y \gg 1/\nu$,
where the DPS approximations used to derive $\sigma_{\text{DPS}}$ are valid.
Also, the 1v1 part of $\sigma_{\text{DPS}}$ becomes clearly larger than its
$\nu$ variation.  As we argued earlier, the $\nu$ variation should be of the
same size as the subtraction term.  This is confirmed in
figure~\ref{fig:DPSLmL-10}, which is analogous to figure \ref{fig:DPSLmL}b but
for the kinematics of the point with $\mu=10\gev$ in
figure~\ref{fig:DPSggvx}a.  Using our argument that $\sigma_{\text{1v1,pt}}$
is of the same magnitude as the order of $\sigma_{\text{SPS}}$ that contains
the lowest-order DPS-type loop, one may in this situation justifiably make
predictions that include the DPS piece but omit the order of SPS just
specified, as well as the associated subtraction term.  This is encouraging,
for instance in the context of four-jet production, where the computation of
the relevant SPS order (namely NNLO) is well beyond the current state of the
art.  Note that lower orders in SPS should be computed and included, if
possible.

Notice that when the $y$ behaviour of $F_{\text{spl}}$ becomes flatter than
$y^{-1}$, the dominant $y$ region in the 1v1 part of
$\mathcal{L}_{\text{DPS}}$ shifts to values $y \sim 1/\Lambda$, where one
cannot compute the splitting DPD in perturbation theory and must rely on a
model.  Likewise, the 2v1 part of $\mathcal{L}_{\text{DPS}}$ becomes
increasingly sensitive to the region $y \sim 1/\Lambda$ as soon as
$F_{\text{spl}}$ becomes flatter than the fixed-order form $y^{-2}$.

\begin{figure}
  \begin{center}
    \subfigure[]{\includegraphics[trim = 1.5cm 0 0 0,
        height=20em]{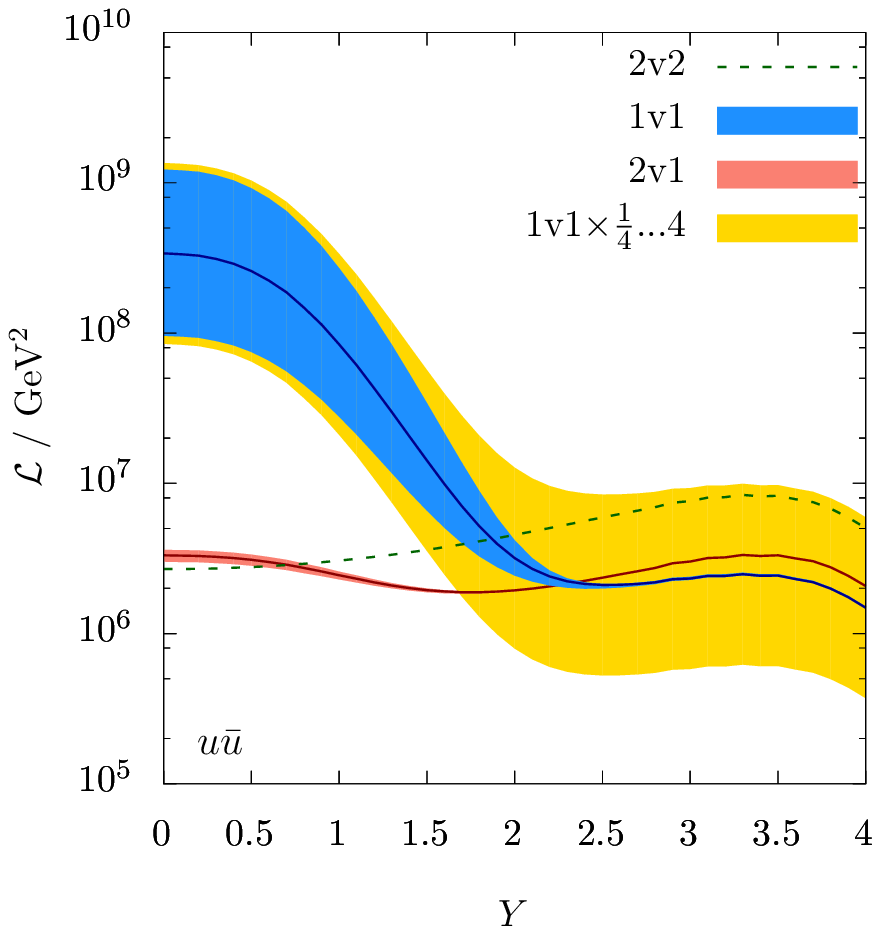}}
    \hspace{1em}
    \subfigure[]{\includegraphics[trim = 1.5cm 0 0 0,
        height=20em]{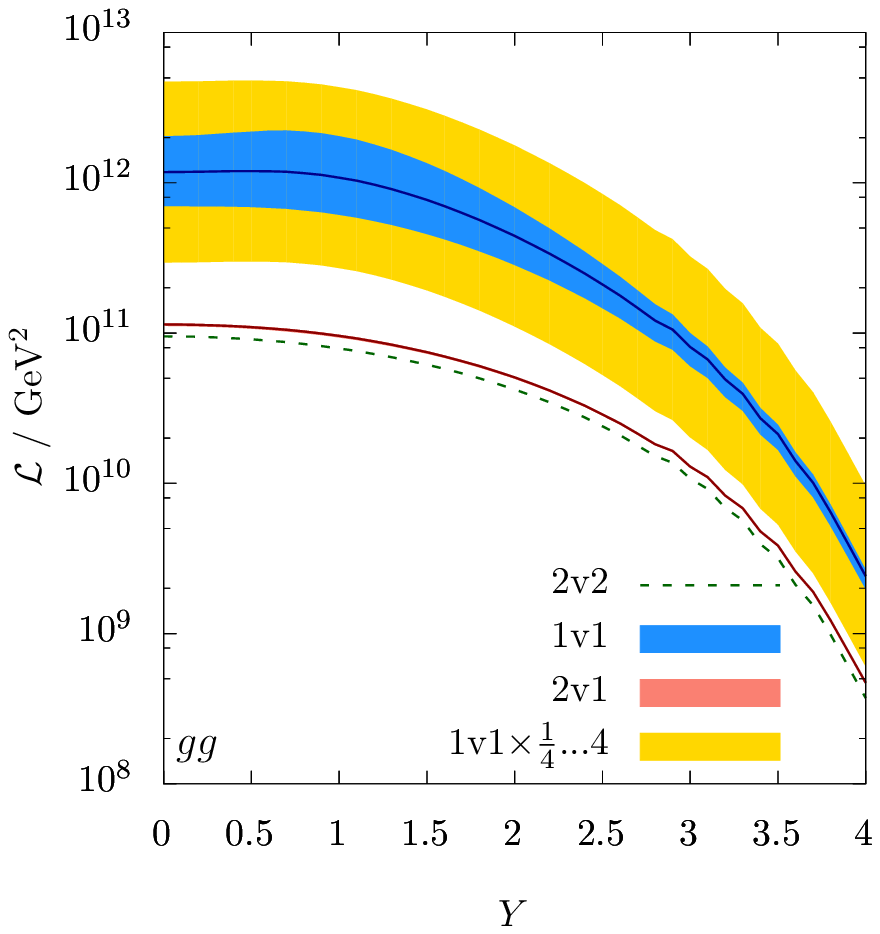}}
      \caption{\label{fig:DPSlumi_opprap} Double parton scattering
        luminosities $\mathcal{L}_{a_1 a_2 b_1 b_2}(Y)$ for the production of
        two systems with $Q_1 = Q_2 = 80\gev$ at $\sqrt{s} = 14\tev$, one with
        rapidity $Y$ and the other with rapidity $-Y$.  The parton
        combinations $a_1 a_2\ms b_1 b_2$ are
        $u\bar{u}\bar{u}u + \bar{u}uu\bar{u}$ (a) and $gggg$ (b).}
  \end{center}
\end{figure}

Another kinematic regime where the 1v1 contribution to
$\mathcal{L}_{\text{DPS}}$ becomes large compared with its $\nu$ variation and
compared with $\mathcal{L}_{\text{1v1,pt}}$ is when the two hard systems have
a large separation in rapidity.  This reduction of the $\nu$ variation can be
seen as $Y$ approaches $4$ in figure \ref{fig:DPSlumi}, but only in the
$u\bar{u}$ and not in the $gg$ channel.  In both channels, the effect becomes
more pronounced once the rapidity separation of the hard systems is increased
beyond $4$.  To illustrate this, we make plots similar to
figure~\ref{fig:DPSlumi} but now with one hard system at rapidity $Y$ and the
other at rapidity $-Y$ (rather than one at $Y$ and the other at $0$), such
that a given value of $Y$ corresponds to a rapidity separation of $2Y$.  The
results are shown in figure \ref{fig:DPSlumi_opprap}, where we see that the
$\nu$ variation in the 1v1 contribution is strongly reduced towards the right
hand side of the plots, becoming hardly visible in the $u\bar{u}$ channel.
Also notable is the fact that for large $Y$, the 2v2 contribution in this
channel exceeds 1v1, which strongly decreases between $Y=0$ and $2$.

This reduction in $\nu$ dependence can be explained by the fact that at large
$Y$ we probe splitting DPDs with one large $x$ and one small $x$ parton.  From
the point of view of small $x$ logarithms, it is preferable to generate such a
configuration by having the $1 \to 2$ splitting at large $x$, generating
directly the large $x$ parton plus a gluon with smaller $x$, the latter of
which splits in a number of stages into smaller $x$ gluons, eventually
yielding the small $x$ parton.  This increasingly happens with increasing $y$,
since the ``evolution distance'' between the initial and final scales, $\mu_y$
and $\mu$, is increased.  Thus, evolution again flattens the DPD compared to
the naive $y^{-2}$ expectation and reduces the $\nu$ dependence of the DPS
luminosities.

\begin{figure}
  \begin{center}
    \includegraphics[trim = 0cm 0 -4cm 0,
        height=17em]{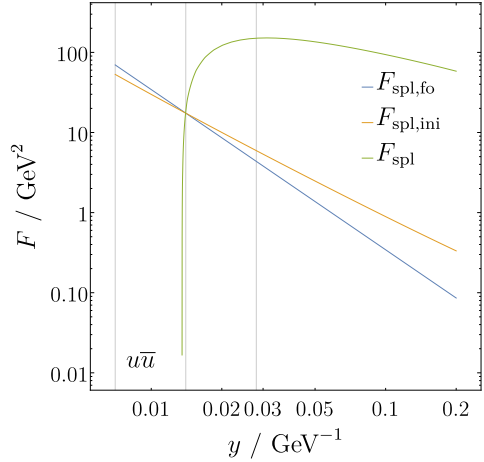}
        \caption{\label{fig:DPDsvsy-highy} As
          figure~\protect\ref{fig:DPDsvsy}a but for asymmetric kinematics
          $x_1 = 10^{-4}$, $x_2 = 0.31$ at $\mu=80\gev$, which is needed for
          the point $Y=4$ in figure~\ref{fig:DPSlumi_opprap}a.}
  \end{center}
\end{figure}

The effect is particularly drastic in the $u\bar{u}$ channel, because the
lowest-order splitting $g \to u\bar{u}$ is inefficient at generating a pair
with very different $x$ values (as $P_{g \to q\bar{q}}(v)$ goes to a constant
at small $v$).  Therefore, the repeated splitting mechanism described in the
previous paragraph is strongly preferred, even though its last step is
penalised by the lack of small-$v$ enhancement in $P_{gq}(v)$.  In
figure~\ref{fig:DPDsvsy-highy} we plot the $u\bar{u}$ DPD in the $x_1\ll x_2$
configuration relevant for $Y=4$ in figure \ref{fig:DPSlumi_opprap}a.  It has
a similar $y$ dependence as the $u\bar{d}$ DPD in equal kinematics, with the
curve for $u\bar{d}$ crossing zero exactly at $y = b_0/\mu$ (by construction),
and the curve for $u\bar{u}$ crossing zero rather close to this point.  In the
former case, the lowest-order $1 \to 2$ splitting process is forbidden,
whereas in the latter it is numerically almost irrelevant.  The situation for
$x_1\gg x_2$ is the same.

To end this section, we study polarised distributions and luminosities.  We
limit ourselves to the splitting part $F_{\text{spl}}$ of the DPD and to the
1v1 contribution to $\mathcal{L}_{\text{DPS}}$.  In fact, we have little
guidance for modelling the intrinsic part $F_{\text{int}}$, where a product
ansatz as in \eqref{full_intrinsic_input} makes no sense.  We note that in
\cite{Diehl:2014vaa}, different ans\"atze were made for polarised DPDs, and
the effects of evolution on these distributions were investigated.  It was
found that, in many cases, evolution to higher scales leads to a suppression
of the polarised DPD with respect to its unpolarised counterpart.

We initialise the polarised $F_{\text{spl}}$ at $\mu_y$ using the expression
in \eqref{full_splitting_input} with the unpolarised $1\to 2$ splitting
functions replaced by their polarised counterparts (given in Appendix B of
\cite{Diehl:2014vaa}).  These distributions are then evolved to the scale
$\mu$ using the appropriate polarised double DGLAP equations (the required
polarised splitting functions are collected in Appendix A of
\cite{Diehl:2013mla}).  The settings we used for the evolution code are the
same as specified in section~\ref{sec:set-up}.

\begin{figure}
  \begin{center}
    \subfigure[]{\includegraphics[trim = 0cm 0 -1.4cm 0,
        height=19em]{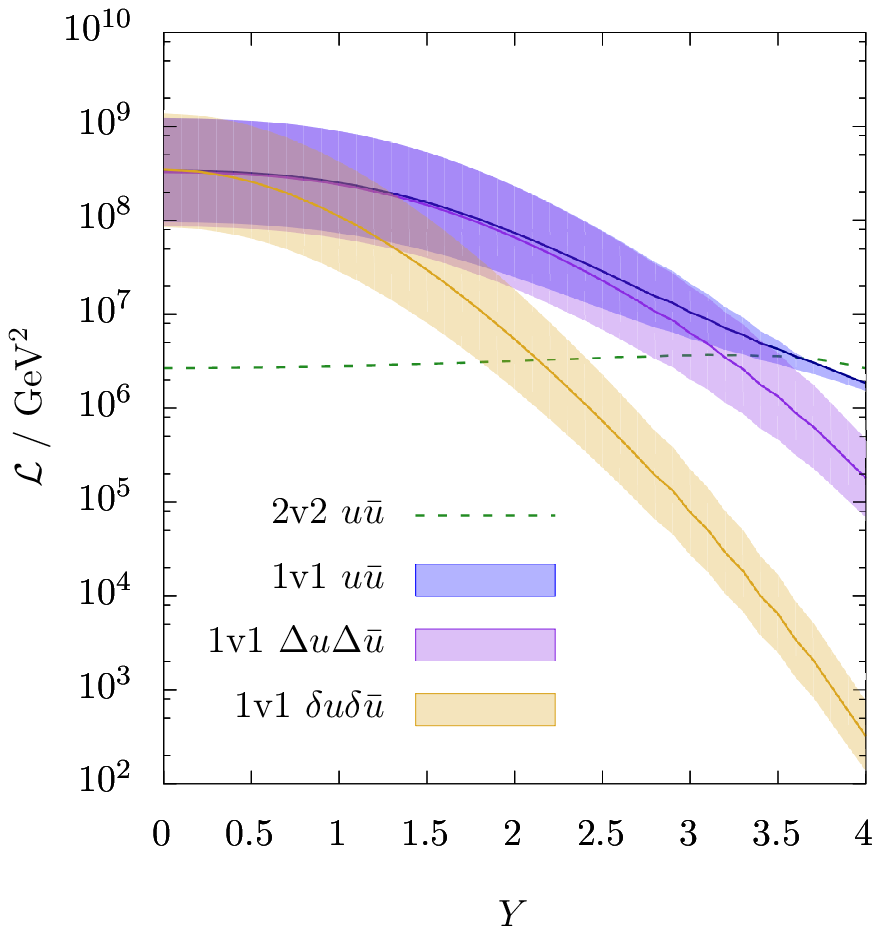}}
    \subfigure[]{\includegraphics[trim = 1.4cm 0 0 0,
        height=19em]{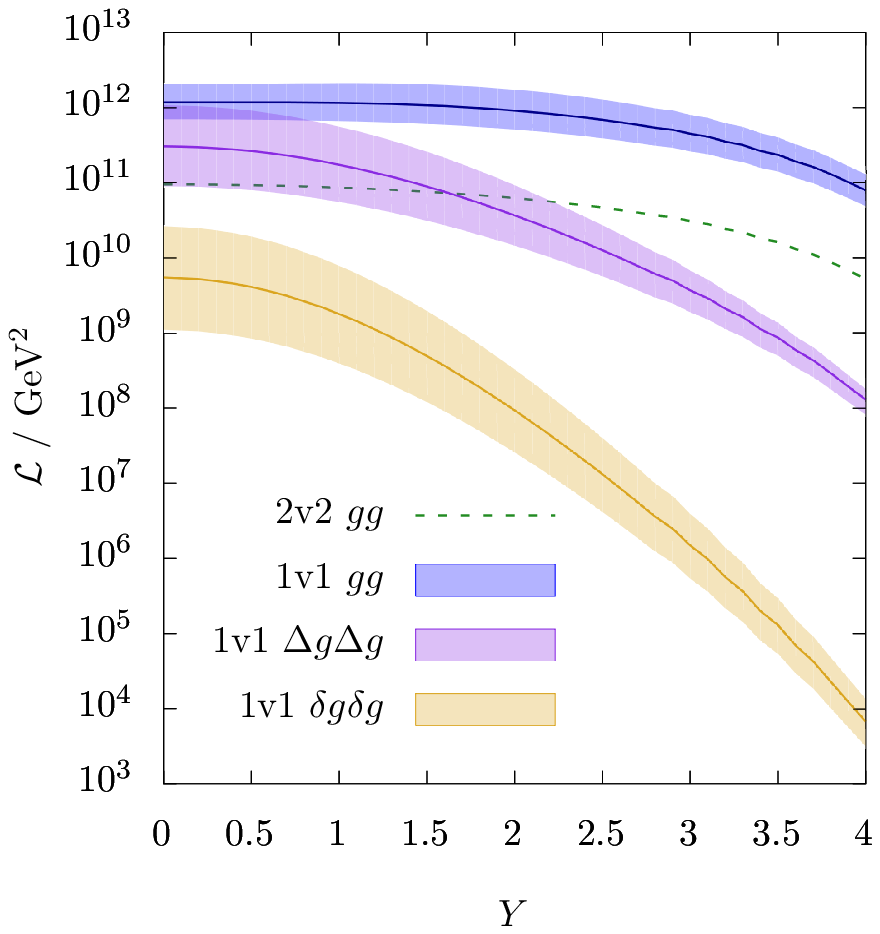}}
      \caption{\label{fig:DPSlumipol} As in
        figure~\protect\ref{fig:DPSlumi}, but with two polarised 1v1
        luminosity bands included in each case (and the unpolarised 2v1
        band omitted).}
  \end{center}
\end{figure}

We consider the same scenario as in figure \ref{fig:DPSlumi}, i.e.\ the
production of two systems with $Q_1 = Q_2 = 80\gev$ at $\sqrt{s} =
14\tev$, one with rapidity $0$ and the other with rapidity $Y$, and now
include polarised 1v1 contributions.  In figure \ref{fig:DPSlumipol}a, we
reproduce figure \ref{fig:DPSlumi}a but include 1v1 luminosities for the
polarised parton combinations $a_1 a_2\ms b_1 b_2 = \Delta u \Delta
\bar{u} \Delta \bar{u} \Delta u + \Delta \bar{u} \Delta u \Delta u \Delta
\bar{u} $ and $\delta u\ms \delta \bar{u}\ms \delta \bar{u}\ms \delta u +
\delta \bar{u}\ms \delta u\ms \delta u\ms \delta \bar{u}$.  For brevity we
refer to these combinations as $\Delta u \Delta \bar{u}$ and $\delta u\ms
\delta \bar{u}$ in the following, \rev{recalling that $\Delta q$ and
  $\delta q$ indicate longitudinal and transverse quark polarisation,
  respectively.}  These polarised luminosities appear in the DPS cross
section for double $Z$ production, see section 3.1 of
\cite{Kasemets:2012pr} (combinations like $u\bar{u}\ms \Delta \bar{u}
\Delta u$ also appear but are not shown here).  To avoid cluttering the
plots, we omit the unpolarised 2v1 luminosity, reminding the reader that
it is of the same magnitude as 2v2.  In figure \ref{fig:DPSlumipol}b we
repeat the exercise for the pure gluon channel, including the 1v1
luminosities for the polarised combinations $a_1 a_2\ms b_1 b_2 = \Delta g
\Delta g \Delta g \Delta g$ and $\delta g\ms \delta g\ms \delta g\ms
\delta g$, referred to as $\Delta g \Delta g$ and $\delta g\ms \delta g$,
\rev{where $\Delta g$ and $\delta g$ respectively denote longitudinal and
linear gluon polarisation.}

We see that the 1v1 luminosities for $\Delta u \Delta \bar{u}$ and
$\delta u\ms \delta \bar{u}$ essentially coincide with the one for $u \bar{u}$
at central rapidities, with $\nu$ variation bands of very similar size.  These
observations can be explained by the fact that at $Y=0$ we initialise all
three DPDs using the same expression up to a minus sign, with
$P_{g \to q\bar{q}} = |P_{g \to \Delta q \Delta \bar{q}}| = |P_{g \to \delta
  q\ms \delta \bar{q}}|$ at $v=1/2$, and that evolution has a rather weak
effect on all three DPDs for central rapidities.  This close agreement extends
out to higher values of $Y$ for the $u\bar{u}$ and $\Delta u \Delta \bar{u}$
curves, owing to the fact that
$P_{g \to q\bar{q}} = |P_{g \to \Delta q \Delta \bar{q}}|$ for general $v$.
By contrast, the $\delta u\ms \delta \bar{u}$ curve clearly falls below the
other two for larger $Y$, because
$P_{g \to q\bar{q}} > |P_{g \to \delta q\ms \delta \bar{q}}|$ for
$v \neq 1/2$.

In figure \ref{fig:DPSlumipol}b, the $\delta g\ms \delta g$ luminosity lies
well below the one for $\Delta g \Delta g$, which in turn lies somewhat below
the one for $gg$.  This is mostly driven by the differences in initialisation
expressions for the DPDs: at $v=1/2$ (relevant at $Y=0$) the relevant
splitting function satisfy for instance $P_{g \to gg}:P_{g \to \Delta g \Delta
  g}:P_{g \to \delta g\ms \delta g} = 9:7:1$.  Ignoring evolution effects, one
would then expect the $gg$ luminosity at $Y=0$ to be roughly $1.5$ times
bigger than the one for $\Delta g \Delta g$, which in turn should be roughly
$50$ times bigger than the one for $\delta g\ms \delta g$.  This expectation
is quite close to the actual luminosity ratios, which are $\sim 4$ and $\sim
60$, respectively.  The remaining difference is due to evolution effects,
which increase $gg$ rather considerably, increase $\Delta g \Delta g$ to a
lesser extent, and hardly change $\delta g \delta g$.

Overall, we see that the polarised 1v1 luminosities can be of the same order
of magnitude as the unpolarised ones, so that one must in general take into
account all possible polarisation combinations in the DPS contribution,
together with the SPS term and subtraction (where the subtraction will also
contain both unpolarised and polarised contributions).


\subsection{Production of two scalars}
\label{sec:ggtoscal}

In this section, we study an explicit example to test our general argument
that $\sigma_{\text{1v1,pt}}$ is of similar size as the corresponding
perturbative order of the SPS cross section.  The process we investigate is an
artificial one, chosen for ease of computation.  We consider the hypothetical
production of two identical massive scalar bosons $\phi$, which couple to
quarks with a Yukawa term $c\ms \phi\ms \bar{q} q$, where $c$ is a constant.
For simplicity we take only one light quark flavour; including further light
flavours just multiplies all following results (SPS and subtraction) by $n_f$.
We will not compare the subtraction term to the full SPS contribution,
but rather to the $gg$ initiated part of $\sigma_{\text{SPS}}$, which
is the piece that contains the lowest order DPS-type $gg \to \phi\phi$
box diagram.  This piece is gauge invariant and can hence be
meaningfully considered by itself.

In keeping with the notation of the paper, let us denote the mass of
each produced boson $\phi$ by $Q$.  Let $Y$ be the rapidity of the
diboson system in the $pp$ centre-of-mass frame, $\hat{s}$ its squared
invariant mass, and $\beta = \sqrt{1 - 4 Q^2 /\hat{s}}$ the velocity
of one boson in the diboson centre-of-mass frame.  The
$gg\to \phi\phi$ part of the SPS cross section can be written as
\begin{align}
  \label{HH-SPS}
  \frac{d\sigma_{\text{SPS}}}{dY d\beta}
  &= x \bar{x} \, g(x)\, g(\bar{x})\,
  \frac{2\beta}{1-\beta^2}\, \hat{\sigma}_{gg \to \phi\phi}
  \nonumber \\
  &= x \bar{x} \, g(x)\, g(\bar{x})\,
  \frac{1}{128\ms \pi}\, \frac{1-\beta^2}{Q^4}\,
  \int_0^{q^2_{\text{max}}} \!
  \frac{d q^2}{\sqrt{1 - q^2 /q^2_{\text{max}}}}\,
  \bigl| \mathcal{A}_{gg\to \phi\phi} \bigr|^2 \,,
\end{align}
where $q_{\text{max}} = \beta Q \big/\sqrt{1-\beta^2}$ and the $gg \to
\phi\phi$ matrix element squared, $|\mathcal{A}_{gg\to \phi\phi}|^2$, is
averaged over the spin and colour of the incoming gluons.  The momentum
fractions $x$ and $\bar{x}$ in the gluon distributions are obtained from
the constraints $\hat{s} = x \bar{x} s$ and $Y = \half\log (x/\bar{x})$.

The matrix element squared for $gg \to \phi\phi$ can be obtained from the
$gg \to HH$ matrix element squared given in \cite{Glover:1987nx} by making
the replacement
\begin{align}
  \frac{\pi \alpha_{\scriptscriptstyle W}^{}
    m_q^2}{M_{\scriptscriptstyle W}^2} \to c^2\,.
\end{align}
In particular, $|\mathcal{A}_{gg\to \phi\phi}|^2$ in \eqref{HH-SPS} is
related to the terms ``gauge1'' and ``gauge2'' in \cite{Glover:1987nx}
according to
\begin{align} \label{MtoGvdB}
  \bigl| \mathcal{A}_{gg\to \phi\phi} \bigr|^2
  &= \dfrac{c^4 \alpha_s^2}{2^{10} \pi^2 m_q^4}
    \left( \left| \text{gauge1} \right|^2
    + \left| \text{gauge2} \right|^2 \right) \,.
\end{align}
The right hand side of \eqref{MtoGvdB} is given in
\cite{Glover:1987nx} for general quark mass $m_q$.  We evaluate this
expression numerically for very small $m_q$, using analytic
$m_q \to 0$ approximations where this is necessary to avoid numerical
instabilities.

Now we turn to the subtraction term. The two elementary $q\bar{q} \to \phi$
cross sections in this term contain the kinematic constraint $2\pi \delta(x_i
\bar{x}_i\ms s - Q^2)$, which fixes the momentum fractions $x_i$ and
$\bar{x}_i$ entering the two hard subprocesses at given $x, \bar{x}$ and
$\beta$ up to a two-fold ambiguity. The DPS subtraction term is given by
\begin{align} \label{ggphphsub1}
  \frac{d\sigma_{\text{1v1,pt}}}{dY d\beta}
  &= \frac{\pi^2}{16}\, \frac{1-\beta^2}{Q^8}\,
       x^2 \bar{x}^2 \sum_{a_1 a_2 b_1 b_2}
  \bigl| \mathcal{A}_{a_1 b_1\to \phi} \bigr|^2\,
  \bigl| \mathcal{A}_{a_2 b_2\to \phi} \bigr|^2
  \int d^2\tvec{y}\; \Phi^2(y \nu) \sum_{R=1,8}
\nonumber \\
  &\quad \times \biggl[
    {}^{R\!}F_{a_1 a_2,\ms \text{spl,pt}}\Bigl( \half x (1+\beta),
    \half x (1-\beta), \tvec{y} \Bigr)\,
        {}^{R\!}F_{b_1 b_2,\ms \text{spl,pt}}\Bigl( \half \bar{x}\ms (1-\beta),
        \half \bar{x}\ms (1+\beta), \tvec{y} \Bigr)
\nonumber \\
  &\qquad  + \{ \beta \to -\beta \} \biggr] + \{ F \to I \} \,,
\end{align}
where we sum over all possible colour, spin, and quark number
interference/correlation possibilities. The index $R$ on $F$ denotes the
colour channel ($R=1$ for colour singlet, $R=8$ for colour octet), and in the
sum over $a_1 a_2\ms b_1 b_2$ we sum over both unpolarised (e.g.\
$q\bar{q}\bar{q}q$) and polarised (e.g.\
$\Delta q \Delta \bar{q} \Delta \bar{q} \Delta q$) combinations. By
$\{ F \to I \}$ we denote the same expression, but with the quark number
diagonal distributions replaced by the quark number interference ones (see
section~2 of \cite{Diehl:2011yj}).

The splitting kernels for the relevant spin combinations
are~\cite{Diehl:2011yj}
\begin{align}
  \label{ggphphkernels}
{}^{1\!}P_{g\to q\bar{q}\,}(v)
   = - \, {}^{1\!}P_{g\to \Delta q \Delta\bar{q}\,}(v)
  &= \frac{1}{2}\ms \bigl[ v^2 + (1-v)^2 \ms\bigr] \,,
&
{}^{1\!}P_{g\to \delta q \delta\bar{q}\,}(v) &= {}- \delta^{jj'} v (1-v)
\end{align}
and ${}^{8\!}P = - \, {}^{1\!}P /\sqrt{N_c^2 - 1}$, where $j, j'$ are
indices for transverse quark polarisation.  The term with
interference DPDs $I$ gives the same contribution as that with $F$,
since the corresponding diagrams are simply related by reversing the
direction of fermion flow in one of the two quark loops, and this does
not change the expression for the diagram.  The squared subprocess
amplitudes, including an average over colour in the initial state,
read
\begin{align} \label{ggphphhard}
  \bigl| \mathcal{A}_{q \bar{q}\to \phi} \bigr|^2\,
  &= \bigl| \mathcal{A}_{\Delta q \Delta\bar{q}\to \phi} \bigr|^2\,
  = \frac{c^2 Q^2}{2 N_c} \,,
&
  \bigl| \mathcal{A}_{\delta q \delta\bar{q}\to \phi} \bigr|^2\,
  &= \delta^{jj'}\, \frac{c^2 Q^2}{2 N_c} \,,
\end{align}
where $j, j'$ are the indices for transverse quark polarisation and where
we have used the spin projection operators given in equation~(2.90) of
\cite{Diehl:2011yj}.

\begin{figure}
  \begin{center}
    \includegraphics[trim = 0cm 0 0 0,
      height=14em]{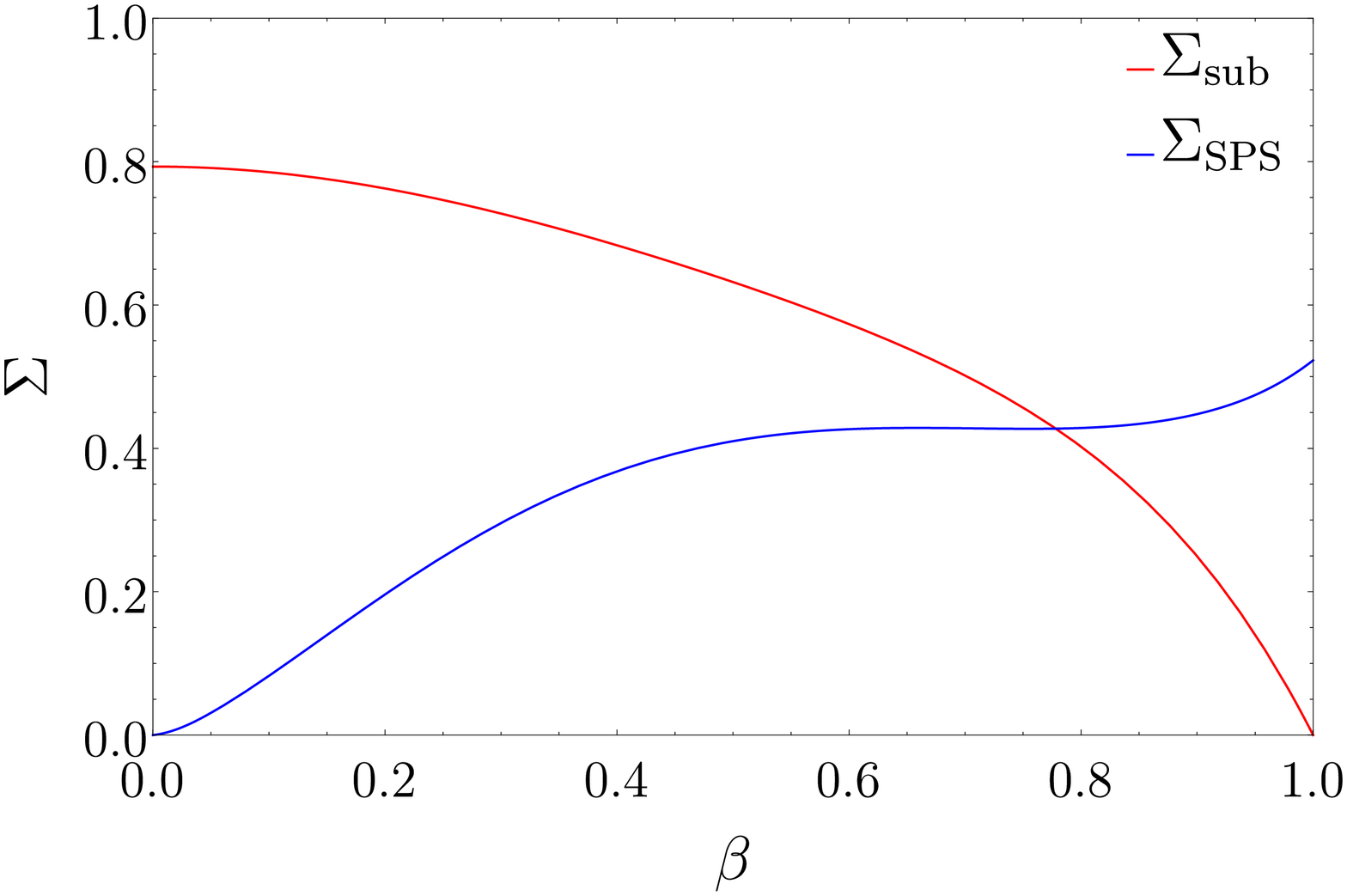}
      \caption{\label{fig:subvsSPS} The scaled cross section $\Sigma$
        defined in \protect\eqref{sigdef}, plotted for the
        $gg \to \phi\phi$ contribution to SPS ($\Sigma_{\text{SPS}}$),
        and the subtraction term ($\Sigma_{\text{1v1,pt}}$).}
  \end{center}
\end{figure}

Inserting \eqref{ggphphkernels} and \eqref{ggphphhard} into
\eqref{ggphphsub1}, we obtain
\begin{align}
  \label{HH-DPS}
  \frac{d\sigma_{\text{1v1,pt}}}{dY d\beta}
  &= x \bar{x} \, g(x) \, g(\bar{x})\,
  \frac{1}{128 \pi}\, \frac{1-\beta^2}{Q^4}\, (1+\beta^4)\,
  \frac{c^4 \alpha_s^2}{N_c^2 -1}\,
  \int_0^\infty \frac{d{y}^2}{y^4}\, \Phi^2(y \nu) \,.
\end{align}
Note that both \eqref{HH-SPS} and \eqref{HH-DPS} contain a product of
gluon PDFs evaluated at the same $x$ values, $g(x)\, g(\bar{x})$.  For
the comparison, we can divide the common PDF factor out of the two
expressions, in order to avoid having to use an explicit
parameterisation.  We also divide out various factors appearing in
both expressions and compare the quantity
\begin{align}
  \label{sigdef}
\Sigma(\beta)
  &= \frac{d\sigma}{dY d\beta}\, \frac{Q^2}{x \bar{x}\, g(x)\, g(\bar{x})}\,
    \frac{128 \pi (N_c^2-1)}{c^4\alpha_s^2} \,.
\end{align}
It is straightforward to show that $\Sigma$ is a function of the variable
$\beta$ only.  We plot $\Sigma$ for SPS and the subtraction term in figure
\ref{fig:subvsSPS}, where the curve for $\Sigma_{\text{1v1,pt}}$ corresponds
to the central choice $\nu = Q$ of the cutoff scale.

We see that $\Sigma_{\text{1v1,pt}}(\beta)$ is indeed generally of the same
order of magnitude as $\Sigma_{\text{SPS}}(\beta)$.  The agreement is perhaps
surprisingly good, given that $\Sigma_{\text{1v1,pt}}(\beta)$ involves
integrating a low~$q$ approximation to the matrix element squared outside the
region where the approximation is valid.  The agreement gets worse towards the
endpoints $\beta \to 0$ and $\beta \to 1$, which correspond to the high energy
limit and the threshold limit, respectively.  It is to be expected that the
agreement is especially bad at these points, since in the subtraction term we
effectively assume that the integration over the squared transverse momentum
$q^2$ of the scalar particles goes from zero to values of order $Q^2$.  For
both $\beta \to 0$ and $\beta \to 1$ this assumption becomes a poor one: near
threshold the phase space in $q^2$ shrinks to zero, whilst in the high energy
limit, $q^2$ can go up to size $\hat{s} \gg Q^2$.

\section{Summary}
\label{sec:summary}

Consistently incorporating the perturbative splitting of one parton into
two is a highly nontrivial problem for the theoretical description of
double parton scattering.  DPS graphs in which such splittings occur in
both protons (1v1 graphs) overlap with loop corrections to single parton
scattering.  Another type of graph, typically referred to as 2v1, in which
one parton pair arises from a perturbative splitting, and the other pair
is an ``intrinsic'' one already existing at the nonperturbative level,
overlaps with twist-four contributions to the cross section.  Finally
there is an overlap between DPS contributions where a splitting occurs in
both protons only in the amplitude or its conjugate, and SPS/DPS
interference graphs.

We have presented a scheme to compute DPS and to consistently merge its
contribution to the cross section with SPS and the other terms just
mentioned.  The scheme works in a similar manner for collinear and for TMD
factorisation.  Ultraviolet divergences that arise from perturbative
splitting in a naive treatment of DPS are regulated by a cutoff function
$\Phi(y \nu)$ in transverse position space.  This avoids modification of
the position space DPDs, which are defined via operator matrix elements in
close analogy to single-parton distributions.  In collinear factorisation,
these DPDs hence evolve according to a homogeneous DGLAP equation, whilst
their TMD counterparts satisfy a generalisation of the renormalisation
group equations for single-parton TMDs.  No modification of
hard scattering cross sections computed for standard collinear or TMD
factorisation is necessary in our scheme.  Collins-Soper type equations
describe the rapidity evolution of transverse-momentum dependent DPDs and
of collinear DPDs in colour non-singlet channels \cite{Buffing:2016wip}.

The problem of double counting between DPS and other contributions ---
notably between DPS and SPS --- is solved by subtraction terms as
specified in \eqref{coll-master} and \eqref{tmd-master}, which are
obtained in a simple way from $\sigma_{\text{DPS}}$ by replacing the DPDs
with their appropriate short-distance limits.  This paves the way for
using the scheme at higher orders in $\alpha_s$, with calculations being
considerably simpler for the subtraction terms compared with the full
hard scattering process at the corresponding order.
With a suitable choice of starting conditions and scales, specified in
section~\ref{sec:dglap-logs}, the DPS part of the cross section correctly
resums DGLAP logarithms that are not included in the fixed order
twist-four contributions.

Our scheme is naturally formulated with position-space DPDs
$F(x_i,\tvec{y})$, but it is possible to relate the Fourier transform of
$F(x_i,\tvec{y})\, \Phi(y \nu)$ to DPDs $F(x_i,\tvec{\Delta})$ that are
renormalised in transverse momentum space and satisfy an inhomogeneous
DGLAP equation rather than a homogeneous one.  This relation has the form
of a perturbative matching equation, see \eqref{matching-small-delta}, and
is somewhat similar to the matching between PDFs defined in different
schemes such as the $\overline{\text{MS}}$ and the DIS scheme.  
The momentum space representation also allows us to show that for the 2v1
contribution to DPS our scheme is equivalent to the ones in
\cite{Blok:2011bu,Blok:2013bpa} and in \cite{Ryskin:2011kk,Ryskin:2012qx}
to leading logarithmic accuracy.

For collinear DPDs, one can make a model ansatz consisting of two
terms which in the limit $y \ll 1/\Lambda$ respectively give the
perturbative splitting and the intrinsic part of the distribution.
With such an ansatz, the DPS cross section naturally splits into 1v1,
2v1 and 2v2 terms, where 2v2 refers to contributions in which the
parton pairs from both protons are intrinsic.  A crucial question is
how large DPS is compared with SPS at the perturbative order where
graphs contribute to both mechanisms.  This is especially acute in
collinear factorisation, where DPS is power suppressed with respect to
SPS.  Note that only in very few channels (notably pair production of
electroweak gauge bosons) SPS calculations are available at the
required order.  We argue in section~\ref{sec:dglap-subtr} that in our
scheme the variation with $\nu$ 
of the 1v1 term in DPS provides an order-of magnitude estimate for the
SPS contribution $\sigma_{\text{SPS}}$ (at the appropriate
perturbative order), as it involves the same PDFs, overall coupling
constants and kinematic region (small $\tvec{y}$, corresponding to
large transverse momenta and virtualities of internal lines).  An
alternative estimate is provided by the double counting subtraction
term $\sigma_{\text{1v1,pt}}$, which by construction is dominated by
small $\tvec{y}$.  For the hypothetical process of scalar boson pair
production from two gluons, we have shown that the latter estimate
works well within about a factor of two, provided that one stays away
from the extreme kinematic regions where the velocity $\beta$ of a
boson in the boson pair rest frame is close to $0$ or $1$.

We constructed explicit (collinear) DPD input forms using the model
ansatz just described, restricting ourselves to three quark flavours for
simplicity and ease of implementation.  These inputs were then
numerically evolved to other scales using a code that implements the
homogeneous double DGLAP equation. We used the resultant DPDs to
compute so-called DPS luminosities (DPS cross sections omitting the
process-dependent hard parts) and plotted these under various
conditions.  We observed that generically, the 1v1 contributions to
the luminosity (both unpolarised and polarised) are comparable to or
larger than the 2v1 and 2v2 contributions, with a large dependence on
the cutoff $\nu$.  This demonstrates that, when including DPS in a
cross section calculation, one must in general include
$\sigma_{\text{SPS}}$ up to the order that contains DPS-like double
box graphs, together with the associated subtraction term (with
unpolarised and polarised partons).  Otherwise, one would have an
uncertainty on the overall cross section that is as large as, or
larger than the DPS term itself.  We also confirmed that the $\nu$
variation of the 1v1 DPS contribution is indeed comparable to the
central value of the associated double counting subtraction term,
so that either of them may be used as an estimate for the SPS
contribution.

We identified several processes and scenarios where the $\nu$ variation of
the 1v1 DPS luminosity is considerably smaller than the central value.  As
we argued above, one may then justifiably neglect the order of
$\sigma_{\text{SPS}}$ containing the first DPS-like double box, as well as
the associated subtraction term, compared with $\sigma_{\text{DPS}}$.  One
scenario of this kind is when the flavour indices in both DPDs are
$u\bar{d}$ (the luminosity with this parton flavour combination appears in
$W^+W^+$ production).  The suppression of the DPS-like double box in
$\sigma_{\text{SPS}}$ is in this case related to the fact that
perturbative splitting in the $u\bar{d}$ DPD starts at order $\alpha_s^2$
rather than at order $\alpha_s^{}$.  Further scenarios in which the $\nu$
variation in the 1v1 contribution is reduced are when $\sqrt{s}$ becomes
very large compared to the hard scales $Q_i$, or when the rapidity
separation between the produced hard systems is large.  Both of these
scenarios involve small $x$ values in the DPDs --- in the first, both $x$
values in each DPD are small, whilst in the second, one $x$ value in each
DPD becomes much smaller than the other.  Such processes and kinematic
regions are the most promising ones to make useful calculations and
measurements for DPS.  In fact, several measurements investigating DPS
have already been made in kinematics with $Q_i \ll \sqrt{s}$ or with large
rapidity differences.  It will be interesting in future work to make more
complete and comprehensive predictions for such processes and kinematic
regions in our framework, including for instance the full flavour
dependence and contributions from all partonic channels for a considered
final state.

\appendix

\section{Fourier integrals}
\label{app:fourier}

In this appendix we collect a number of results for Fourier transforms
in $2$ or $2-2\epsilon$ di\-men\-sions.

\paragraph{Fourier transform of a fractional power.}  To compute the
Fourier transform of $(\tvec{k}^2)^{-\lambda}$ in $2-2\epsilon$ dimensions
we write $(\tvec{k}^{2})^{-\lambda} = \Gamma^{-1}(\lambda) \int_0^\infty
d\alpha\, \alpha^{\lambda-1} e^{-\alpha \tvec{k}^2}$.  One can then perform
the Gaussian integration over $\tvec{k}$, and subsequently the integral
over $\alpha$.  The result is
\begin{align}
  \label{power-FT}
\int d^{2-2\epsilon}\tvec{k}\; e^{-i \tvec{y}\tvec{k}}\,
  (\tvec{k}^2)^{-\lambda}
 &= \frac{\Gamma(1-\epsilon-\lambda)}{\Gamma(\lambda)}\;
    \pi^{1-\epsilon}\,
    \biggl( \frac{\tvec{y}^2}{4} \biggr)^{-1+\epsilon+\lambda}
\end{align}
As a corollary we obtain
\begin{align}
  \label{vector-FT}
\int d^{2-2\epsilon}\tvec{k}\; e^{i \tvec{y}\tvec{k}}\,
   \frac{\tvec{k}^j}{k^2}
 &= \frac{i}{2}\, \Gamma(1-\epsilon)\, (4\pi)^{1-\epsilon}\,
    \frac{\tvec{y}^j}{y^{2-2\epsilon}}
\end{align}
using $\tvec{k}^j\ms e^{i \tvec{y}\tvec{k}} =
-i \partial/(\partial \tvec{y}^j)\, e^{i \tvec{y}\tvec{k}}$.


\paragraph{Integrals involving $\Phi$.}  We now compute the integrals
in \eqref{tmd-sub-ints}, which are also needed for \eqref{dpd-mom-lo}
and \eqref{dpd-mom-hi}, If $\Phi$ is the step function $\Theta(u-b_0)$
then
\begin{align}
  \label{bessel-step}
2 \int_{b_0}^\infty \frac{du}{u}\, J_0(u r)
 &= \log\frac{1}{r^2} + \frac{b_0^2\, r^2}{4}\,
       {}_{2}F_{3} \biggl(1,1; 2,2,2; - \frac{b_0^2\, r^2}{4} \biggr) \,,
\end{align}
where $b_0$ is defined in \eqref{step-fct} and
${}_{2}F_{3}(1,1; 2,2,2;z)$ is a generalised hypergeometric function,
which can be expanded as $1 + \mathcal{O}(z)$ for small arguments.
For $\Phi(u) = 1 - e^{-u^2/4}$ we can proceed as follows:
\begin{align}
  \label{bessel-exp}
2 \int_0^\infty \frac{du}{u}\, J_0(u r)\ms \Bigl[ 1 - e^{-u^2/4} \Bigr]
&= 2 \int_0^\infty \frac{du}{u}\, J_0(u r)\, u^2
     \int_0^{1/4} d\tau\, e^{-\tau u^2}
 = \int_0^{1/4} \frac{d\tau}{\tau}\,
     \exp\biggl[ - \frac{r^2}{4 \tau} \biggr]
\nonumber \\[0.2em]
&= E_1(r^2) \,,
\end{align}
where $E_1$ is the exponential integral.  For small arguments one has
$E_1(r^2) = \log(1/r^2) - \gamma + \mathcal{O}(r^2)$.  The corresponding
integrals with $J_2$ instead of $J_0$ are given by
\begin{align}
  \label{bessel-2-step}
2 \int_{b_0}^\infty \frac{du}{u}\, J_2(u r)
 &= \frac{2 J_1(b_0\ms r)}{b_0\ms r}
\end{align}
and
\begin{align}
  \label{bessel-2-exp}
2 \int_0^\infty \frac{du}{u}\, J_2(u r)\ms \Bigl[ 1 - e^{-u^2/4} \Bigr]
&= \int_0^\infty du\; \frac{J_1(u r)}{u r}\, u e^{-u^2/4}
 = \frac{1 - e^{-r^2}}{r^2} \,,
\end{align}
where in the second case we have used integration by parts.
Both \eqref{bessel-2-step} and \eqref{bessel-2-exp} behave like $1
+ \mathcal{O}(r^2)$ for small $r$.


\paragraph{Connection between the Fourier-Bessel transform and a cutoff.}

Consider the integral \eqref{mom-dpd-spl} with one term of the
series \eqref{I-series}.  We will show that for $\Delta \ll \nu$
\begin{align}
   \label{must-show}
\int_{b_0/\nu}^{\infty} \frac{dy}{y}\, J_0(y \Delta)\,
   \log^{n} \biggl( \frac{y \nu}{b_0} \biggl)
 &= \int_{b_0/\nu}^{b_0/\Delta} \frac{dy}{y}\,
   \log^{n} \biggl( \frac{y \nu}{b_0} \biggl)
\nonumber \\
 &\quad
   + \mathcal{O}\biggl( \frac{\Delta^2}{\nu^2} \biggr)
   + \mathcal{O}\biggl( \log^{n-2} \frac{\nu}{\Delta} \biggl) \,.
\end{align}
The following argument is similar to the derivation given in appendix~B
of \cite{Bacchetta:2008xw}.
The integral on the r.h.s.\ is readily performed and gives
\begin{align}
    \label{int-rhs}
\int_{1}^{\nu/\Delta} dv\,
    \frac{1}{v} \log^{n} v
 &= \frac{1}{1+n} \log^{n+1} \biggl( \frac{\nu}{\Delta} \biggl) \,.
\end{align}
The expression on the l.h.s.\ of \eqref{must-show} can be rewritten as
\begin{align}
\int_{b_0 \Delta/\nu}^{\infty} \!\! du\, J_0(u)\,
    \frac{1}{u} \log^{n} \biggl( \frac{u \nu}{b_0 \Delta} \biggl)
 &= \frac{1}{n+1} \int_{b_0 \Delta/\nu}^{\infty} \!\! du\, J_1(u)
          \log^{n+1} \biggl( \frac{u \nu}{b_0 \Delta} \biggl)
    + \mathcal{O}\biggl( \frac{\Delta^2}{\nu^2} \biggr)
\end{align}
using integration by parts.  Since $J_1(u) = \mathcal{O}(u)$ at small
$u$, the integral on the r.h.s.\ can be extended down to $u=0$ with an
accuracy of $\Delta^2/ \nu^2$.  Rewriting the logarithm using the
binomial series, we obtain
\begin{align}
& \frac{1}{n+1} \sum_{k=0}^{n+1} \binom{n+1}{k}\,
    \log^{n+1-k} \biggl( \frac{\nu}{\Delta}  \biggl)\;
    \int_{0}^{\infty} du\, J_1(u)\, \log^{k} \biggl( \frac{u}{b_0} \biggl)
\nonumber \\
& \qquad = \frac{1}{n+1} \Biggl[\,
        \log^{n+1} \biggl( \frac{\nu}{\Delta} \biggl)
    + \sum_{k=3}^{n+1} \binom{n+1}{k}\, d_k\,
      \log^{n+1-k} \biggl( \frac{\nu}{\Delta} \biggl) \,\Biggr] \,.
\end{align}
In the last step we performed the integrals for $k=0,1,2$ explicitly
(the ones for $k=1,2$ are zero) and replaced the remaining ones by
coefficients $d_k$ whose values are not important here.  Comparison
with \eqref{int-rhs} gives the desired result~\eqref{must-show}.


\acknowledgments

We gratefully acknowledge discussions with V.~Braun, K.~Golec-Biernat,
D.~Boer, A.~Sch\"afer and F.~Tackmann.  Special thanks are due to
T.~Kasemets for valuable remarks on the manuscript.  J.G.\ acknowledges
financial support from the European Community under the FP7 Ideas program
QWORK (contract 320389).  Two of us (M.D.\ and J.G.) thank the Erwin
Schr{\"o}dinger International Institute for Mathematics and Physics (ESI)
for hospitality during the programme ``Challenges and Concepts for Field
Theory and Applications in the Era of LHC Run-2'', when portions of this
work were completed.  The Feynman diagrams in this paper were produced
with JaxoDraw~\cite{Binosi:2003yf,Binosi:2008ig}.


\phantomsection
\addcontentsline{toc}{section}{References}

\bibliographystyle{JHEP}
\bibliography{double}

\end{document}